# THE INTERNATIONAL LINEAR COLLIDER

Technical Design Report | Volume 1: Executive Summary

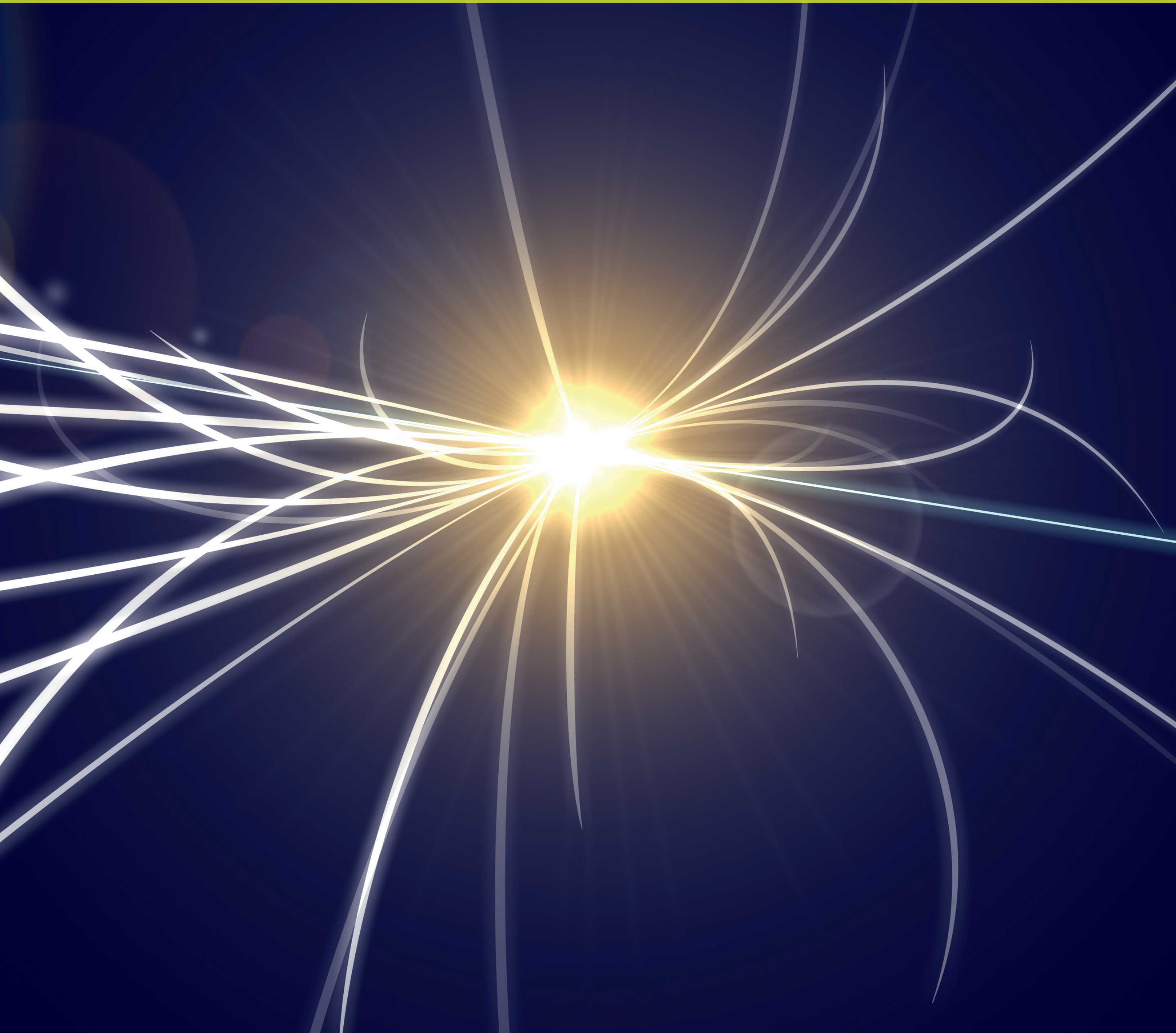

# The International Linear Collider

# Technical Design Report

2013

international linear collider

Volume 1: Executive Summary

Volume 2: Physics

Volume 3: Accelerator

Part I: R&D in the Technical Design Phase

Volume 3: Accelerator

Part II: Baseline Design

Volume 4: Detectors

# Volume 1

# Executive Summary

Editors

Ties Behnke, James E. Brau, Brian Foster, Juan Fuster, Mike Harrison, James McEwan Paterson, Michael Peskin, Marcel Stanitzki, Nicholas Walker, Hitoshi Yamamoto

# Acknowledgements

We acknowledge the support of BMWF, Austria; MinObr, Belarus; FNRS and FWO, Belgium; NSERC, Canada; NSFC, China; MPO CR and VSC CR, Czech Republic; Commission of the European Communities; HIP, Finland; IN2P3/CNRS, CEA-DSM/IRFU, France; BMBF, DFG, Helmholtz Association, MPG and AvH Foundation, Germany; DAE and DST, India; ISF, Israel; INFN, Italy; MEXT and JSPS, Japan; CRI(MST) and MOST/KOSEF, Korea; FOM and NWO, The Netherlands; NFR, Norway; MNSW, Poland; ANCS, Romania; MES of Russia and ROSATOM, Russian Federation; MON, Serbia and Montenegro; MSSR, Slovakia; MICINN-MINECO and CPAN, Spain; SRC, Sweden; STFC, United Kingdom; DOE and NSF, United States of America.

# Contents

# Chapter 1
# Introduction

The International Linear Collider (ILC) is a 200–500 GeV (extendable to 1 TeV) centre-of-mass high-luminosity linear electron-positron collider, based on 1.3 GHz superconducting radio-frequency (SCRF) accelerating technology. Its parameters have been set by physics requirements first outlined in 2003, updated in 2006, and thoroughly discussed over many years with the physics user community. The physics parameters have been reviewed continuously and found to be robust to advances in the science, including the recent discovery of a Higgs boson at the Large Hadron Collider at CERN.

The collider design is the result of nearly twenty years of R&D. The heart of the ILC, the superconducting cavities, is based on over a decade of pioneering work by the TESLA collaboration in the 1990s. Some other aspects were based on the R&D carried out for the JLC/GLC and NLC projects, which were based on room-temperature accelerating structures. Since 2005, the design of the ILC accelerator has continued as a worldwide international collaboration coordinated by the Global Design Effort (GDE) under a mandate from the International Committee for Future Accelerators (ICFA). Drawing on the resources of over 300 national laboratories, universities and institutes worldwide, the GDE produced the ILC *Reference Design Report* (RDR) in August 2007. The work done by the GDE during the RDR phase identified several high-risk challenges that required R&D, which have since been the focus of the worldwide activity during the Technical Design Phase. In parallel with the accelerator effort, detailed baseline designs of two detectors have been developed by large international teams as a result of intense detector R&D under the coordination of the Research Directorate, also established by a mandate of ICFA.

These efforts have now successfully culminated in the publication of this *Technical Design Report*, completing the mandates of both the GDE and the Research Directorate. The report describes a conceptual design for the ILC project that reflects an international consensus.

This Executive Summary first discusses the physics goals of the ILC. It then addresses the design and the associated value cost estimate of the accelerator, followed by the R&D results which underpin its technical design. It concludes with a discussion of the ILC detector design and development outcomes and their cost estimates.

# Chapter 2
# Physics at the International Linear Collider

Today the search for new particles and forces at energies of hundreds or thousands of GeV plays a central role in the field of elementary particle physics. Particle physicists have established a "Standard Model" for the strong, weak, and electromagnetic interactions that passes tests at both low and high energies. The model is extremely successful, and yet it is incomplete in many important respects. New particles and interactions are needed to fill the gaps.

Some of the difficulties of the Standard Model are deep and abstract; their explanations may be found only in the distant future. The Standard Model does not explain how gravity is connected to the other forces of nature. It does not explain why the basic particles of matter are the quarks and leptons, or how many of these there should be.

However, the Standard Model also fails to explain three phenomena that, by rights, should be accounted for at the energies now being probed with particle accelerators. Astronomers believe that the dominant form of matter in the universe is a neutral, weakly interacting species called "dark matter" that cannot be composed of any particle present in the Standard Model. The Standard Model cannot explain why the universe contains atomic matter made of electrons, protons and neutrons but no comparable amount of antimatter.

Behind these two striking problems, there is a third. The equations of the Standard Model are based on a symmetry principle, electroweak symmetry, that forbids the generation of mass for any of its fundamental particles. The universe contains an element that chooses a direction with respect to this symmetry. This asymmetrical force creates the masses of the quarks, leptons, and bosons of the Standard Model and also drives many other essential properties of the laws of nature. The Standard Model postulates a field, called the *Higgs field*, that gives rise to this force. However, it does not explain the properties of this field. The idea that the asymmetry comes from a single Higgs field is just a guess among many other possibilities.

The problem of the Higgs field is likely to be connected to the earlier questions about the matter content of the universe. Explanatory models of the Higgs field often contain particles with the correct properties to make up the dark matter. There are also strong, independent, arguments that the mass of the dark-matter particle is comparable to the masses—of the order of 100 GeV—of the heaviest particles that receive mass from the Higgs field. The excess of baryons over antibaryons in the universe could arise from interactions among Higgs fields that violate the space-time symmetry $CP$. More generally, any model of fundamental physics at energies above 100 GeV must contain the Higgs field or some generalisation and must account for the place of this field within its structure.

A way to prove the existence of the Higgs field and to study its interactions is to find and study the quantum of this field, called the *Higgs boson*. The International Linear Collider was designed to study this particle and other new particles that might be associated with it. It provides an ideal setting for detailed exploration of the origin and nature of the Higgs field. The physics program of the ILC is described in detail in the Physics Volume, Volume 2 of this report.

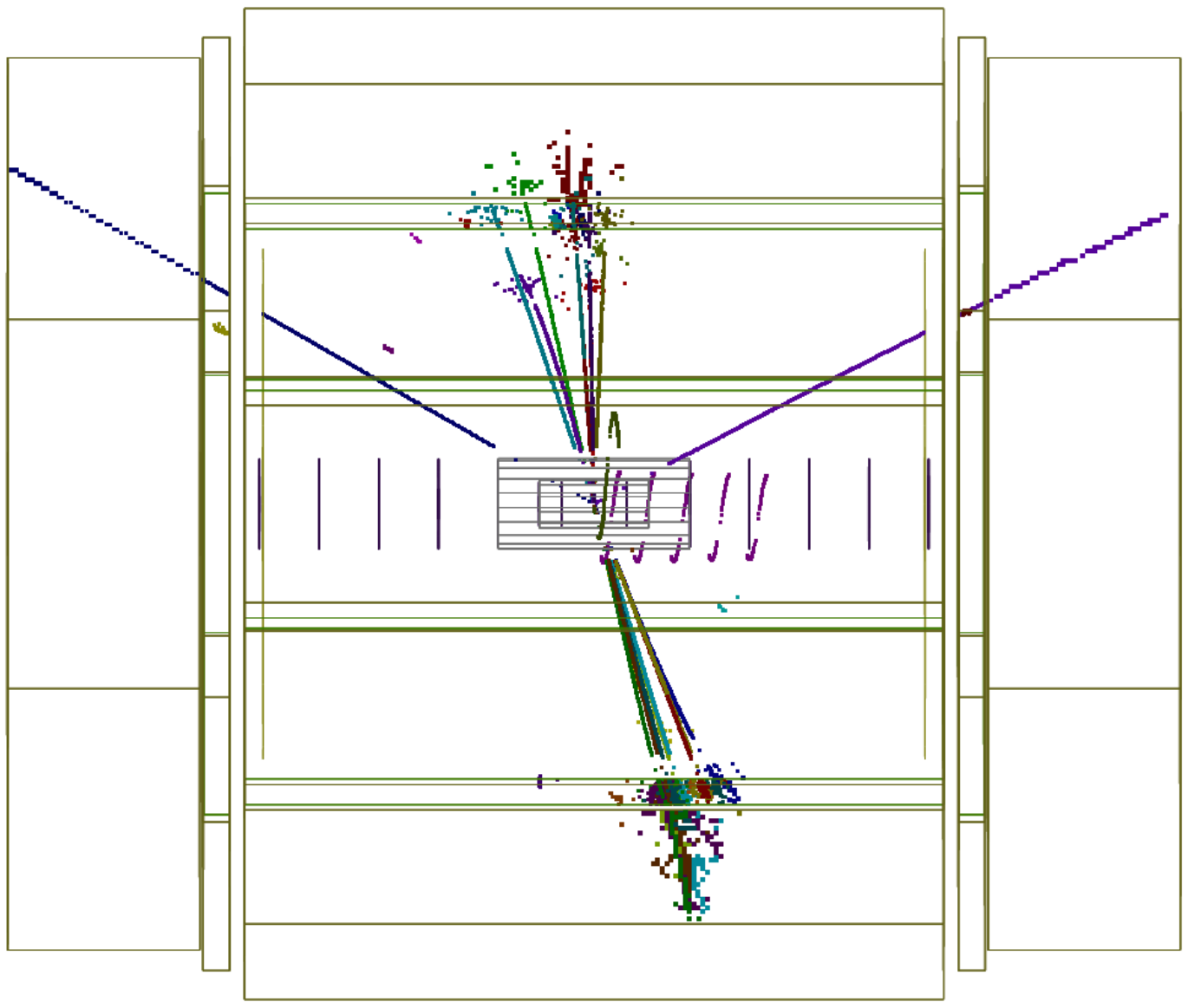

**Figure 2.1**
An event of reaction $e^+e^- \to Zh$, with $Z \to \mu^+\mu^-$, $h \to b\bar{b}$, as it would be observed in the ILD detector at the ILC.

In July, 2012, the ATLAS and CMS experiments at the CERN Large Hadron Collider announced the discovery of a new particle with a mass of 125 GeV and many properties of the Higgs boson as postulated in the Standard Model. The LHC experiments also exclude the possibility that the Higgs boson has higher mass, up to masses beyond 600 GeV and close to the theoretical upper bound. The ILC is an ideal machine to study this Higgs particle at 125 GeV. If this particle is one of several Higgs bosons and a different boson is the one that makes the main contributions to the W and Z boson masses, that particle must also appear at the ILC. Thus, the ILC definitely offers a direct path to the study of the Higgs field and its implications for particle physics.

The initial program of the ILC for a 125 GeV Higgs boson will be centred at an energy of 250 GeV, which gives the peak cross section for the reaction $e^+e^- \to Zh$. In this reaction, the identification of a $Z$ boson at the energy appropriate to recoil against the Higgs boson tags the presence of the Higgs boson. In this setting, it is possible to measure the rates for all decays of the Higgs boson—even decays to invisible or unusual final states—with high precision. An example of an event with a Higgs boson decay to quarks is shown in Fig. 2.1. Such decays are very difficult to separate from Standard Model background events at the LHC. The precision measurement of the rates of decay of the Higgs boson to the various types of quarks, leptons, and bosons will give evidence on whether the Higgs field operates alone to create the masses of these particles, or whether it has partners that are additional new particles addressing the other questions raised above.

The study of the Higgs boson will continue, with additional essential elements, at higher energies. At 500 GeV, the full design energy of the ILC, measurement of the process $e^+e^- \to \nu\bar{\nu}h$ will give the absolute normalisation of the underlying Higgs coupling strengths, needed to determine the individual couplings to the percent level of accuracy. Raising the energy further allows the ILC experiments to make precise measurements of the Higgs boson coupling to top quarks and to determine the strength of the Higgs boson's nonlinear self-interaction.

The ILC also will make important contributions to the search for new particles associated with the Higgs field, dark matter, and other questions of particle physics. For many such particles with only electroweak interactions, searches at the LHC will be limited by low rates relative to strong-interaction induced processes, and by large backgrounds. The ILC will identify or exclude these particles unambiguously up to masses at least as high as the ILC beam energy. The Physics Volume discusses a variety of models for new physics in which such particles are required and illustrates the crucial role of the ILC for these models. For example, many models contain partners of the Higgs boson. The very attractive theory of supersymmetry, a symmetry between matter and force-carrying particles, requires matter-type Higgs particles. These particles are difficult to detect at the LHC because of their expected low energy release and low production rate. The ILC, however, can search for these particles in a direct and definitive way. If new particles are discovered, either at the LHC or at the ILC, the ILC will be able to measured their quantum numbers unambiguously and determine their couplings to the percent level. The motivations for this study include exploration of the question of whether a new neutral, weakly interacting particle discovered at particle accelerators actually might be the constituent of the dark matter of the universe.

The ILC will also constrain or discover new interactions at higher mass scales through pair-production of quarks and leptons, $W$ and $Z$ bosons, and top quarks. Much of our detailed knowledge of the current Standard Model comes from the precision measurement of the properties of the $Z$ boson at $\mathrm{e}^+\mathrm{e}^-$ colliders. The ILC will extend this level of precision to the $W$ boson and the top quark. The ILC will measure the mass of the top quark in a direct way that is not possible at hadron colliders, fixing a crucial input to particle physics calculations. The top quark is the heaviest particle of the Standard Model, and, as such, must have especially strong coupling to the Higgs field. The precision study of the electroweak couplings of the top quark can reveal the presence of composite structure in the Higgs particle. Characteristic effects are expected in models with strong interactions among the Higgs fields, and in models where the asymmetrical forces associated with the Higgs fields are signs of extra, hidden dimensions of space.

**Table 2.1**
Major physics processes to be studied by the ILC at various energies. The table indicates the various Standard Model reactions that will be accessed at increasing collider energies, and the major physics goals of the study of these reactions. A reaction listed at a given energy will of course be studied at all higher energies.

| Energy | Reaction | Physics Goal |
|---|---|---|
| 91 GeV | $\mathrm{e}^+\mathrm{e}^- \to Z$ | ultra-precision electroweak |
| 160 GeV | $\mathrm{e}^+\mathrm{e}^- \to WW$ | ultra-precision $W$ mass |
| 250 GeV | $\mathrm{e}^+\mathrm{e}^- \to Zh$ | precision Higgs couplings |
| 350–400 GeV | $\mathrm{e}^+\mathrm{e}^- \to t\bar{t}$ | top quark mass and couplings |
| | $\mathrm{e}^+\mathrm{e}^- \to WW$ | precision $W$ couplings |
| | $e^+e^- \to \nu\bar{\nu}h$ | precision Higgs couplings |
| 500 GeV | $\mathrm{e}^+\mathrm{e}^- \to f\bar{f}$ | precision search for $Z'$ |
| | $\mathrm{e}^+\mathrm{e}^- \to t\bar{t}h$ | Higgs coupling to top |
| | $\mathrm{e}^+\mathrm{e}^- \to Zhh$ | Higgs self-coupling |
| | $\mathrm{e}^+\mathrm{e}^- \to \tilde{\chi}\tilde{\chi}$ | search for supersymmetry |
| | $\mathrm{e}^+\mathrm{e}^- \to AH, H^+H^-$ | search for extended Higgs states |
| 700–1000 GeV | $\mathrm{e}^+\mathrm{e}^- \to \nu\bar{\nu}hh$ | Higgs self-coupling |
| | $\mathrm{e}^+\mathrm{e}^- \to \nu\bar{\nu}VV$ | composite Higgs sector |
| | $\mathrm{e}^+\mathrm{e}^- \to \nu\bar{\nu}t\bar{t}$ | composite Higgs and top |
| | $\mathrm{e}^+\mathrm{e}^- \to \tilde{t}\tilde{t}^*$ | search for supersymmetry |

One of the advantages of a linear collider is its ability to operate, with minimal modification, at any energy within a wide range that its technology makes available. Table 2.1 lists the major reactions that will be studied at the ILC in the various stages of its program, with collider energies from 90 GeV to 1000 GeV. These include Higgs boson reactions, studies of pair production, and searches for new particles, as described in the previous paragraphs. A linear collider also makes it possible to collide electron and positron beams with high spin polarisation. This makes available many new observables that cannot be measured at colliders of other types. Because the Standard Model violates parity in a maximal way, the quantum numbers of new particles and the full set of couplings for $W$, $t$, and other known particles can only be measured unambiguously through the use of these spin observables. In the Physics Volume, the key role of beam polarisation in the ILC measurements is explained for each of the topics discussed.

There is another advantage of a linear electron-positron collider over a proton-proton collider such as the LHC. The total rate of high-energy collisions and the complexity of these collisions are both much lower at the electron-positron accelerator. This allows the design of particle detectors with higher intrinsic resolution than those at proton colliders, and with less obscuring material between the particle collision point and the sensitive elements. At an electron-positron collider, silicon sensors can be brought within 2 cm of the interaction point to produce a high-precision image of the particle tracks emerging from the collision. This enables very efficient identification of short-lived particles such as the heavy quarks and leptons $b$, $c$, and $\tau$. The proposed designs of the detectors for the ILC, and the evaluation of the capabilities of these detectors in relation to the goals of the ILC physics program, is described in the Detector Volume, Volume 4 of this report.

**Table 2.2.** Key precision Standard Model measurements that will be achieved with the ILC. Estimates for Higgs assume $m_{\rm h} = 125$ GeV. These analyses require no significant model-dependent assumptions. ILC will also search systematically for all particles with electroweak couplings up to the masses close to the $\rm e^+$ and $\rm e^-$ beam energies. The last section of the table shows the precision that the ILC can achieve in measuring properties of some illustrative new particles that might be discovered at the ILC. Measurement accuracies are quoted for ILC event samples of 250 fb$^{-1}$ at 250 GeV, 500 fb$^{-1}$ at 500 GeV, and 1000 fb$^{-1}$ at 1000 GeV, with electron/positron polarization of 80%/30% in the first two cases and 80%/20% in the third case. Dedicated threshold scans require less luminosity at energies specifically chosen for the task.

| Topic | Parameter | Accuracy $\Delta X/X$ | |
|---|---|---|---|
| Higgs | $m_{\rm h}$ | 0.03% | $\Delta m_{\rm h} = 35$ MeV, 250 GeV |
| | $\Gamma_{\rm h}$ | 1.6% | 250 GeV and 500 GeV |
| | $g(hWW)$ | 0.24% | |
| | $g(hZZ)$ | 0.30% | |
| | $g(hb\bar{b})$ | 0.94% | |
| | $g(hc\bar{c})$ | 2.5% | |
| | $g(hgg)$ | 2.0% | |
| | $g(h\tau^+\tau^-)$ | 1.9% | |
| | $BR(h \to \text{invis.})$ | < 0.30% (95% conf.) | |
| | $g(ht\bar{t})$ | 3.7% | 1000 GeV |
| | $g(hhh)$ | 26% | |
| | $g(h\mu^+\mu^-)$ | 16% | |
| Top | $m_{\rm t}$ | 0.02% | $\Delta m_{\rm t} = 34$ MeV, threshold scan |
| | $\Gamma_{\rm t}$ | 2.4% | |
| | $\tilde{F}^{\gamma}_{\rm 1V}$ | 0.2% | 500 GeV |
| | $\tilde{F}^{\rm Z}_{\rm 1V}$ | 0.3% | |
| | $\tilde{F}^{\rm Z}_{\rm 1A}$ | 0.5% | |
| | $\tilde{F}^{\gamma}_{\rm 2V}$ | 0.3% | |
| | $\tilde{F}^{\rm Z}_{\rm 2V}$ | 0.6% | |
| $W$ | $m_{\rm W}$ | 0.004% | $\Delta m_{\rm W} = 3$ MeV, threshold scan |
| | $g_1$ | 0.16% | 500 GeV |
| | $\kappa_\gamma$ | 0.03% | |
| | $\kappa_{\rm Z}$ | 0.03% | |
| | $\lambda_\gamma$ | 0.06% | |
| | $\lambda_{\rm Z}$ | 0.07% | |
| $H^0$, $A^0$ | $m_{\rm H}$, $m_{\rm A}$ | 1.5% | |
| | $\tan\beta$ | 20% | |
| $\widetilde{\chi}^+$ | $m(\widetilde{\chi}^+)$ | 1% | |
| | $m(\widetilde{\chi}^0)$ | 1% | |
| $\widetilde{t}$ | $m(\widetilde{t})$ | 1% | |
| | $\cos\theta_{\rm t}$ | 0.4% | |

Table 2.2 gives an illustrative list of quantities that can be measured at ILC, and the measurement accuracies that the ILC experiments expect to achieve. Any unfamiliar terminology in this table is defined in the appropriate chapter of the Physics Volume. The ILC study of the Higgs boson gives the complete profile of Higgs boson couplings. The ILC studies of the top quark and the $W$ boson probe the electroweak couplings of these heavy elementary particles with measurement accuracies qualitatively superior to those expected from the LHC. As shown in the Physics Volume, measurements at this level probe the effects expected from the coupling of top quarks and $W$ bosons to composite and strongly interacting Higgs fields.

The final elements of the table give some examples of the ability of the ILC to follow up the discovery of new particles with precision measurements of their properties. In experiments at hadron colliders, mixing angles, such as $\tan\beta$ for Higgs bosons and $\cos\theta_{\mathrm{t}}$ for supersymmetric partners of top quarks, are measured indirectly through multi-variable fits that depend on detailed properties of the model under study. In contrast, the ILC measurements are unambiguously interpreted and lead to model-independent determinations of the parameters. These mixing angles appear ubiquitously in the predictions of extended Higgs and supersymmetry models. Through these measurements, the ILC will not only produce its own discoveries but also will enable sharper and more informative analysis of the data collected at the LHC.

The ILC then offers many opportunities for measurements that will address the most important current problems of particle physics. It will give unique views of the Higgs boson, the top quark, and possible new particles relevant to the mysteries of the matter content of the universe. The collider enables incisive measurements of very high precision. The ILC is thus an essential tool that will advance our understanding of the basic laws of nature.

# Chapter 3
# The International Linear Collider Accelerator

## 3.1 The ILC Technical Design

### 3.1.1 Overview

The International Linear Collider (ILC) is a high-luminosity linear electron-positron collider based on 1.3 GHz superconducting radio-frequency (SCRF) accelerating technology. Its centre-of-mass-energy range is 200–500 GeV (extendable to 1 TeV). A schematic view of the accelerator complex, indicating the location of the major sub-systems, is shown in Fig. 3.1:

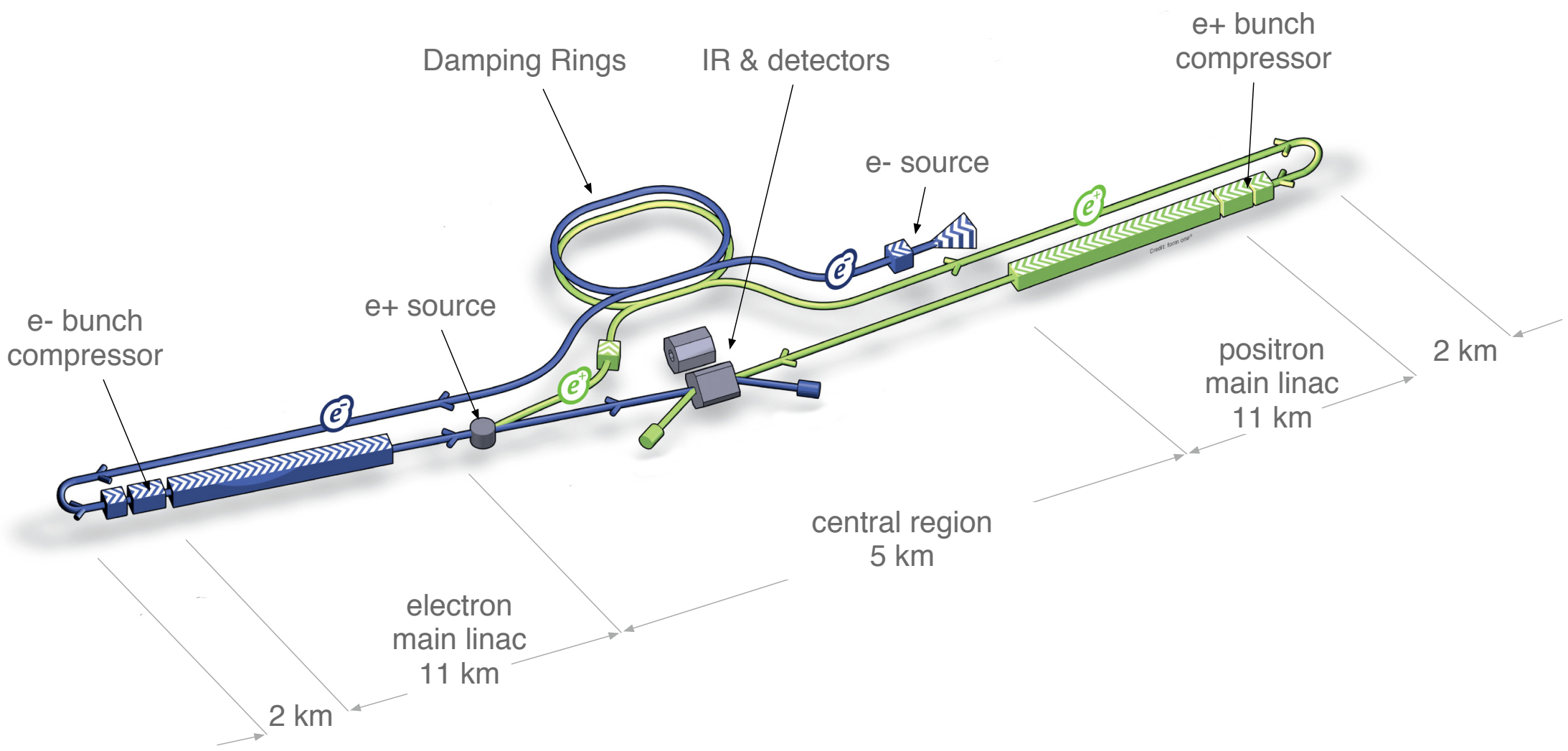

**Figure 3.1.** Schematic layout of the ILC, indicating all the major subsystems (not to scale).

- a polarised electron source based on a photocathode DC gun;
- a polarised positron source in which positrons are obtained from electron-positron pairs by converting high-energy photons produced by passing the high-energy main electron beam through an undulator;
- 5 GeV electron and positron damping rings (DR) with a circumference of 3.2 km, housed in a common tunnel;
- beam transport from the damping rings to the main linacs, followed by a two-stage bunch-compressor system prior to injection into the main linac;
- two 11 km main linacs, utilising 1.3 GHz SCRF cavities operating at an average gradient of 31.5 MV/m, with a pulse length of 1.6 ms;

- two beam-delivery systems, each 2.2 km long, which bring the beams into collision with a 14 mrad crossing angle, at a single interaction point which can be occupied by two detectors in a so-called "push-pull" configuration.

The total footprint of the ILC complex is $\sim$ 31 km long. The electron source, positron source (including an independent low-powered auxiliary source), and the electron and positron damping rings are centrally located around the interaction region (IR) in the Central Region. The damping-ring complex is displaced laterally to avoid interference with the detector hall. The electron and positron sources themselves are housed in the same (main accelerator) tunnels as the beam-delivery systems, which reduces the overall cost and size of the central-region underground construction.

### 3.1.2 Machine parameters

The top-level parameters for the baseline operational range of centre-of-mass energies from 250 to 1000 GeV were set in close discussion with the physics community that will exploit the ILC. The baseline performance requirements thus obtained have been optimised with respect to cost, physics performance and risk. All have been either directly demonstrated, or represent justifiable extrapolations from the current state of the art. Table 3.1 shows the parameters for several centre-of-mass energies, including possible upgrades and staging.

The parameters in Table 3.1 represent relatively conservative operating points resulting from optimisation subject to the constraints imposed by the various accelerator sub-systems. For example, the bunch charge, bunch spacing and the total number of bunches in the damping rings are limited by various instability thresholds (most notably the electron cloud in the positron ring), realistic rise-times for the injection and extraction kickers, and the desire to minimise the circumference of the rings. Secondly, the maximum length of the beam pulse is constrained to $\sim$ 1.6 ms, which is routinely achieved in the available 1.3 GHz 10 MW multi-beam klystrons and modulators. The beam current is further constrained by the need to minimise the number of klystrons (peak power) and higher-order modes (cryogenic load and beam dynamics). Dynamic cryogenic load (refrigeration) is also a cost driver, which limits the repetition rate of the machine. Thirdly, both the electron and positron sources constrain the achievable beam current and total charge: For the laser-driven photocathode polarised electron source, the limits are set by the laser; for the undulator-based positron source, the limits are set by the power deposition in the photon target. The beam pulse length is further constrained by the achievable performance of the warm RF capture sections (both sources). Finally, at the interaction point, single-bunch parameters are limited by the strong beam-beam effects and requirements on both the beam-beam backgrounds and beam stability.

**Table 3.1.** Summary table of the 250–500 GeV baseline and luminosity and energy upgrade parameters. Also included is a possible 1st stage 250 GeV parameter set (half the original main linac length)

| | | | Baseline 500 GeV Machine | | | 1st Stage | L Upgrade | $E_{\mathrm{CM}}$ Upgrade | |
|---|---|---|---|---|---|---|---|---|---|
| | | | | | | | | A | B |
| Centre-of-mass energy | $E_{\mathrm{CM}}$ | GeV | 250 | 350 | 500 | 250 | 500 | 1000 | 1000 |
| Collision rate | $f_{\mathrm{rep}}$ | Hz | 5 | 5 | 5 | 5 | 5 | 4 | 4 |
| Electron linac rate | $f_{\mathrm{linac}}$ | Hz | 10 | 5 | 5 | 10 | 5 | 4 | 4 |
| Number of bunches | $n_{\mathrm{b}}$ | | 1312 | 1312 | 1312 | 1312 | 2625 | 2450 | 2450 |
| Bunch population | $N$ | $\times 10^{10}$ | 2.0 | 2.0 | 2.0 | 2.0 | 2.0 | 1.74 | 1.74 |
| Bunch separation | $\Delta t_{\mathrm{b}}$ | ns | 554 | 554 | 554 | 554 | 366 | 366 | 366 |
| Pulse current | $I_{\mathrm{beam}}$ | mA | 5.8 | 5.8 | 5.8 | 5.8 | 8.8 | 7.6 | 7.6 |
| Main linac average gradient | $G_{\mathrm{a}}$ | MV m$^{-1}$ | 14.7 | 21.4 | 31.5 | 31.5 | 31.5 | 38.2 | 39.2 |
| Average total beam power | $P_{\mathrm{beam}}$ | MW | 5.9 | 7.3 | 10.5 | 5.9 | 21.0 | 27.2 | 27.2 |
| Estimated AC power | $P_{\mathrm{AC}}$ | MW | 122 | 121 | 163 | 129 | 204 | 300 | 300 |
| RMS bunch length | $\sigma_{\mathrm{z}}$ | mm | 0.3 | 0.3 | 0.3 | 0.3 | 0.3 | 0.250 | 0.225 |
| Electron RMS energy spread | $\Delta p/p$ | % | 0.190 | 0.158 | 0.124 | 0.190 | 0.124 | 0.083 | 0.085 |
| Positron RMS energy spread | $\Delta p/p$ | % | 0.152 | 0.100 | 0.070 | 0.152 | 0.070 | 0.043 | 0.047 |
| Electron polarisation | $P_{-}$ | % | 80 | 80 | 80 | 80 | 80 | 80 | 80 |
| Positron polarisation | $P_{+}$ | % | 30 | 30 | 30 | 30 | 30 | 20 | 20 |
| Horizontal emittance | $\gamma\epsilon_{\mathrm{x}}$ | µm | 10 | 10 | 10 | 10 | 10 | 10 | 10 |
| Vertical emittance | $\gamma\epsilon_{\mathrm{y}}$ | nm | 35 | 35 | 35 | 35 | 35 | 30 | 30 |
| IP horizontal beta function | $\beta^*_{\mathrm{x}}$ | mm | 13.0 | 16.0 | 11.0 | 13.0 | 11.0 | 22.6 | 11.0 |
| IP vertical beta function | $\beta^*_{\mathrm{y}}$ | mm | 0.41 | 0.34 | 0.48 | 0.41 | 0.48 | 0.25 | 0.23 |
| IP RMS horizontal beam size | $\sigma^*_{\mathrm{x}}$ | nm | 729.0 | 683.5 | 474 | 729 | 474 | 481 | 335 |
| IP RMS veritcal beam size | $\sigma^*_{\mathrm{y}}$ | nm | 7.7 | 5.9 | 5.9 | 7.7 | 5.9 | 2.8 | 2.7 |
| Luminosity | $L$ | $\times 10^{34}$ cm$^{-2}$s$^{-1}$ | 0.75 | 1.0 | 1.8 | 0.75 | 3.6 | 3.6 | 4.9 |
| Fraction of luminosity in top 1% | $L_{0.01}/L$ | | 87.1% | 77.4% | 58.3% | 87.1% | 58.3% | 59.2% | 44.5% |
| Average energy loss | $\delta_{\mathrm{BS}}$ | | 0.97% | 1.9% | 4.5% | 0.97% | 4.5% | 5.6% | 10.5% |
| Number of pairs per bunch crossing | $N_{\mathrm{pairs}}$ | $\times 10^3$ | 62.4 | 93.6 | 139.0 | 62.4 | 139.0 | 200.5 | 382.6 |
| Total pair energy per bunch crossing | $E_{\mathrm{pairs}}$ | TeV | 46.5 | 115.0 | 344.1 | 46.5 | 344.1 | 1338.0 | 3441.0 |

## 3.2 Accelerator Layout & Design

### 3.2.1 Superconducting RF Main Linacs

The ILC Main Linacs accelerate the beam from 15 GeV (after acceleration in the upstream bunch compressors) to a maximum energy of 250 GeV. Beam acceleration in each linac is provided by approximately 7,400 1 m-long superconducting nine-cell niobium cavities (see Fig. 3.2) operating at 2 K, assembled into $\sim$ 850 cryomodules. The average accelerating gradient of the cavities is 31.5 MV/m (for 500 GeV centre-of-mass beam energy), with a corresponding quality factor $Q_0 \geq 10^{10}$. A random cavity-to-cavity gradient spread of $\pm$20% is tolerated to accommodate expected mass-production variations in the maximum achievable gradient. The choice of these key parameters is the result of over a decade of extensive R&D. The GDE recognised the need to establish expertise in this technology in all three regions of the world and established the high-gradient programme as its highest priority during the Technical Design Phase. As a result, extensive worldwide experience both in the labs and in industry now gives high confidence that these requirements can be routinely achieved. (See Section 3.3.1 for details.)

**Figure 3.2**
A 1.3 GHz superconducting niobium nine-cell cavity.

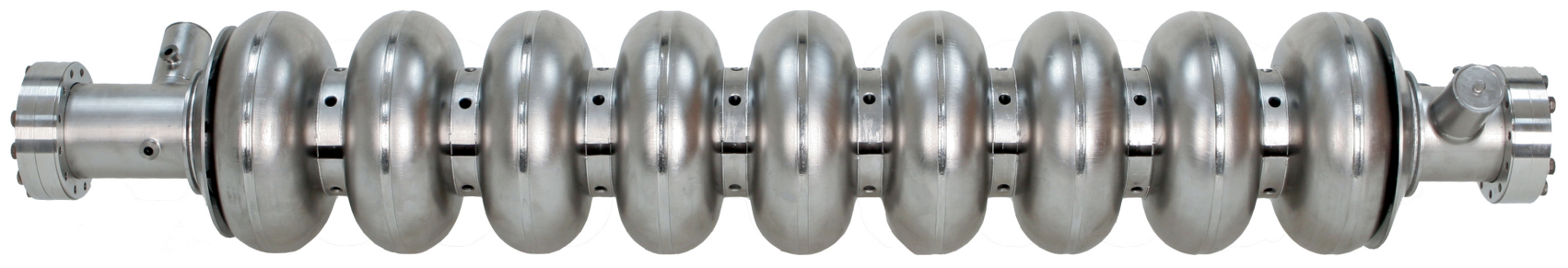

For an average of 31.5 MV/m operation with the nominal beam current of 5.8 mA, the optimal matched $Q_{\mathrm{L}}$ is $\sim 5.4\times10^6$. This corresponds to a cavity fill time of 925 μs, which, added to the nominal beam pulse width of 727 μs, gives a total RF pulse length of 1.65 ms.

The cavity package includes the cavity mechanical tuner, which is integrated into the titanium helium vessel of the cavity, and an adjustable high-power coupler. In addition to a slow mechanical tuner (used for initial tuning and slow drift compensation), a fast piezo-driven tuner is also included to compensate dynamically for the mechanical deformation of the cavity caused by the RF pulse, known as "Lorentz-force detuning".

The cryomodules (Fig. 3.3) that make up the Main Linacs are 12.65 m long. There are two types: a Type A module with nine 1.3 GHz nine-cell cavities and Type B with eight nine-cell cavities and one superconducting quadrupole package located at the centre of the module.

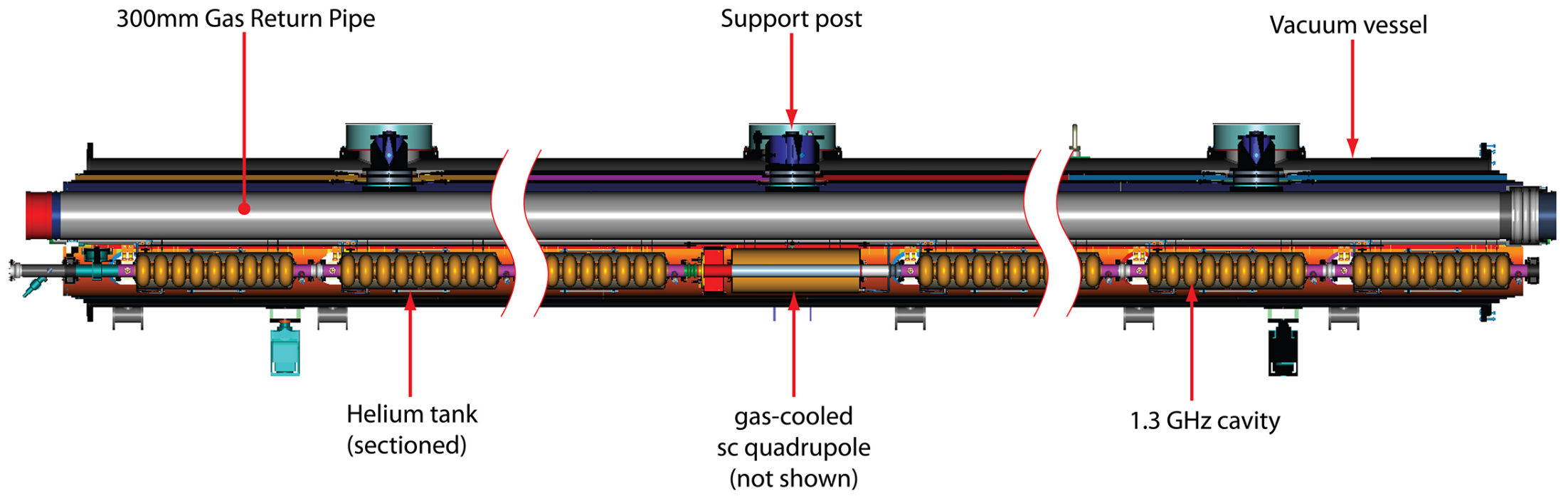

**Figure 3.3.** Longitudinal cross section of an ILC cryomodule (Type B).

The cryomodule design is a modification of the Type-3 version developed and used at DESY in the FLASH accelerator as well as the 100 cryomodules currently being produced by industry for the European X-Ray FEL (XFEL), also based at DESY. A 300 mm-diameter helium-gas return pipe serves as the primary support for the nine cavities and other beamline components in the Type A module. For Type B, the central cavity package is replaced by a superconducting quadrupole package that includes the quadrupole itself, a cavity BPM, and superconducting horizontal- and vertical-corrector dipole magnets. The quadrupoles establish the magnetic lattice for the Main Linac, which is a weak-focusing FODO optics with an average beta function of $\sim$ 80 m. Every cryomodule also contains a 300 mm-long assembly that removes energy from beam-induced higher-order modes above the cavity cut-off frequency through the 40–80 K cooling system.

To operate the cavities at 2 K, they are immersed in a saturated He-II bath. Shields cooled with Helium gas intercept thermal radiation and provide a heat sink for conduction at 5–8 K and at 40–80 K. Each cryomodule has an estimated 2 K static cryogenic heat load of 1.3 W while the 2 K dynamic heat load is approximately 9.8 W. Liquid helium for the main linacs and the bunch compressor RF is supplied from a total of 10-12 large cryogenic plants, each of which has an installed equivalent cooling power of $\sim$ 20 kW at 4.5 K. The plants are located in pairs separated by 5 km along the linacs, with each plant cooling $\sim$ 2.5 km of continuous linac. The main linacs follow the Earth's average curvature to simplify the liquid-helium transport and tunnel construction.

The RF power is provided by 10 MW multi-beam klystrons (MBK) each driven by a 120 kV Marx modulator. The 10 MW MBK has achieved the ILC specifications and is now a well established technology with several vendors worldwide. The 120 kV Marx-modulator prototypes have achieved the required specifications and are now undergoing design optimisation for transfer to industrial vendors. (See Section 3.3.2.)

Two alternative methods of transporting the RF microwave power to the accelerating structures are considered in the baseline. The first is a Distributed Klystron Scheme (DKS), where each klystron drives 39 cavities; the klystrons and modulators are distributed along the entire length of the SCRF linacs, in the same tunnel but shielded from the accelerator itself, which allows personnel access to make repairs as necessary while beam is on. The second is a novel Klystron Cluster Scheme (KCS), where all the klystrons are located in 'clusters' in surface buildings located periodically along the linacs. The power from a single cluster of 19 klystrons ($\sim$ 190 MW) is combined into an over-moded waveguide, which then transports the power down into the tunnel and along an approximately 1 km section of linac. A Coaxial Tap-Off extracts $\sim$ 6.7 MW of power to a local power-distribution system feeding three cryomodules containing 26 cavities.

The advantages of KCS are primarily that most of the heat load is on the surface, where it can be more cost-effectively removed, at the same time as reducing the required underground volume. The disadvantages are the need for additional surface buildings and shafts (one every 2 km of linac), and additional losses in the long waveguide distribution systems. In addition significant R&D is still required compared to the mature and tested distributed system (see Section 3.3.2). Nonetheless, the estimated cost savings associated with KCS make it an attractive solution for flatter terrains with sufficient space for the required surface infrastructure. For more mountainous terrains or sites where surface access is at a premium, DKS is the preferred solution.

For both KCS and DKS, the in-tunnel power-distribution system to the cavities themselves is essentially identical. A key requirement is the ability to tune both the phase and forward power to each cavity remotely, in order to support the allowed $\pm$20% gradient spread among the cavities, thus maximising the accelerating gradient.

### 3.2.2 Electron source

The polarised electron source shares the central region accelerator tunnel with the positron Beam Delivery System. The beam is produced by a laser illuminating a strained GaAs photocathode in a DC gun, providing the necessary bunch train with 90% polarisation. Two independent laser and gun systems provide redundancy. Normal-conducting structures are used for bunching and pre-acceleration to 76 MeV, after which the beam is accelerated to 5 GeV in a superconducting linac. Before injection into the damping ring, superconducting solenoids rotate the spin vector into the vertical, and a separate Type-A superconducting RF cryomodule is used for energy compression.

### 3.2.3 Positron source

The major elements of the ILC positron source are shown in Fig. 3.4. After acceleration in the main linac, the primary electron beam is transported through a 147 m superconducting helical undulator that generates photons with maximum energies from $\sim$ 10 MeV up to $\sim$ 30 MeV depending on the electron beam energy. The electron beam is then separated from the photon beam and displaced horizontally by 1.5 m using a low-emittance-preserving chicane. The photons from the undulator are directed onto a rotating 0.4 radiation-length Ti-alloy target $\sim$ 500 m downstream, producing a beam of electron-positron pairs. This beam is then matched using an optical-matching device (a pulsed flux concentrator) into a normal conducting (NC) L-band RF and solenoidal-focusing capture system and accelerated to 125 MeV. The electrons and remaining photons are separated from the positrons and dumped. The positrons are accelerated to 400 MeV in a NC L-band linac with solenoidal focusing. Similar to the electron beam, the positron beam is then accelerated to 5 GeV in a superconducting linac which uses modified Main Linac cryomodules, the spin is rotated into the vertical, and the energy spread compressed before injection into the positron damping ring.

The target and capture sections are high-radiation areas which will require shielding and remote-handling facilities.

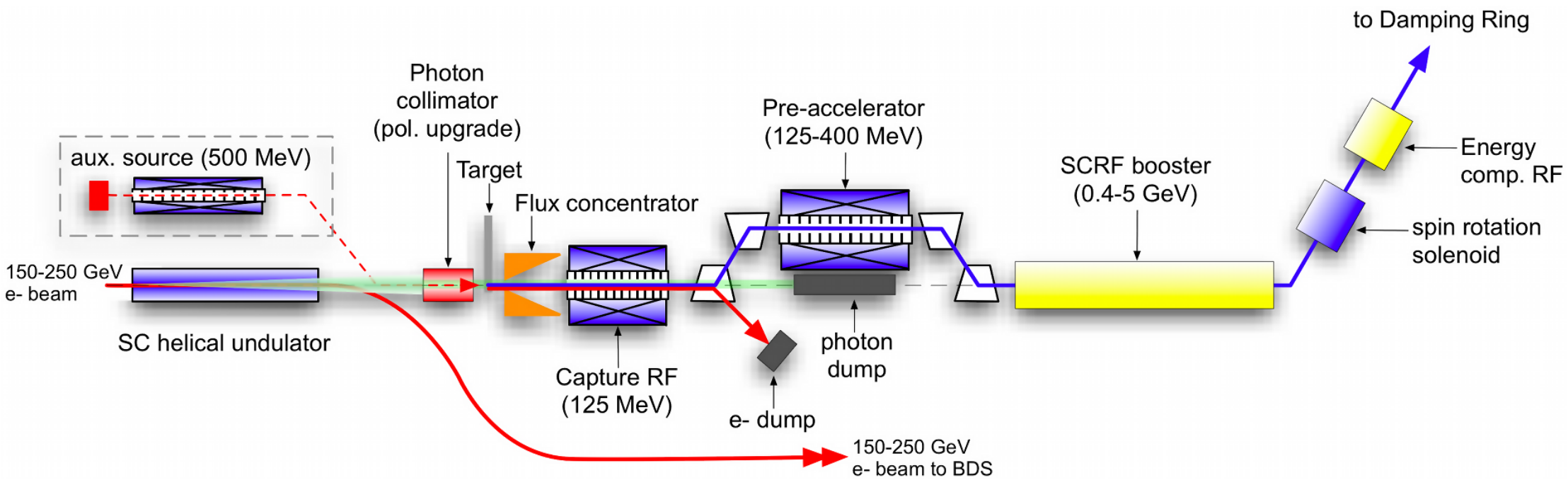

**Figure 3.4.** Overall Layout of the Positron Source, located at the end of the electron Main Linac.

The baseline design provides a polarisation of 30%. Space for a $\sim$ 220 m undulator has been reserved for an eventual upgrade to 60% polarisation, which would also require a photon collimator upstream of the target.

A low-intensity auxiliary positron source supports commissioning and tuning of the positron and downstream systems when the high-energy electron beam is not available. This is effectively a conventional positron source, which uses a 500 MeV NC linac to provide an electron beam that is directed onto the photon target, providing a few percent of the nominal positron current.

To accommodate the 10 Hz operation required to produce the required number of positrons at centre-of-mass energies below 300 GeV (see Section 3.2.8), a separate pulsed extraction line is required immediately after the undulator, to transport the 150 GeV electron pulse for positron-production to the high-powered tune-up dump, located downstream in the Beam Delivery System.

### 3.2.4 Damping rings

The ILC damping rings must accept $e^-$ and $e^+$ beams with large transverse and longitudinal emittances and damp them to the low emittances required for luminosity production. The specification for the extracted normalised vertical emittance of 20 nm represents a reduction of five orders of magnitude for the positron bunch. This reduction must be achieved within the 200 ms between machine pulses (100 ms for 10-Hz mode). In addition, the $\sim$ 1 ms beam pulse must be compressed on injection by roughly a factor of 90 to fit into the ring circumference of 3.2 km; a corresponding decompression is required on extraction. For the baseline parameters, the bunch spacing within trains is approximately 8 ns, which determines the rise and fall time of the injection and extraction kicker systems. (For the luminosity upgrade this number reduces to $\sim$ 4 ns.) Individual bunch injection and extraction is accomplished in the horizontal plane using a total of 42 fast kickers switching 10 kV pulses with rise/fall times of $\sim$ 3 ns.

One electron and one positron ring are included in the baseline, operating at a beam energy of 5 GeV. Both rings are housed in a single tunnel with one ring positioned directly above the other. The damping ring complex is located in the central region, horizontally offset from the interaction region by approximately 100 m to avoid the detector hall. The damping rings are connected to the electron and positron main accelerators by transfer lines.

The damping-ring lattice follows a race-track design. The two arc sections are constructed from 75 Theoretical Minimum Emittance (TME) cells. One of the two 712 m-long straight sections accommodates the RF cavities, damping wigglers, and a variable path length to accommodate changes in phase (phase trombone), while the other contains the injection and extraction systems, and a circumference-adjustment chicane.

Damping is accomplished by approximately 113 m of superferric wigglers (54 units $\times$ 2.1 m) in each damping ring. The wigglers operate at 4.5 K, with a peak-field requirement of 2.16 T.

The superconducting RF system is operated in continuous-wave (CW) mode at 650 MHz, and provides a maximum of 20 MV for each ring, required for the positron ring in 10 Hz mode (nominal 5 Hz operation requires 14 MV for both electron and positron). The frequency is chosen to be half the linac RF frequency to maximise the flexibility for different bunch patterns. The single-cell cavities operate at 4.5 K and are housed in twelve 3.5 m-long cryomodules. The RF section of the lattice can accommodate up to 16 cavities, of which 12 are assumed to be installed for the baseline.

The momentum compaction of the lattice is relatively large, which helps to maintain single-bunch stability, but requires a relatively high RF voltage to achieve the design RMS bunch length (6 mm). The dynamic aperture of the lattice is sufficient to allow the large-emittance injected beam to be captured with minimal loss.

Mitigation of the fast ion instability in the electron damping ring is achieved by limiting the pressure in the ring to below 1 nTorr and by the use of short gaps in the ring fill pattern and a fast transverse feedback system, similar to those used in B-factories.

The performance of the damping rings was noted as one of the biggest challenges facing the post-RDR R&D programme. In particular, intensive studies were carried out on the electron-cloud effect, in which electrons emitted from the vacuum-pipe walls by synchrotron-radiation photons are attracted to the positron beam, resulting in a perturbing electromagnetic field that increases the beam emittance. These studies have resulted in mitigation methods that have been included in a major redesign of the vacuum systems for the baseline damping rings (see Section 3.3.3).

### 3.2.5 Ring to Main Linac (RTML)

The electron and positron Ring to Main Linac (RTML) systems are the longest continuous beamlines in the ILC. The layout of the RTML is identical for both electrons and positrons. The RTML consists of five subsystems, representing the various functions that it must perform: a $\sim 15$ km long 5 GeV transport line; betatron and energy collimation systems; a 180° turn-around, which enables feed-forward beam stabilisation; spin rotators to orient the beam polarisation to the desired direction; and a two-stage bunch compressor to compress the beam bunch length from several millimetres to a few hundred microns, as required at the IP.

The two-stage bunch compressor includes acceleration from 5 GeV to 15 GeV in order to keep the increase in relative energy spread associated with bunch compression small. The acceleration is provided by sections of SCRF main-linac technology. A primary challenge for the RTML systems is the preservation of the damped emittance extracted from the damping rings; the combination of the long uncompressed bunch from the damping ring and large energy spread (after compression) make the tuning and tolerances particularly demanding. However, tuning techniques developed through detailed simulations have demonstrated acceptable emittance growth.

In addition to the beam-dynamics challenges, acceptable jitter in bunch arrival time at the IP requires an RMS phase jitter of $\leq 0.24°$ between the electron and positron bunch-compressor RF systems. Beam-based feedback systems integrated into the bunch-compressor low-level RF system should be able to limit the phase jitter to this level.

### 3.2.6 Beam Delivery System (BDS)

The ILC BDS is responsible for transporting the $e^+e^-$ beams from the exit of the high-energy linacs, focusing them to the sizes required to meet the ILC luminosity goals, bringing them into collision, and then transporting the spent beams to the main beam dumps. In addition, the BDS must perform several other critical functions, including characterising the incoming (transverse) beam phase-space and matching it into the final focus; removing any beam halo from the linac to minimise background in the detectors; and measuring and monitoring the key physics parameters such as energy and polarisation before and after the collisions.

There is a single collision point with a 14 mrad total crossing angle. The 14 mrad geometry provides space for separate extraction lines but requires crab cavities to rotate the bunches in the horizontal plane for effective head-on collisions. There are two detectors in a common interaction region (IR) hall in a so-called 'push-pull' configuration.

The geometry of the BDS has been designed to accommodate the upgrade to 1 TeV centre-of-mass energy, in particular to minimise the emittance growth due to synchrotron radiation to less than 10%. The baseline lattice uses fewer magnets (predominantly dipoles) for the lower-energy operation.

There are five main subsystems of the BDS. First, in the direction of the beam, is a section containing emittance measurement and matching (correction), trajectory feedback, polarimetry and energy diagnostics; secondly, a collimation section removes beam-halo particles that would otherwise generate unacceptable background in the detector and also contains magnetised iron shielding to deflect muons generated in the collimation process; thirdly, the final focus (FF) uses strong compact superconducting quadrupoles to focus the beam at the IP, with sextupoles providing local chromaticity correction; the final pair of quadrupoles closest to the IP is integrated into each particle physics detector to facilitate rapid exchanges of the detectors in 'push-pull'; fourthly, the interaction region contains the detectors; finally, the extraction line has a large-enough bandwidth to cleanly transport the heavily disrupted beam to a high-powered water-cooled dump, and also contains important polarisation and energy diagnostics.

The beam-delivery optics provides a demagnification factor of typically several hundreds in the beam size, resulting in very large beta functions (several thousand kilometres) at critical locations, leading to the tightest alignment tolerances in the entire machine. In addition, correction of the strong chromaticity and geometric aberrations requires a delicate balance of higher-order optical terms. The tight tolerances on magnet motion (tens of nanometres), makes continuous trajectory correction and the use of fast beam-based feedback systems mandatory. Furthermore, several critical components (e.g. the final-focusing doublet) may well require mechanical stabilisation. Beam-based alignment and beam phase-space tuning algorithms are necessary to adjust and tune the optical aberrations that would otherwise significantly degrade the luminosity. The ability to tune the beams to the required levels relies extensively on precision remote mechanical adjustment of the magnets and diagnostics with matching precision. Many of the techniques and instrumentation were successfully developed at the Final Focus Test Beam experiment at SLAC, and currently at the ATF2 facility at KEK (see Section 3.3.3).

The tight tolerance on the relative uncorrelated phase jitter between the electron and positron superconducting crab-cavity systems requires timing precisions to the level of tens of femtoseconds. Although this tolerance is tight, it is comparable to that achieved at modern linac-driven FELs.

Control of machine-generated backgrounds is performed by careful optics control and tuning of an extensive collimation system, as well as by the use of non-linear elements ('tail-folding' octupoles). Wakefield effects at the small apertures are taken into account in the design of the mechanical collimators.

The main beam dumps, which use a high-pressure high-velocity water design, are a major installation. Since the dumps will be significantly activated during operation, they are designed and rated for the full upgrade average beam power of 14 MW, in order to avoid having to replace them during the energy upgrade to 1 TeV.

### 3.2.7 Conventional facilities and site-dependent designs

Together with the Main Linac systems, conventional facilities and siting (CFS) represent the two largest elements of the total project cost. The CFS design and costs can be broken down into three main areas: civil construction, including underground and surface structures, shafts and access tunnels; electrical systems (AC power distribution etc.); and mechanical systems (water cooling and air handling etc.).

The CFS solutions and associated cost are developed based on the requirements defined by the accelerator layout and parameters outlined in the previous sections. Minimisation of the total CFS cost requires an understanding of how it depends on the accelerator design, and if necessary modifying and iterating that design. Reduction of the scope of the underground civil construction (for example) was considered a primary design goal during the Technical Design Phase and resulted in significant modifications to the accelerator design and parameter space. In addition, the criteria for the electrical and mechanical systems, access shafts and tunnels have been scrutinised to reduce costs.

The solutions to implement the accelerator functionality are heavily influenced by regional considerations of site topography and geology, as well as local legislation (such as safety requirements). Geology will determine the most cost-effective approach to tunnelling method, while topography can influence the surface structures, access tunnels and shafts. All of these factors can shift the balance of the cost-optimisation and influence the accelerator design. As a result, the final machine will be influenced by the choice of site.

In the absence of a definitive site for the ILC, it was necessary to evaluate different sites with different characteristics. To this end, several sample sites have been developed. As noted in Section 3.2.1, two site-specific design variants emerged: one suitable for a flatter terrain (flat topography) using the Klystron Cluster Scheme (KCS); and one for a more mountainous terrain (mountainous topography) using the Distributed Klystron Scheme (DKS).

The Americas site in Northern Illinois near Fermilab is essentially flat and provides a range of locations to position the ILC in a north-south orientation; that chosen has approximately one-quarter of the machine on the Fermilab site with the tunnels bored in a contiguous dolomite rock strata ('Galena Platteville'), at a typical depth of 30–100 m below the surface. For Asia, two possible ILC candidate sites have been identified: Kitakami in the Tohoku district in northern Japan; and Sefuri in the Kyushu district in the south. Both provide a uniform terrain located along a mountain range, with a tunnel depth ranging from 40 m to 600 m in uniform granite geology highly suited to modern tunnelling methods. The European site is located at CERN, Geneva, Switzerland, and runs parallel to the Jura mountain range, close to the CERN site, mostly in the 'Molasse' (a local impermeable sedimentary rock), at a typical depth of 370 m.

The Americas and European sample sites are relatively similar and are examples of 'flat topography' sites using KCS. The geology lends itself well to the use of Tunnel Boring Machines (TBM) which provide a round tunnel cross section. The Japanese sites are examples of mountainous topology using DKS, where available space for surface infrastructure is severely limited, requiring most of the accelerator infrastructure to be housed underground. The most cost-effective tunnelling solution in this geology and topography is "drill and blast" as exemplified by the New Austrian Tunnelling Method.

### 3.2.8 Special considerations for running at low centre-of-mass energy

While performance requirements at the maximum energy dictate many of the key parameters and the overall geometry and cost of the machine, running at lower energies is also an important part of the physics potential of the ILC. This is especially true given the recent discovery of a Higgs boson at the LHC. Sufficient luminosity needs to be produced to enable the study of important physics quantities at the threshold for producing it in association with a $Z^0$, around 220 GeV. Two issues limit the possible performance at these lower energies: first, the undulator-based positron source must have an electron-beam energy of at least 150 GeV to produce the requisite positron intensity; secondly, the beam divergence at the interaction point is constrained by the allowable synchrotron radiation fan generated by the final doublet. This results in luminosity falling off more rapidly than the normal scaling as the inverse of the beam energy.

The solution adopted for the first issue for the current baseline is to interleave the normal "physics" pulse (which at low energies cannot produce the required number of positrons) with an additional electron pulse at 150 GeV. Having produced the positrons, this "extra" pulse is extracted to the beam dump. This scheme doubles the frequency of electron (but not positron) beam pulses to 10 Hz. This so-called "10 Hz" operation mode implies several design criteria for the baseline. First, both electron and positron damping rings must now damp the beam in 100 ms instead of the nominal 200 ms. This requires additional wigglers and RF in the ring. Secondly, the positron damping ring is 'empty' for 100 ms, after which the current is ramped up in $\sim 1$ ms (and similarly ramped down during extraction). Dealing with transient beam loading requires an additional RF power overhead of approximately 15%. Thirdly, all the machine elements of the electron machine before the positron source must run at 10 Hz. Fourthly, the positron-production pulse (150 GeV beam) must be safely extracted after the undulator and dumped, requiring an additional pulsed magnet and extraction beamline system. Finally, a pulsed-magnet steering system is required upstream of the source undulator (downstream of the main electron linac) to compensate for the difference in trajectory between the 150 GeV positron production and 150 GeV luminosity pulses. The 10 Hz mode is cost effective because the total RF power and cryoload for the main (electron) linac do not exceed that at 250 GeV beam energy at 5 Hz when the beam energy (and therefore the main linac gradient) is reduced below 150 GeV. However, for the electron bunch compressor and source linacs, the AC power requirement effectively doubles for the 10 Hz operation mode, which also drives the design criteria and power requirements for the damping rings.

To mitigate the second problem related to the beam-divergence constraint at the IP, a shorter final-doublet (FD) arrangement can be used for energies less than 300 GeV. This increases the collimation depth and the allowed IP beam divergence (by up to 30% in the horizontal plane), thus increasing the luminosity. The FD will be implemented in a modular design to accommodate both high- and low-energy running configurations, thus avoiding the need to exchange the magnet cryostat.

There are no issues with running the main SCRF linacs at reduced gradient for the low centre-of-mass running. The lower average gradient results in a shorter fill time and overall higher RF-to-beam power efficiency. Simulations of the beam dynamics have indicated no significant additional degradation of vertical emittance.

### 3.2.9 Luminosity and energy upgrade options

This TDR has been optimised for the 500 GeV baseline scenario. Although considered in significantly less detail, two upgrade scenarios are foreseen.

The first upgrade is in luminosity, by up to a factor of two, accomplished by doubling the number of bunches per beam pulse (doubling the beam power). This requires increasing the number of klystrons and modulators by approximately 50%. A second positron damping ring can be installed in the same tunnel should electron-cloud effects at these higher currents prove problematic. All other accelerator systems are already rated for the higher beam power. Basic (minimum) provisions for the required conventional facilities are included in the baseline design to support the luminosity upgrade, although modifications to the cooling systems would be required.

Secondly, an upgrade of the centre-of-mass energy up to 1 TeV is foreseen, by approximately doubling the length of the main linacs. The current preferred concept is to retain the original linacs and use them to accelerate a higher energy beam (265–500 GeV), while the new upgrade linac extensions would then provide acceleration from the bunch compressor energy of 15 GeV up to 265 GeV. In addition to the major construction required for the linac extensions, a relocation of the bunch compressors and 180-degree turn-around is required, as well as an extension of the RTML long transport line. The upgrade to the main linacs themselves is based on a forward-looking SCRF technology compatible with the original linac installation (RF pulse length etc.), but with higher-performance specifications (45 MV/m with $Q_0 = 2 \times 10^{10}$). These ambitious goals will require extensive R&D, which is assumed to continue in parallel with the construction and operation of the baseline accelerator. The overall site requirement for the TeV machine is approximately 50 km, with a site power requirement of approximately 300 MW. In order to minimise the impact on the existing machine during the upgrade construction, the baseline BDS geometry and high-power beam dumps are already compatible with 500 GeV beam operation. All of the candidate sites can accommodate such an upgrade.

### 3.2.10 Value estimate and methodology

The new baseline design for the ILC given in the TDR provides the basis for an updated cost estimate. A value estimate approach has been adopted, which is independent of any particular national accounting system but compatible with them all and thus is suitable for a collaborative venture. It has been used by both ITER and the LHC.

The value estimate covers the construction cost for a machine of 500 GeV centre-of-mass energy. It includes a small number of items rated for 1 TeV to enable a later upgrade. Typically these elements are in places difficult to retrofit, such as beam dumps. The value estimate omits a number of items such as pre-construction, taxes, contingency, escalation, spare equipment, beam commissioning, etc. that would be included in some regionally specific estimates.

The value estimate has two parts: component value and explicit labour. The value of a component is defined as the lowest reasonable estimate of the procurement cost of an item that meets the technical specifications in the required quantity in a major industrial nation. It includes the material costs and the implicit labour required to fabricate the component.

Explicit labour, which is accounted for in person-hours, is required manpower which could be provided by collaborating institution or the host laboratory, or could be purchased from industrial vendors.

The estimate should allow funding agencies that are considering contributions to the project to assess the nature and scope of the required resources. It also provides detailed information on cost drivers which can be used during the pre-construction phase for further cost optimisation through either value engineering or additional R&D.

In order to minimise regional price distortions related to exchange rates, currency conversions were based on purchasing-power-parity (PPP) indices as published by the Organization for Economic Co-operation and Development (OECD). The currency unit for the value estimate was the ILCU, defined as a 2012 US dollar. Estimates performed at dates different from the TDR reference date of January 2012 were escalated using regional indices for either civil engineering or equipment. Volume 3 Part-II contains a detailed discussion on the use of PPP indices.

The estimates for each cost element are median estimates that correspond to the 50% probability point in the cost-distribution curve. All major procurements are estimated assuming a minimum of two vendors with specifications and quantities as given in the TDR.

Uncertainty estimates have been made for each cost element in the TDR. The uncertainties are based on the design maturity of the item, the level of technical risk, the source or quality of the cost or labour information, and the extent, if any, of the scaling to large quantities. Based on this information, the fractional cost increase required to reach an 84% confidence level has been computed. This cost increase is called the "cost premium", and it is provided as part of the value estimate.

To account for the economies of scale when the cost basis explicitly corresponds to a significantly smaller number of units than that required for the ILC, the unit estimate has been obtained by applying a discount derived from a learning curve. Based on previous HEP projects, learning curves in the range of 85-95% would be expected. In the absence of any additional vendor information a relatively conservative 95% was typically adopted.

Approximately 75% of the TDR estimate has a new cost basis compared to the previous costing exercise in 2006. This includes the major project cost drivers, conventional facilities and the SCRF systems and components. Significant information was accumulated during the GDE R&D program related to SCRF cavity costs in addition to recent data obtained from European XFEL contracts. The other 25% was based on estimates made during the 2007 Reference Design study, adjusted for both inflation at the component level and the unit counts reflecting the updated TDR design. The breakdown of the cost basis is shown in Fig. 3.5.

The majority of the cost basis involves industrial estimates of various types. The labour estimate is principally based on prior experience of national laboratory staff.

**The value estimate for the TDR design, averaged over three regional sites, is 7.8 billion ILCU**. This is dominated by the SCRF components and related systems, together with the conventional facilities. These two elements account for 73% of the total. The main linac itself corresponds to 67% of the total project. **The total explicit labour estimate is 22.6 million person-hours.** Installation and laboratory management/administration are the biggest single elements. The cost premiums for both value and labour are approximately 25%. The sub-system value breakdown is shown in Fig. 3.6.

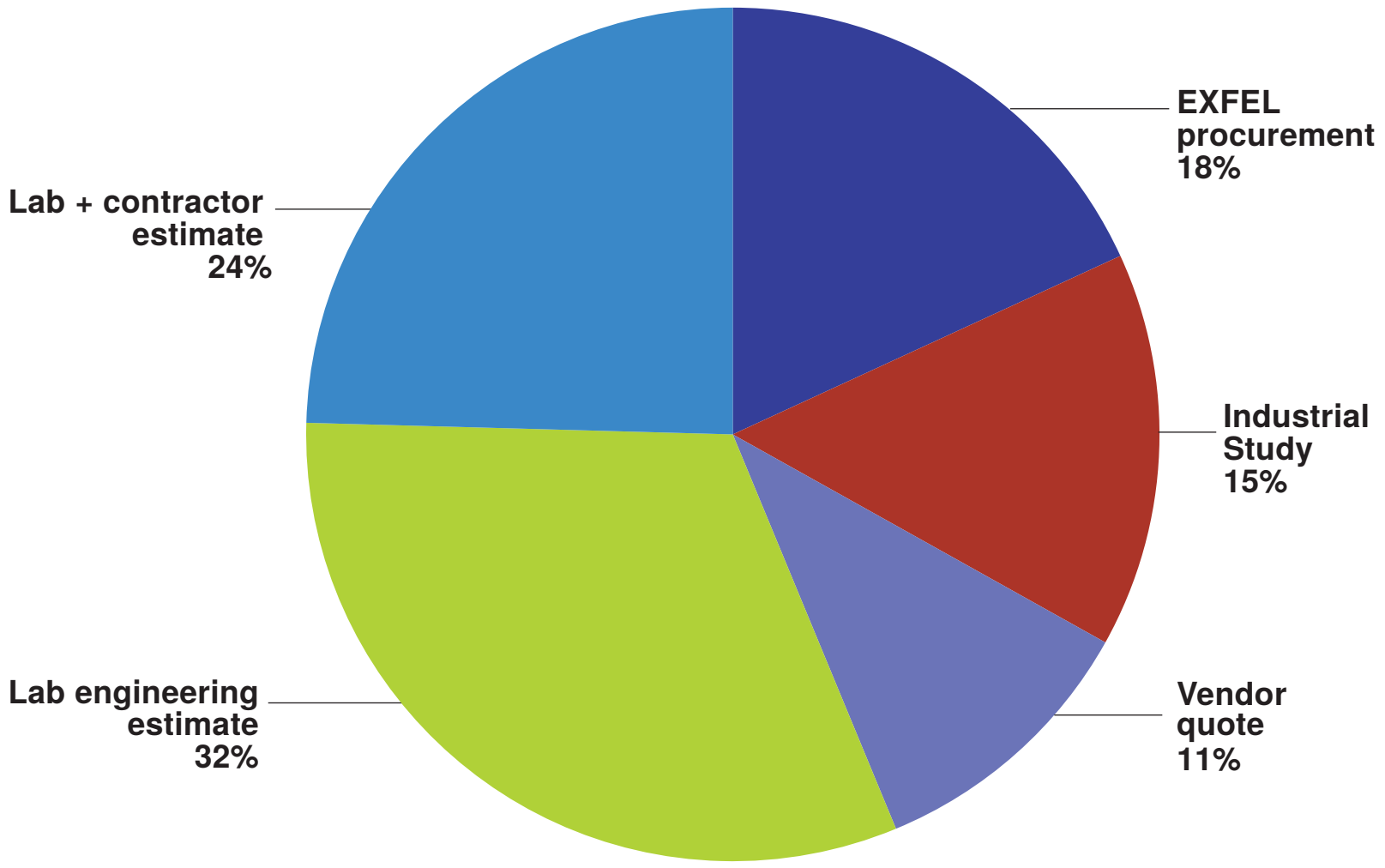

**Figure 3.5**
ILC TDR Value estimate cost basis.

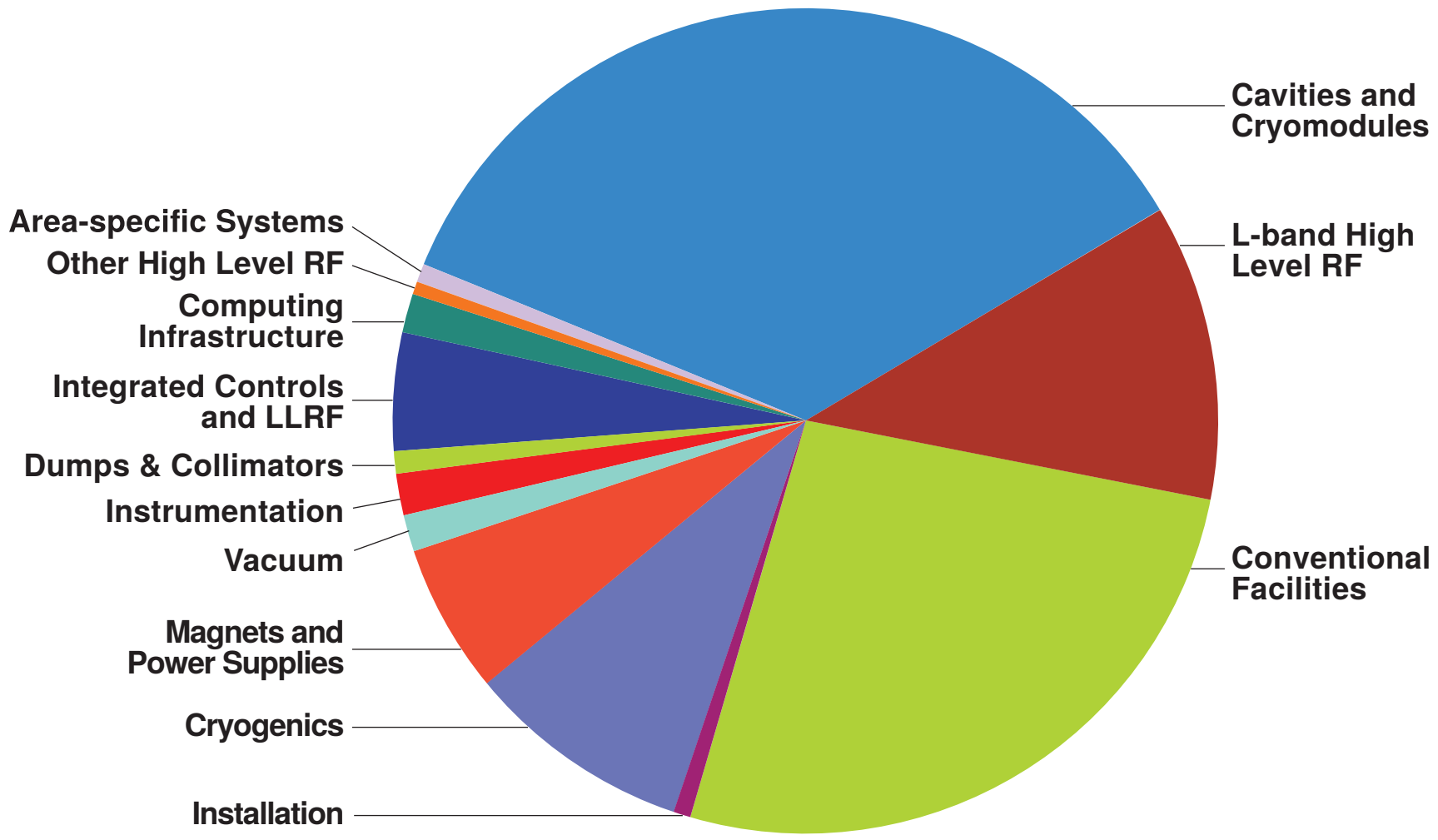

**Figure 3.6**
Distribution of cost by sub-system.

These totals represent an increase of 7% in value and a reduction of 8% in explicit labour relative to the estimates made for the 2007 *Reference Design Report* (after adjustment for inflation from 2007 to 2012). The major contribution to the increase was the cryomodule cost which was based on current industrial studies and actual European XFEL contracts extrapolated to ILC quantities, rather than older industrial studies and engineering estimates. This increase was offset in several areas due in large part to the more efficient TDR design.

Any schedule for a project such as the ILC is determined by the availability of resources and the ability to utilise them efficiently. Without knowledge of the chosen Governance and Project Management structure and funding profiles, a more accurate schedule cannot be formulated. Nonetheless, making some reasonable assumptions in these areas, it appears that the overall construction schedule is determined by the civil construction activities in the central campus region covering the detector halls, the damping rings, and the injectors. These elements are site dependent. The Main Linac schedule is determined by the delivery of the SCRF cryomodules, which are the technical components with the longest lead time. A funding profile which peaks at 15% of the total project cost in year four is consistent with a nine-year period between ground breaking and the start of beam commissioning. Machine installation starts in year seven. A representative schedule for a mountainous site is shown in Fig. 3.7.

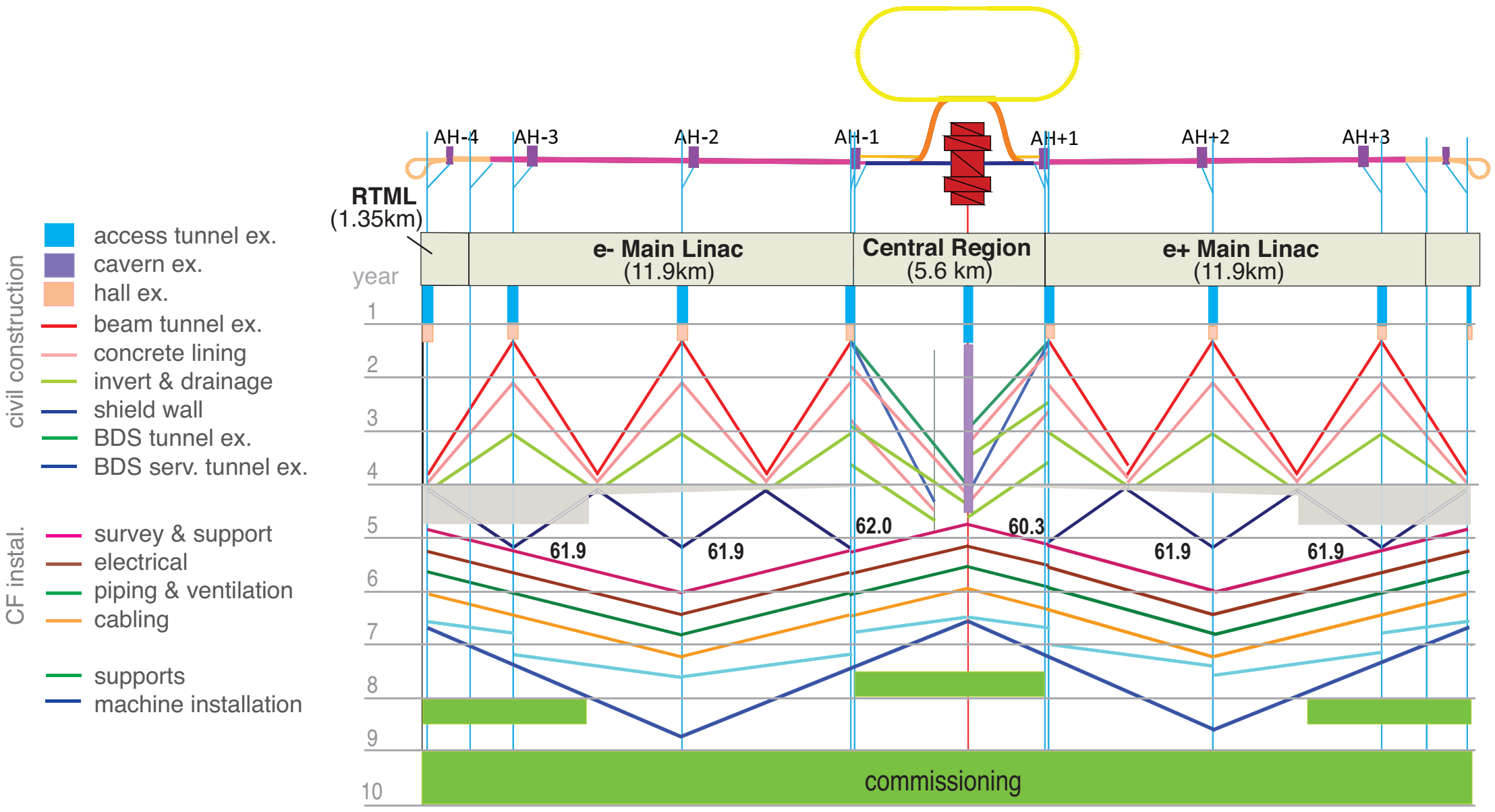

**Figure 3.7.** A possible ILC project construction schedule. Years from construction start are represented vertically, while progress along the machine footprint is shown horizontally (not to scale). Vertical lines represent access tunnel locations.

## 3.3 ILC R&D during the technical design phase

The ILC technical design, layout and parameters presented in Section 3.1 reflect the results of the significant R&D that has been performed during the technical design phase. The global R&D programme itself was prioritised based on an evaluation of the remaining technical risks in the published 2007 Reference Design. At that time, the five highest identified priority items were:

1. SCRF cavities capable of reproducibly achieving at least 35 MV/m;
2. a cryomodule consisting of nine cavities, operating at an average of gradient of 31.5 MV/m;
3. linac string test (or integration test) of more than one cryomodule with beam;
4. development of models and mitigation techniques for electron-cloud effects in the positron damping ring;
5. demonstration of the small spot size and stability produced using the final focus optics that is used in the baseline design of the beam delivery system.

Other R&D areas (for example the sources) were also identified. The first three priority R&D items all relate to the SCRF linear-accelerator technology, the primary cost driver of the machine. Although it was noted by the International Technology Review Panel that TESLA SRF technology was ‘mature’, the ILC gradient goal had only been achieved in a handful of cavities (one of which had accelerated beam at 35 MV/m in the TESLA Test Facility at DESY – a proof of principle). One of the major technical aims has been the demonstration of large-scale production of reproducible high-gradient SCRF cavities, which required a detailed fundamental understanding of the physics involved in the technology. During the R&D programme from 2007–2012, more than 200 cavities have been successfully manufactured and processed.

The successful development of industrial capacity in each of the three regions (Asia, the Americas and Europe) resulted in multiple vendors capable of producing high-performance ILC cavities. In the USA, these were tested at Fermilab, Argonne National Laboratory and Jefferson Lab; in Japan at KEK; and in Europe at DESY, where development has been driven by the design and construction of the European X-ray free-electron laser (XFEL). The 17.5 GeV SCRF linac of the European XFEL represents the largest deployment of the technology to date. In many ways it provides an excellent large-scale prototype for the ILC.

Cryomodule development and integrated systems testing is being pursued at all the primary SCRF sites. The FLASH FEL facility at DESY has successfully accelerated an ILC-like electron beam through high-gradient cryomodules, and has demonstrated many of the tuning techniques required by the ILC. Further system-test accelerators at Fermilab (NML) and at KEK (STF) have been under development and will see beam operation in 2013. The global cryomodule integration (S1-Global) programme at KEK successfully integrated cavities and auxiliary components delivered from DESY, FNAL, INFN and KEK into a single cryomodule, allowing direct comparative tests of different technologies as well as demonstrating the key concept of "plug compatibility", simplifying integration of parallel design-efforts worldwide. Other main-linac components developed by the R&D program include a tuneable high-power RF delivery system and associated low-level RF controls, and a next generation solid-state modulator.

During development of the 2007 *Reference Design Report*, emittance dilution in the positron damping ring arising from the electron-cloud effect was identified as one of the primary major technical risks to the design luminosity. As a result, a multi-year study was launched at the CESR accelerator at Cornell, later named the Cornell Electron-Positron Storage Ring Test Accelerator (CesrTA). The well understood machine characteristics and highly flexible operating parameters of the CesrTA facility supported an R&D program that has culminated in a definitive report on the physics of high-intensity, positively charged beams. The study also developed and identified techniques to reduce the electron-cloud effect to an acceptable level that are now included in the TDR.

Some highlights of the R&D programmes whose success is the foundation of the TDR baseline design are given below.

### 3.3.1 Main achievements of the SCRF R&D effort

The performance of superconducting cavities is primarily limited by two effects: field emission and quench-causing surface defects. Improvements in surface treatments have essentially mitigated the onset of field emission at gradients below 35 MV/m. The invention and deployment of tools to identify and repair quench-causing defects at low cavity gradient has led to the establishment of a baseline set of procedures for cavity fabrication and surface preparation which minimise surface defects. These techniques were fully implemented during the final phase of the R&D program giving the results shown in Fig. 3.8 a two-pass production yield of 94% for cavities satisfying 35 MV/m $\pm$ 20%, with an average gradient of 37.1 MV/m. These results exceed the 2006 GDE R&D goal of 90% yield and 35 MV/m average gradient.

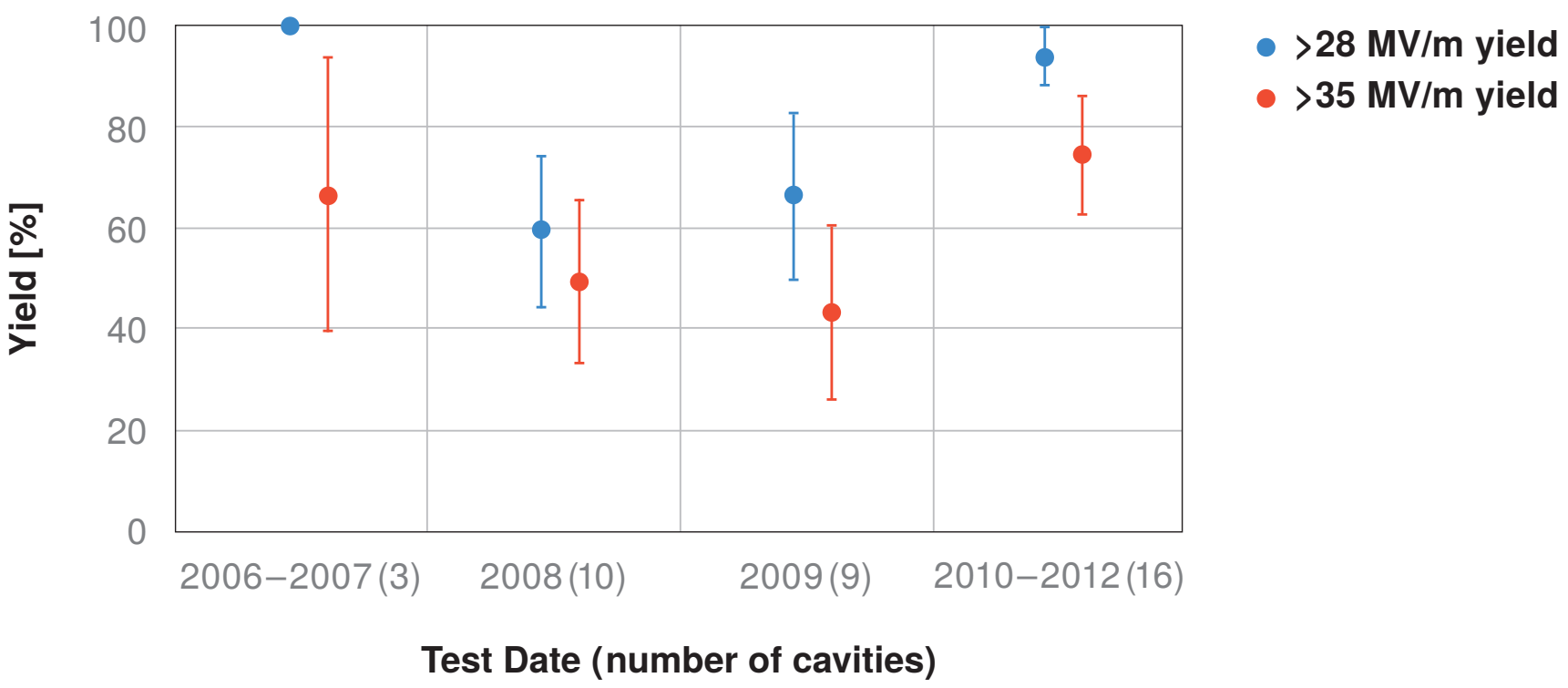

**Figure 3.8** Cavity yield for the years indicated during the GDE R&D programme. The results include cavities having had up to two surface chemistry treatment.

In addition to the performance of individual cavities an average field gradient of 32 MV/m has been achieved in a prototype cryomodule for the European XFEL program (pictured in Fig. 3.9a).

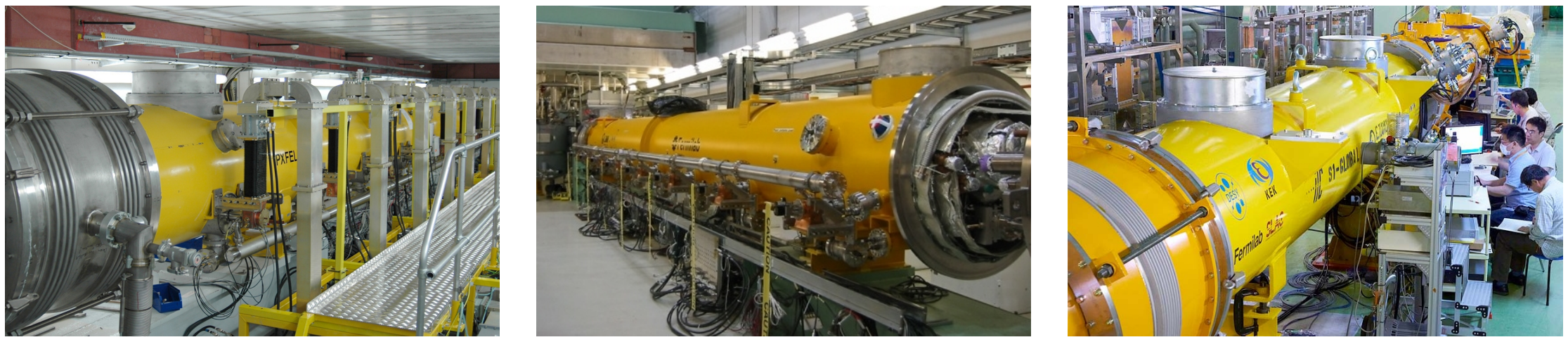

(a) PXFEL1 at DESY (b) CM2 at FNAL (c) S1-Global at KEK

**Figure 3.9.** Cryomodules assembled in all three regions.

The S1-Global programme (see Fig. 3.10) successfully demonstrated "plug compatibility" by building one cryomodule from cavities and couplers supplied from several different national laboratories. Specifically the half-size cryomodule-C was built from an INFN cryostat, DESY cavities and couplers, and FNAL cavities and couplers in cooperation with SLAC. The cryomodule power system is another good example of plug compatibility. The ability to incorporate and test several different component designs within a single integrated test setup was another important aspect of the S1-Global programme, providing input to many of the baseline technology decisions for the TDR.

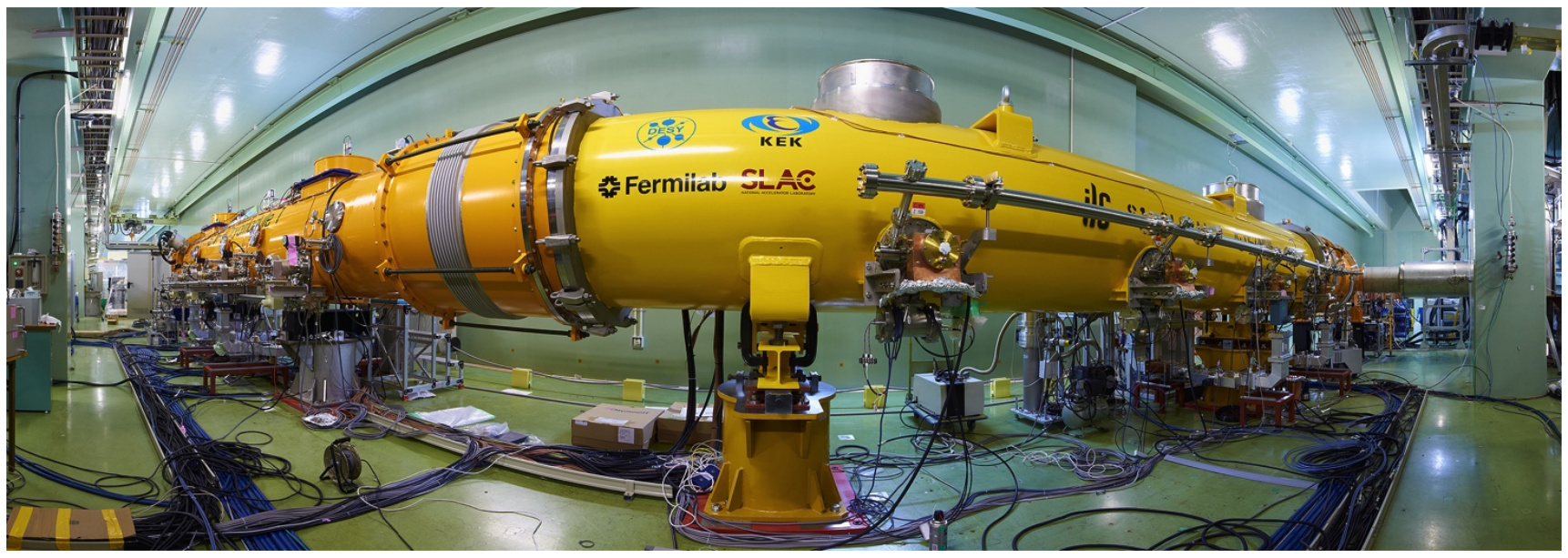

**Figure 3.10**
Fish-eye view of the S1-Global cryomodule at KEK.

## 3.3.2 R&D on RF power generation and distribution

*Klystron and modulators* The RF system features a 10 MW, high-efficiency (65%), Multiple-Beam Klystron (MBK) that produces 1.6 ms pulses at a rate up to 10 Hz. For this application, three vendors developed tubes of somewhat different designs, of which two were successful to the extent that they have been adopted for use in the European XFEL linac where they will run nominally at 5 MW but have 10 MW capabilities.

The klystron modulator is another large component in the linac RF power system. A new type of modulator, the all-solid-state Marx modulator, was proposed and developed at SLAC National Accelerator Laboratory and with industry. This type of modulator has many advantages over the 'traditional' kind: compactness, lower cost, operational flexibility and lack of oil-filled transformers inside tunnel enclosures. The Marx modulator design has been adopted as baseline for all ILC RF power systems.

*RF distribution* The ILC baseline linac will need to accelerate a nominal 6 mA of beam current, but needs to be capable of accelerating up to 9 mA to allow scope for future upgrades. Multiple cavities (26 or 39 depending on the configuration and beam current) are effectively driven by one 10 MW power source. A $\pm$20% spread in cavity-to-cavity performance is expected. The power-distribution system driving the cavities has therefore been developed to provide remotely adjustable power ratios, allowing each cavity to be driven at its maximum gradient while minimising the power losses in the waveguide components.

For the flat-topography single-tunnel design, a new high-power distribution system was proposed and developed. The Klystron Cluster System (KCS) has all klystrons and associated power supplies located in a series of ten surface buildings along the Main Linacs. In each building there are two sets of typically 19 klystrons (for the 5 mA baseline) or 29 klystrons (for the 9 mA upgrade) whose power is combined and then transported in a 0.5 m diameter, pressurised aluminum over-moded circular waveguide. Up to 10 MW of power is tapped off periodically to feed each group of 26 cavities (3 cryomodules). This required the development of several waveguide components, such as combiners, bends, adjustable tap-offs and the long circular over-moded waveguides. Development and testing of these components will continue as required.

### 3.3.3 Beam test facilities

Beam test facilities are required for critical technical demonstrations, including accelerating gradient, beam dynamics, and precision beam handling. The scope of test-facility activity needed to mitigate technical risks was assessed during the development of the 2007 Reference Design Report. Purpose-built test facilities were then either constructed by collaborative teams or provided through adaptation of existing facilities. The primary goals of these beam test facilities were: the demonstration of ILC linac performance and evaluation of realistic cavity performance with beam acceleration; demonstration of a number of cavities operated in an accelerator showing repeatable performance and providing an estimate of reliability; studies of instabilities in the damping rings, such as electron cloud, and mitigation techniques; and demonstrations of the generation and handling of low-emittance beams using precision optics and stabilisation tools.

*The 9 mA Experiment at FLASH* The FLASH (Free electron LASer in Hamburg) accelerator became a user facility operating as a soft X-ray free-electron laser in 2005. The FLASH linac is a 1.25 GeV linac based on Tesla-type technology and operates 5000 hours per year on average. The '9-mA' program was proposed by the GDE in 2008 with the goals of demonstrating reliable operation of the linac with ILC-like bunch trains and to characterise the limits of operation of gradient and RF power. Typical beam properties for FEL user operation (charge, number of bunches, average beam power) are less challenging than those required for the 9 mA studies. For DESY, however, these studies have been important for integration and operational issues associated with running long bunch trains and high bunch charge, both for FLASH itself and for the European XFEL.

Much can be learnt from FLASH but the European XFEL, currently under construction, offers an even greater opportunity for gaining invaluable experience with constructing, commissioning, and operating a large-scale superconducting high-power linac. The superconducting linac facilities at KEK and FNAL (see Fig. 3.9) are in the early stages of commissioning and they will further expand operating experience around the world.

*CesrTA (Cornell) and electron-cloud R&D* One of the principal R&D issues for the positron damping ring of the ILC was to ensure that the build-up of the electron cloud (EC) in the vacuum chambers can be kept below the levels at which EC-induced emittance growth and beam instabilities occur. During Phase I (2008-2010), a focused effort to study methods of suppressing the EC as well as measuring its impact on ultra-low emittance beams was undertaken at CesrTA. In addition, a complementary R&D program has continued at various laboratories around the world to develop better techniques to mitigate the build-up of the electron cloud. As part of this coordinated global programme, a major emphasis was placed on developing and benchmarking simulation tools and measurement techniques.

The results from the first three years (Phase I) of the CesrTA R&D programme have been incorporated into the design of the ILC damping-ring vacuum chamber; the findings of the programme are documented in the CesrTA Phase I Report. In particular, the observed efficacy of grooved chamber surfaces in the dipoles as well as that of the clearing electrode in the high-field wigglers provide confidence that practical electron-cloud mitigation measures can be prepared for the arc and wiggler straight regions of the ILC positron damping ring. The importance of cloud mitigation in the damping ring quadrupole chambers has also been demonstrated. New coating technologies to suppress the secondary-electron yield show great promise. However, there is still the issue of studying the long-term performance and durability of these coatings. This will be a subject of study during Phase II of the CesrTA programme. Perhaps most importantly, the flexibility of CESR operations has enabled a systematic programme of electron-cloud build-up and electron-cloud-induced beam-dynamics studies. By benchmarking physics models and simulations against these studies, confidence in being able to make valid projections of the expected ILC positron-damping performance has been significantly enhanced.

*ATF2 final-focus experiment* The challenge of colliding nanometre-sized beams at the interaction point (IP) involves three distinct issues: creating small emittance beams; preserving emittance during acceleration and transport; and focusing the beams to nanometre sizes before colliding them.

The Accelerator Test Facility (ATF) at KEK is a prototype damping ring to deal with the first issue and has succeeded in obtaining the emittances that almost satisfy ILC requirements; these have also already been achieved in two third-generation light sources. The ATF is now used as an injector for the ATF2 final-focus test beam line, which was constructed in 2008 to study the third issue. The ATF2 beam line stretches about 90 meters from the beam-extraction point in the ATF damping ring to the focal point. Anticipating gradual movements of supports and magnets due to thermal variations or slow ground motion, quadrupoles and sextupoles were mounted on remote-controlled three-axis movers. Overall alignment precisions of 0.1 mm (displacement) and 0.1 mrad (rotations) have been achieved using conventional alignment/metrology techniques. The final alignment of the magnets is achieved via beam-based alignment (BBA) techniques.

ATF2 is a follow-up to the Final-Focus Test Beam (FFTB) experiment at SLAC, but with different beam-line optics based on a scheme of local chromaticity correction, which facilitates a shorter design and a larger energy bandwidth. As with FFTB, the vertical beam size at the focal point has been chosen to have the same demagnification as the ILC and similar levels of chromatic aberration. The primary goals for ATF2 are to achieve a 37 nm vertical beam size at the IP, stabilise it at the nanometre level, and then maintain the beam size and stability over a period of time. In December 2012, ATF2 achieved a beam size within a factor 2 of the first goal. The R&D towards successful completion of the goals will continue following many months of delay after recovery from the Great East Japan Earthquake.

### 3.3.4 Future engineering and R&D

The successful conclusion of the GDE R&D programme establishes technical viability but there are many other aspects of the ILC that will require additional work. Cost containment through value engineering of the cryomodule remains to be performed prior to production. Couplers, tuners, and other components would benefit from further detailed engineering studies. Minimising the loss of gradient performance from individual cavities in vertical tests to cavity strings in cryomodules is an important goal. The RF power systems are also good candidates for additional cost engineering. At some point an actual site will be proposed and adapting the TDR baseline to a site-specific design will be necessary. Continued operation of the test beam facilities is envisaged for the next several years. The transport and stabilisation of high brightness beams at the ATF, developing high brightness techniques at CESR-TA, and beam acceleration tests at FLASH will all continue. The XFEL project at DESY will provide a unique opportunity to monitor high-volume cryomodule production. When this facility is operational in 2015, it will also test ILC beam-control techniques.

# Chapter 4
# ILC Detectors

## 4.1 Introduction and Physics Performance Requirements

In order to realise the physics program, the ILC detectors face challenges requiring significant advances in collider detector performance. The machine environment is benign by LHC standards, enabling designs and technologies that are unthinkable at the LHC. However, the ILC environment poses its own set of background issues that must be overcome. The "Detailed Baseline Design" of the SiD and ILD detectors have been developed to achieve the requirements for all considered physics programs, over the full range of centre-of-mass energies from 200 GeV up to 1 TeV, as well as the possibility of special running at the Z-pole.

The ILC physics opportunities place a premium on high-resolution jet energy reconstruction and di-jet mass performance. Event reconstruction techniques based on the Particle Flow Algorithm (PFA) have been developed to meet this challenge. This motivates highly granular electromagnetic and hadron calorimeters and highly efficient tracking systems. New detector technologies and new reconstruction algorithms based on the PFA approach achieve the needed precision in the reconstruction of jets of 3 to 4 percent for 100 GeV jets, set by the requirement to separate W and Z di-jet final states. The requirements on momentum resolution for charged tracks ($\Delta p/p^2$ of $5 \times 10^{-5}$ (GeV/c)$^{-1}$) are driven by reconstruction of a Higgs boson recoiling from the associated Z boson decaying to a lepton pair in the Higgs-strahlung process. Flavour and quark-charge tagging will be available at an unprecedented level of performance as a result of the development of a new generation of vertex detectors. Particle identification is achieved by the highly granular calorimeters and muon identification is aided by the instrumented iron return yoke.

A very important element of the detector design work has been the common effort to develop and apply simulation tools to realise realistic detector-performance estimates. A small group of experts from both SiD and ILD have cooperated closely on this critical work.

To preserve this unprecedented performance, the inner detectors must accommodate very low-mass detectors and supports. This is a significant challenge. The detector designs have achieved the required light-weight support structures with minimal dead spaces. This was greatly simplified by the ILC time structure of 1 millisecond bunch trains at 5 Hertz. This very sparse filling allows power for many of the detector subsystems to be switched off between bunch trains (so-called power pulsing), reducing the heat load and the need for cooling. The design of these power-pulsed systems presents a significant challenge, including the need for quiescent currents.

## 4.2 Backgrounds

While benign by LHC standards, ILC machine backgrounds must be understood and taken into account to give optimal performance of the detectors. A variety of processes create beam-induced backgrounds in the detectors. The main sources are beamstrahlung, which is induced by the interaction of one beam with the bunch with which it collides, synchrotron radiation, muons, and neutrons. Significant numbers of low-energy $\mathrm{e}^+\mathrm{e}^-$ pairs are produced at the interaction point from beamstrahlung; the interaction region layout has been designed to guide these background charged particles out of the detector by adding a dipole field to the conventional solenoidal field (so-called *anti-DID*). These background pairs must be dealt with in the vertex detector design, but they primarily affect the very forward detectors, which need to be able to withstand the significant radiation loads, and at the same time maintain sensitivity to single high-energy particles. An optimised collimation system is used to control the synchrotron radiation generated upstream. Various schemes are employed to minimise the impact of muons and neutrons.

Another important background comes from photon-photon collisions. These events produce high-transverse-momentum particles that overlap with the particles from the less-common hard-scattering events. Time stamping at the single-bunch level can help in reducing the number of overlapping events. In addition sophisticated algorithms have been developed to identify and subtract these events based on topology and detailed properties.

## 4.3 Beam Instrumentation

Precise knowledge of beam parameters is critical to the ILC physics program. Luminosity, beam energy, and polarisation are measured by instrumentation close to the main detectors. Low-angle Bhabha scattering detected by dedicated calorimeters can provide the necessary precision for the integrated luminosity. Acollinearity and energy measurements of Bhabha events in the polar-angle region from 120–400 mrad can be used to extract the luminosity as a function of energy, $\mathrm{d}L/\mathrm{d}E$. Beam-energy measurements with an accuracy of (100–200) parts per million are done both upstream and downstream of the collision point to provide redundancy and reliability of the results. Precise measurements of parity-violating asymmetries require polarisation measurements with a precision of 0.25% or better. High statistics Giga-Z running requires polarimetry at the 0.1% level. The primary polarisation measurement comes from dedicated Compton polarimeters detecting backscattered electrons and positrons. The best accuracy is achieved by implementing polarimeters both upstream and downstream of the interaction region (IR).

## 4.4 Two Detectors

The ILC has been designed to enable two experiments (SiD and ILD) sharing one interaction region using a push-pull approach. This two-detector design is motivated by the enhanced scientific productivity of past collider facilities which benefited from independent operation of multiple experiments, providing complementary strengths, cross-checking and confirmation of results, reliability, insurance against mishaps, competition between collaborations, as well as increased number of involved scientific personnel. Figure 4.1 shows the arrangement of the two detectors in the detector hall.

Both detector designs are conceived as multi-purpose detectors, optimised for the broad range of physics opportunities at the ILC. SiD is a compact, cost-constrained detector made possible with a 5 Tesla magnetic field and silicon tracking. Silicon enables time-stamping on single bunch crossings to provide robust performance, derived from immunity to spurious background bursts. The highly granular calorimeter is optimised for particle-flow analysis. The ILD group has designed a large detector with robust and stable performance over a wide range of energies. The concept uses a tracking system based on a continuous-readout time-projection chamber combined with silicon tracking for excellent efficiency and robust pattern-recognition performance. A granular calorimeter system contained inside a 3.5 T magnetic field provides very good particle-flow reconstruction. Both detectors provide flexibility for operation at energies up to the TeV range.

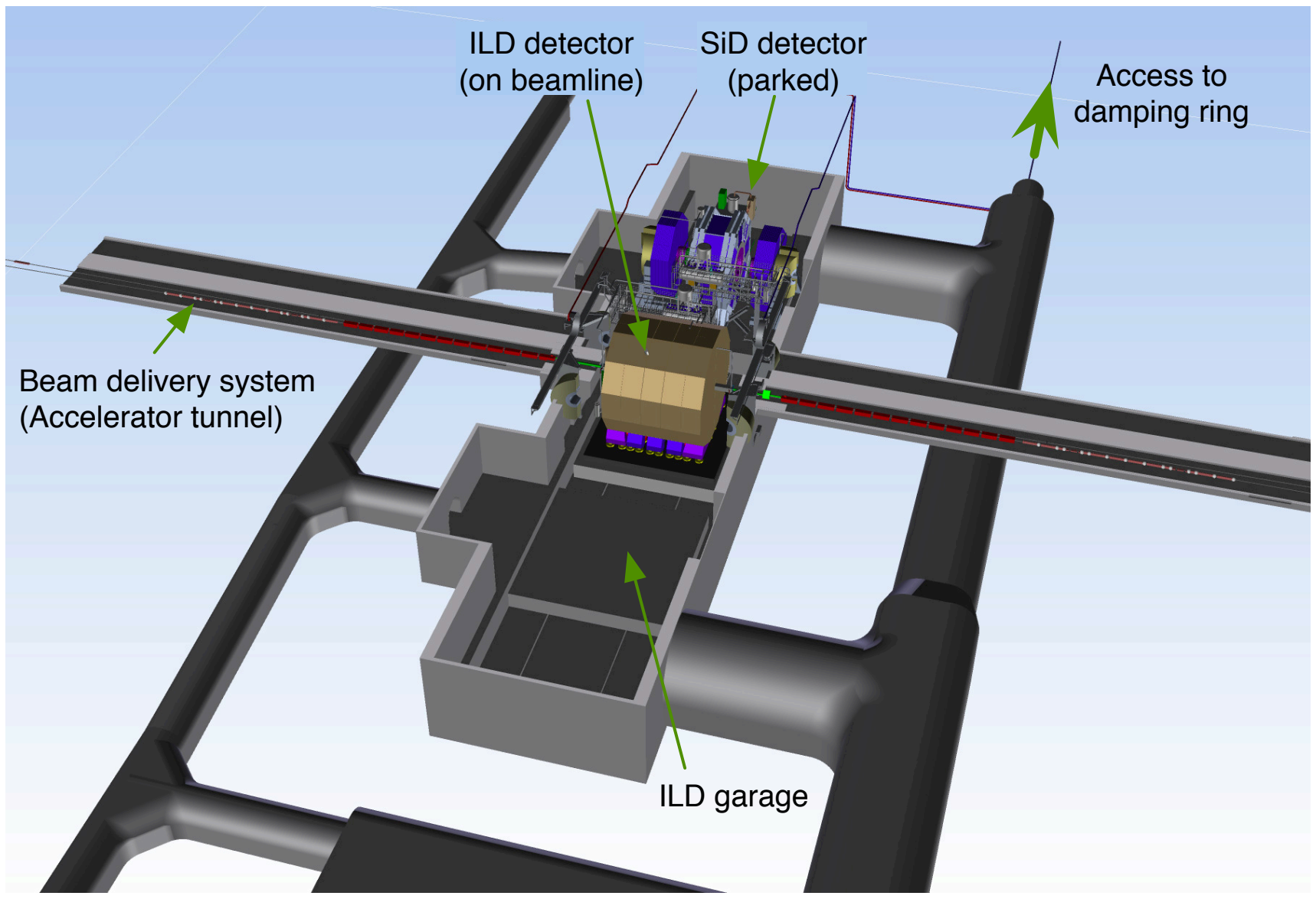

**Figure 4.1** Example layout of the detector hall for the Mountain Topography site, showing the location of the two detectors in a push-pull arrangement.

The push-pull operation scheme calls for one detector taking data while the other is out of the beam in a close-by maintenance position. At regular intervals, the data-taking detector is pushed laterally out of the interaction region, while the other detector is being pulled in. These intervals are short enough to ensure both acquire data on any potential discovery. The time for transition must be on the order of one day to maximise ILC integrated luminosity.

A time-efficient implementation of the push-pull operation sets specific requirements and challenges for many detector and machine systems, in particular the IR magnets, the cryogenics, the alignment system, the beamline shielding, the detector design, and the overall integration. The functional requirements and interface specifications for the push-pull IR have been developed, with a detailed design of technical systems and the experimental area from detailed engineering specifications.

The detector motion and support system has to ensure reliable push-pull operation for roughly one hundred moves over the life of the experiment, while preserving alignment of the detector's internal components and ensuring accuracy of detector positioning. The motion system must preserve structural integrity of the collider hall floor and walls, be compatible with vibration stability of the detector at the level of tens of nanometers and be compatible with earthquake-safety standards.

The detectors will be placed on platforms that preserve the detector alignment and distribute the load evenly onto the floor. Details of the design of such platforms and the interfaces between the platform and the detectors have been developed. The ILC detectors are self-shielding with respect to ionising radiation from maximum credible beam-loss scenarios. Additional shielding must fill the gap between the detector and the wall in the beam position.

The stray magnetic fields outside the iron return yokes of each detector must be small enough not to disturb the other detector during operation or maintenance. A limit for the magnetic fields is 5 mT at a lateral distance of 15 m from the beam line. Fringe fields from the detector return yokes have been carefully simulated and designs for both SiD and ILD have been developed to meet these requirements.

The installation schemes for the detectors and the layout of the experimental areas on the surface and underground depend on the geographical situation of the possible ILC sites. While the European and American sample sites assume a flat surface area, the Asian sample sites in Japan are located in mountains where the requirements for the conventional facilities and buildings are different.

## 4.5 Detector R&D

The high levels of performance in the ILC detector designs result from more than a decade of R&D conducted in the Americas, Asia, and Europe, by a number of R&D collaborations. Technological solutions for the key systems of the SiD and ILD designs have been developed and have been validated in extensive test-beam campaigns. Through internationally coordinated R&D groups, a successful, well-focused programme of innovative instrumentation development has been performed even with rather limited resources. A notable achievement has been the combined work of simulation and beam tests to verify the understanding and precision of the PFA calorimetry technique. These efforts have achieved the critical goals on all subsystems, and continue to advance the technology, further optimising the performance.

## 4.6 SiD

SiD is a general-purpose detector designed to perform precision measurements at a Linear Collider. It satisfies the challenging detector requirements for physics at the ILC. SiD is the result of many years of creative design by physicists and engineers, backed up by a substantial body of past and ongoing detector research and development. While each component has benefitted from continual development, the SiD design integrates these components into a complete system for excellent measurements of jet energies, based on the Particle Flow Algorithm (PFA) approach, as well as of charged leptons, photons and missing energy. The use of robust silicon vertexing and tracking makes SiD applicable to a wide range of energies from a Higgs factory to beyond 1 TeV. SiD has been designed in a cost-conscious manner, with the compact design that minimises the volumes of high-performing, high-value, components, while maintaining critical levels of performance. The restriction on dimensions is offset by the relatively high central magnetic field from a superconducting solenoid.

### 4.6.1 The SiD detector

SiD is a compact detector based on a powerful silicon pixel vertex detector, silicon tracking, silicon-tungsten electromagnetic calorimetry (ECAL) and highly segmented hadronic calorimetry (HCAL). SiD also incorporates a high-field solenoid, iron flux return, and a muon identification system (see Fig. 4.2).

The choice of silicon detectors for tracking and vertexing ensures that SiD is robust with respect to beam backgrounds or beam loss, provides superior charged-particle momentum resolution, and eliminates out-of-time tracks and backgrounds. The main tracking detector and calorimeters are "live" only during each single bunch crossing, so beam-related backgrounds and low-$p_\mathrm{T}$ backgrounds from $\gamma\gamma \to \mathrm{hadrons}$ processes will be reduced to the minimum possible levels. The SiD calorimetry is optimised for excellent jet-energy measurement using the PFA technique. The complete tracking and calorimeter systems are contained within a superconducting solenoid, which has a 5 T field strength, enabling the overall compact design. The coil is located within a layered iron structure that returns the magnetic flux and is instrumented to allow the identification of muons.

The tracking system is a key element as the particle-flow algorithm requires excellent tracking with superb efficiency and good two-particle separation. The requirements for precision measurements, in particular in the Higgs sector, place high demands on the momentum resolution at the level of $\delta(1/p_\mathrm{T}) \sim 2$–$5 \times 10^{-5}\,(\mathrm{GeV/c})^{-1}$ and the material budget of the tracking system. Highly efficient tracking is achieved using the pixel detector and main tracker to recognise and measure prompt tracks.

The SiD vertex detector uses a barrel and disk layout. The barrel section consists of five silicon pixel layers with a pixel size of $20\times20\,\mu\mathrm{m}^2$. The forward and backward regions each have four silicon pixel disks. In addition, there are three silicon pixel disks at a larger distance from the interaction point to provide uniform coverage for the transition region between the vertex detector and the outer tracker. This configuration provides for very good hermeticity with uniform coverage and guarantees excellent charged-track pattern-recognition capability and impact-parameter resolution over the full solid angle. The vertex detector design relies on power pulsing during bunch trains to minimise heating and uses forced air for its cooling. The main tracker technology of choice is silicon-strip sensors arrayed in five nested cylinders in the central region with an outer cylinder radius of 1.25 m and four disks in each of the endcap regions. The geometry of the endcaps minimises the material budget to enhance forward tracking. The detectors are single-sided silicon sensors with a readout pitch of 50 μm.

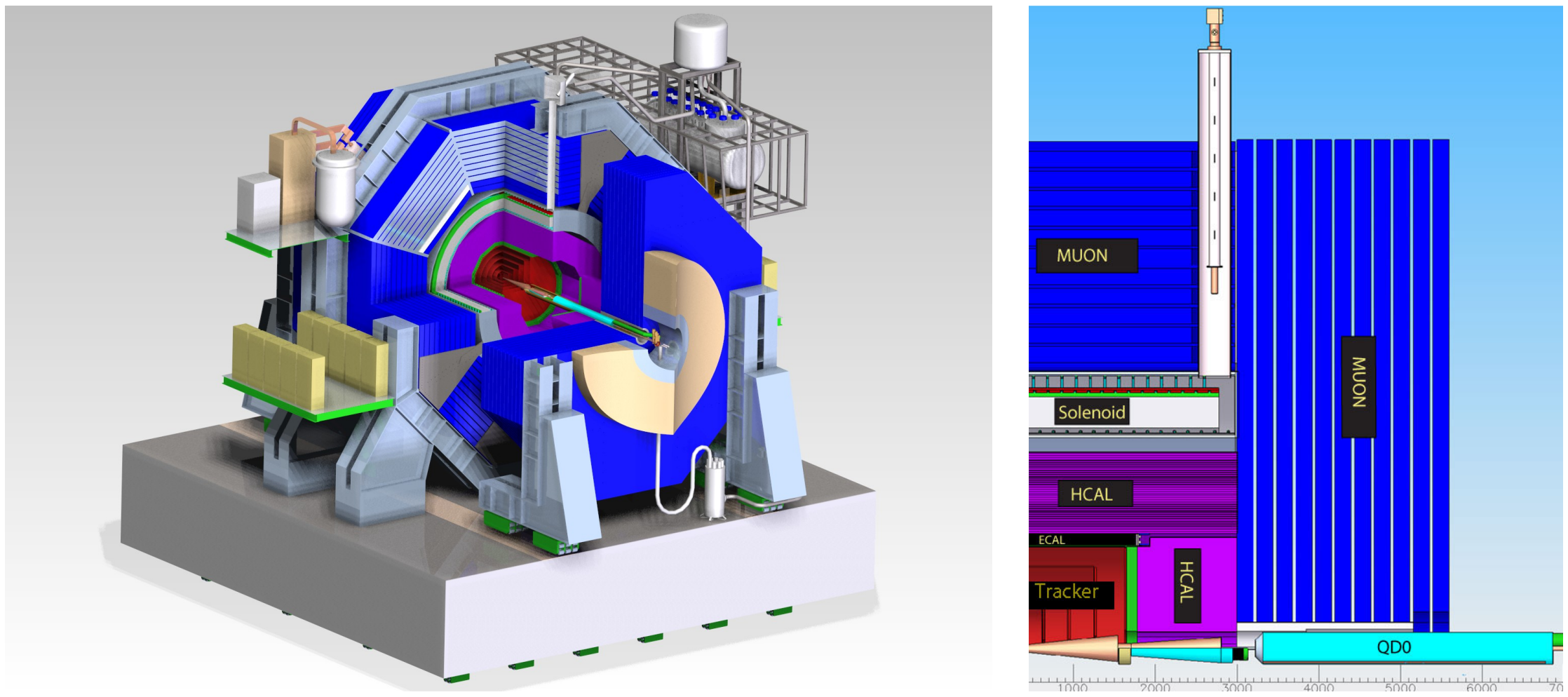

**Figure 4.2.** The SiD detector, showing (left) an isometric view on the platform, and (right) a quadrant section. Colour coding: tracking (red), ECAL (green), HCAL (violet) and the flux return (blue).

The choice of PFA imposes a number of basic requirements on the calorimetry. The central calorimeter system must be contained within the solenoid in order to reliably associate tracks to energy deposits. The electromagnetic and hadronic sections must have imaging capabilities that allow both efficient track-following and correct assignment of energy clusters to tracks. These requirements imply that the calorimeters must be finely segmented both longitudinally and transversely.

The combined ECAL and HCAL systems consist of a central barrel part and two endcaps, nested inside the barrel. The entire barrel system is contained within the volume of the cylindrical superconducting solenoid. The electromagnetic calorimeter has silicon active layers between tungsten absorber layers. The active layers use $3.5{\times}3.5\,\text{mm}^2$ hexagonal silicon pixels, which provide excellent spatial resolution. The structure has 30 layers in total, the first 20 layers having a thinner absorber than the last ten layers. This configuration is a compromise between cost, electromagnetic shower radius, sampling frequency, and shower containment. The total depth of the electromagnetic calorimeter is 26 radiation lengths ($\mathrm{X}_0$) and one nuclear interaction length. The hadronic calorimeter has a depth of 4.5 nuclear interaction lengths, consisting of alternating steel plates and active layers. The baseline choice for the active layers is the glass resistive-plate chamber with an individual readout segmentation of $10{\times}10\,\text{mm}^2$. Two special calorimeters are foreseen in the very forward region: LumiCal for precise measurement, and BeamCal for fast estimation, of the luminosity.

The SiD superconducting solenoid is based on the CMS solenoid design philosophy and construction techniques, using a slightly modified CMS conductor as its baseline design. Superconducting strand count in the coextruded Rutherford cable was increased from 32 to 40 to accommodate the higher 5 T central field. The flux-return yoke is instrumented with position sensitive detectors to serve as both a muon filter and a tail catcher. The SiD Muon System baseline design is based on scintillator technology, using extruded scintillator readout with wavelength-shifting fibre and SiPMs. Simulation studies have shown that nine or more layers of sensitive detectors yield adequate energy measurements and good muon-detection efficiency and purity.

### 4.6.2 SiD detector performance

A large fraction of the software for the generation, simulation and reconstruction is shared between the detector concepts. The SiD detector is fully implemented and simulated using SLIC, which is based on GEANT4. The background originating from incoherent pair interactions and from $\gamma\gamma \to \mathrm{hadrons}$ for one bunch crossing is fully taken into account by the simulation. The events are then passed through the reconstruction software suite, which encompasses digitisation, tracking, vertexing and the Pandora PFA algorithm. The material budget of the simulated tracker and the simulated tracking performance for single particles are shown in Fig. 4.3.

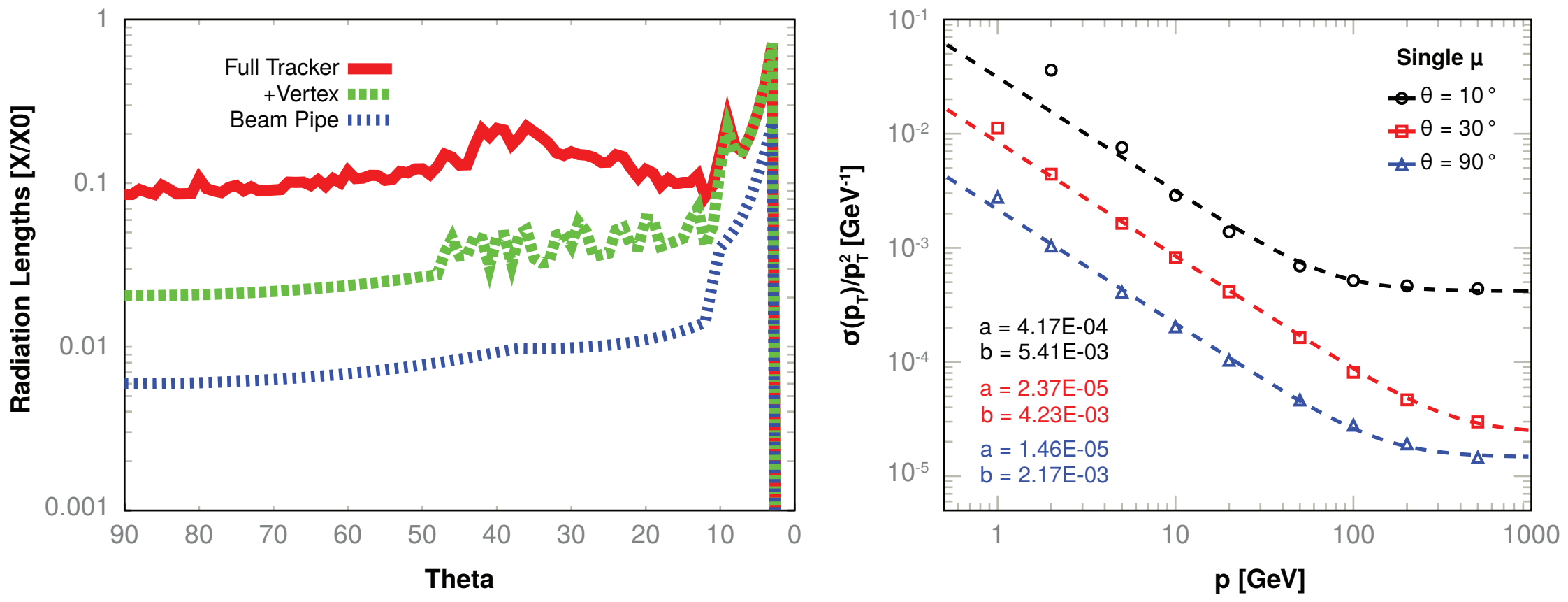

**Figure 4.3.** Left: SiD Tracker Material budget in terms of $\mathrm{X}_0$. Right: the normalised transverse momentum resolution for single-muon events.

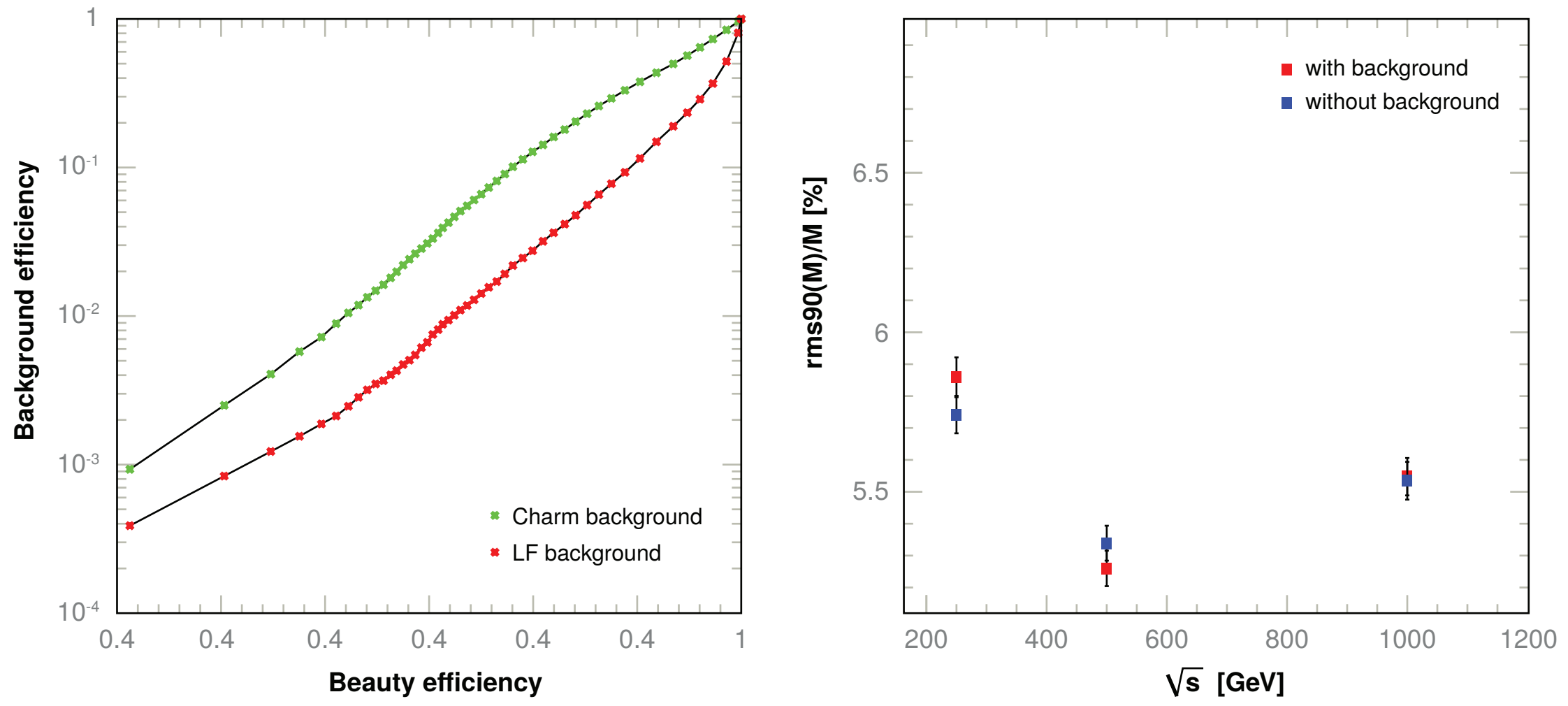

**Figure 4.4.** Left: Mis-identification efficiency of light quark (red points) and c quark events (green points) as b quark jets versus the b identification efficiency in di-jet events at $\sqrt{s} = 91$ GeV including background from $\gamma\gamma \to \mathrm{hadrons}$ and incoherent pairs. Right: Mass resolution of reconstructed $\mathrm{ZZ}$ events with and without the backgrounds from $\gamma\gamma \to \mathrm{hadrons}$ and incoherent pairs at different values of $\sqrt{s}$.

The material budget of the entire tracking system is less than 0.2 $\mathrm{X}_0$ down to very low angles. The current design achieves an asymptotic momentum resolution of $\delta(1/p_\mathrm{T}) = 1.46 \times 10^{-5}\,(\mathrm{GeV/c})^{-1}$ and an transverse impact parameter resolution better than 2 µm. The ability to tag bottom and charm decays with high purity has been a driving factor in the design of the vertex detector. Figure 4.4 (left) illustrates the capability of the SiD to separate b-quarks also in the presence of the full beam background.

Besides the detector performance, sophisticated reconstruction algorithms are necessary to obtain a jet-energy resolution that allows the separation of hadronic W and Z decays. To avoid a bias from possible tails, the $\mathrm{rms}_{90}$ value is computed to describe the energy or mass resolution of a particle-flow algorithm. It is defined as the standard deviation of the distribution in the smallest range that contains 90% of the events. Figure 4.4 (right) shows the mass resolution of reconstructed $\mathrm{Z}$ bosons in $e^+e^- \to \mathrm{ZZ}$ events at different collision energies, where one $\mathrm{Z}$ decays to neutrinos, the other to two light quarks that give rise to two jets.

### 4.6.3 SiD costing

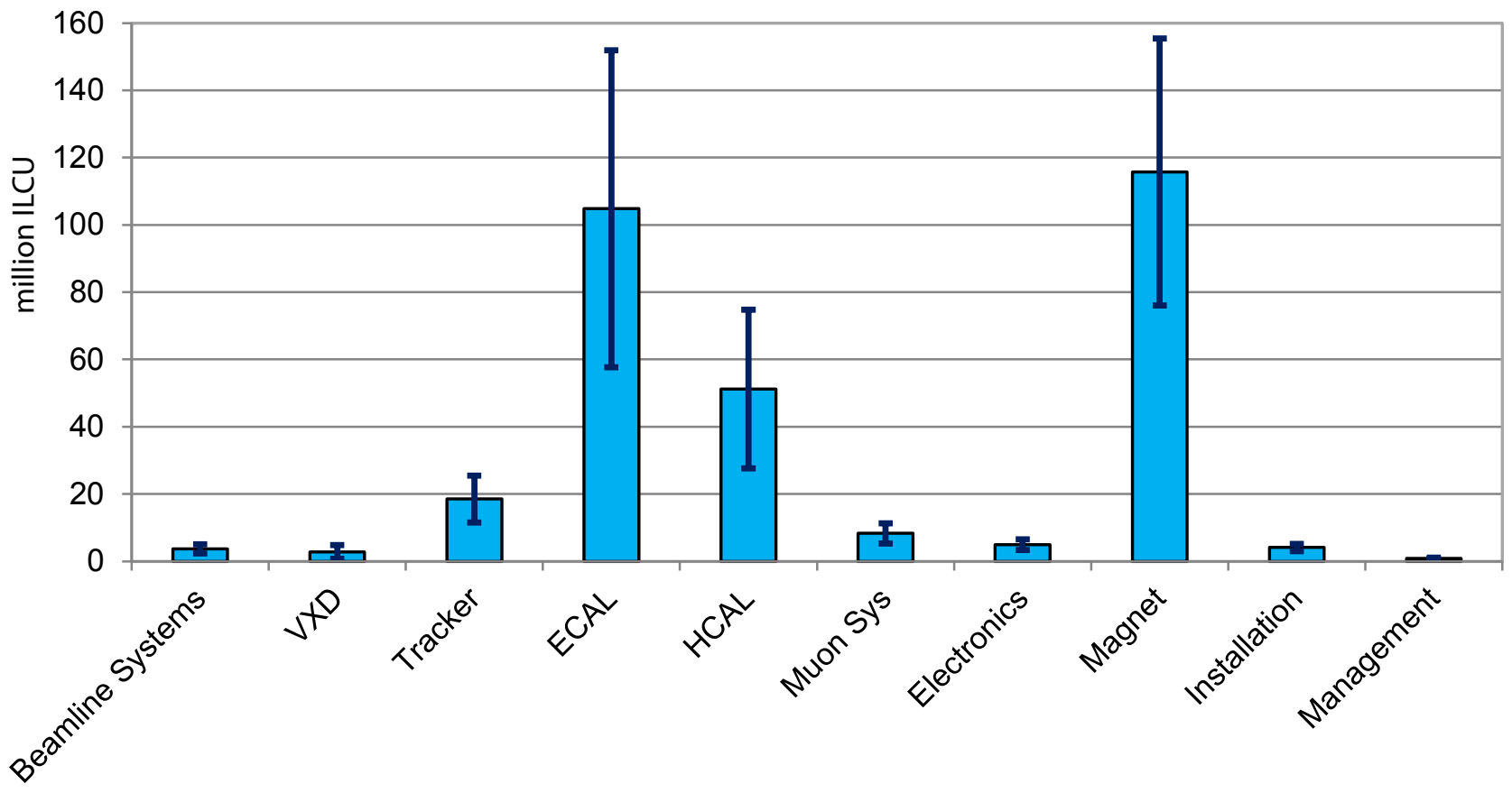

**Figure 4.5** Subsystem M&S Costs in million ILCU. The error bars indicate the subsystem contingency.

The SiD cost estimate is a construction cost estimate; it does not include R&D, commissioning, operating costs, or physicist salaries. The parametric model of the detector maintains a self-consistent model of SiD. It is also straightforward to vary the SiD detector parameters. For each system, the cost-driving component count is calculated. The model then estimates both M&S and labour costs that are associated with the actual scale of SiD. Contingency is estimated for each quantity to estimate the uncertainties in the costs of the detector components (see Fig. 4.5). A set of unit costs for some basic commodity items, which has been agreed upon for ILC and CLIC detector cost estimates, is used in the SiD and ILD cost estimates.

The SiD cost is 315 million ILCU for M&S, 316 thousand person-hours engineering, 904 thousand person-hours technical, and 51 thousand person-hours administrative labour. The estimated M&S contingency, reflecting uncertainty in unit costs and some estimate of the maturity of this study, is 127 million ILCU.

## 4.7 ILD

The ILD concept has been designed as a multi-purpose detector. It has been designed for optimal particle-flow (PFA) performance. A high-precision vertex detector is followed by a hybrid tracking system, realised as a combination of silicon tracking with a time-projection chamber, and a calorimeter system. The complete system is located inside a 3.5 T solenoid. The inner-detector system is highly granular, and provides a robust and detailed three-dimensional imaging capability of the events. On the outside of the coil, the iron return yoke is instrumented as a muon system and as a tail-catcher calorimeter. The detector is shown in Fig. 4.6

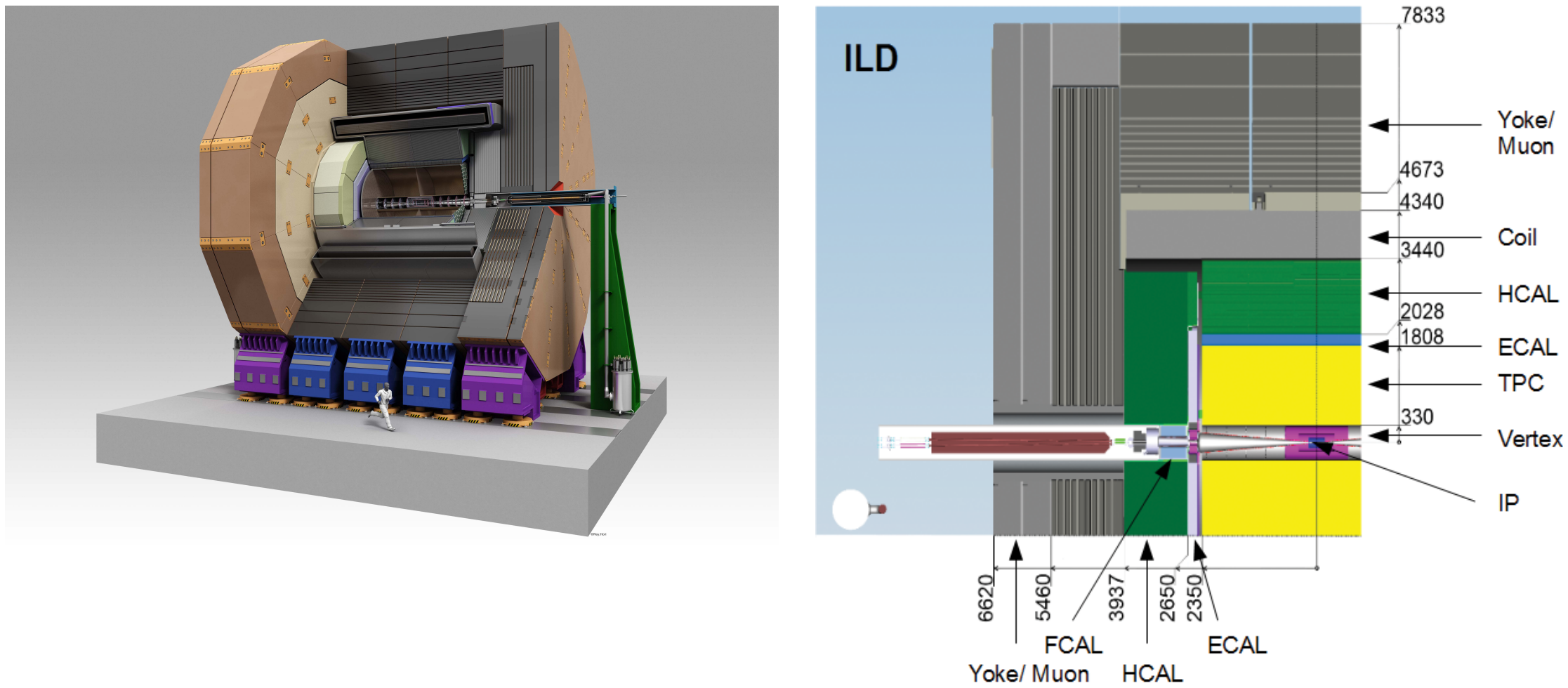

**Figure 4.6.** Views of the ILD detector concept. The interaction point in the quadrant view (right) is in the lower right corner of the picture. Dimensions are in mm.

## 4.7.1 The ILD detector

The vertex detector is realised as a multi-layer pixel vertex detector (VTX), with three superlayers each comprising two layers, or as a 5 layer geometry. In either case the detector has a pure barrel geometry. To minimise the occupancy from background hits, the first superlayer is only half as long as the outer two. Whilst the underlying detector technology has not yet been decided, the VTX is optimised for point resolution and minimum material thickness.

A system of silicon strip and pixel detectors surrounds the VTX detector. In the barrel, two layers of silicon strip detectors (SIT) are arranged to bridge the gap between the VTX and the TPC. In the forward region, a system of two silicon-pixel disks and five silicon-strip disks (FTD) provides low angle tracking coverage.

A distinct feature of ILD is a large-volume time-projection chamber (TPC) with up to 224 points per track. The TPC is optimised for 3-dimensional point resolution and minimum material in the field cage and in the end-plate. It also allows $\mathrm{d}E/\mathrm{d}x$-based particle identification.

Outside the TPC a system of Si-strip detectors, one behind the end-plate of the TPC (ETD) and one in between the TPC and the ECAL (SET), provide additional high-precision space points which improve the tracking performance and provide additional redundancy in the regions between the main tracking volume and the calorimeters.

A highly segmented electromagnetic calorimeter (ECAL) provides up to 30 samples in depth and small transverse cell size, split into a barrel and an end-cap system. Tungsten has been chosen as absorber; for the sensitive area, silicon diodes, scintillator strips or a combination are considered.

The ECAL is followed by a highly segmented hadronic calorimeter (HCAL) with up to 48 longitudinal samples and small transverse cell sizes. Two options are considered, both based on a steel-absorber structure. One option uses scintillator tiles of $3{\times}3\,\mathrm{cm}^2$, which are read out with an analogue system. The second uses a gas-based readout which allows a $1{\times}1\,\mathrm{cm}^2$ cell geometry with a binary or semi-digital readout of each cell.

At very forward angles, below the coverage provided by the ECAL and the HCAL, a system of high-precision and radiation-hard calorimetric detectors (LumiCAL, BeamCAL, LHCAL) is foreseen. These extend the calorimetric solid-angle coverage to almost $4\pi$, measure the luminosity, and monitor the quality of the colliding beams.

A large volume superconducting coil surrounds the calorimeters, creating an axial $B$-field of nominally 3.5 Tesla. An iron yoke, instrumented with scintillator strips or resistive plate chambers (RPCs), returns the magnetic flux of the solenoid, and, at the same time, serves as a muon filter, muon detector and tail-catcher calorimeter.

To maximise the sensitivity of the detector to the physics at the ILC, the detector will be operated in a continuous readout mode, without a traditional hardware based trigger.

In close collaboration with the detector R&D groups, all key components of the ILD proposal have been evaluated and the key performance criteria have been demonstrated in test-beam experiments. This is of particular importance for the very ambitious calorimeter proposal.

A first engineering study of the integration of the ILD detector has been performed. This study has included a detailed review of the detector components, their sizes, and in particular the support structures needed to mount the detector. Assembly and maintenance procedures have been simulated to validate the integration concept. Estimates of services needed have been included as far as possible, and realistic tolerances have been inserted into the designs. A detailed model of the complete detector has been built using modern CAD tools, and compared to the simulation model used for the performance evaluation.

## 4.7.2 ILD detector performance

The performance of the ILD concept has been extensively studied using a detailed GEANT4 based simulation model and sophisticated reconstruction tools. Backgrounds have been taken into account to the best of current knowledge.

A key characteristic of the detector is the amount of material within it. Particle flow requires a thin tracker, to minimise interactions before the calorimeters, and thick calorimeters, to fully absorb the showers. Figure 4.7 (left) shows the material in the detector in radiation lengths, for the different parts of the tracking system. The amount of material up to the end of the tracking is mostly below 10% for the full solid angle. The right-hand plot shows the total interaction length including the calorimeter system. Clearly visible is the increase of material in the end-cap due to the support structure for the readout in the TPC. Since this material however is located very close to the end-cap calorimeter it has only a very small negative impact on the performance, as can be seen from the top curve in the left plot, the total number of radiation length until the first active layer inside the ECAL.

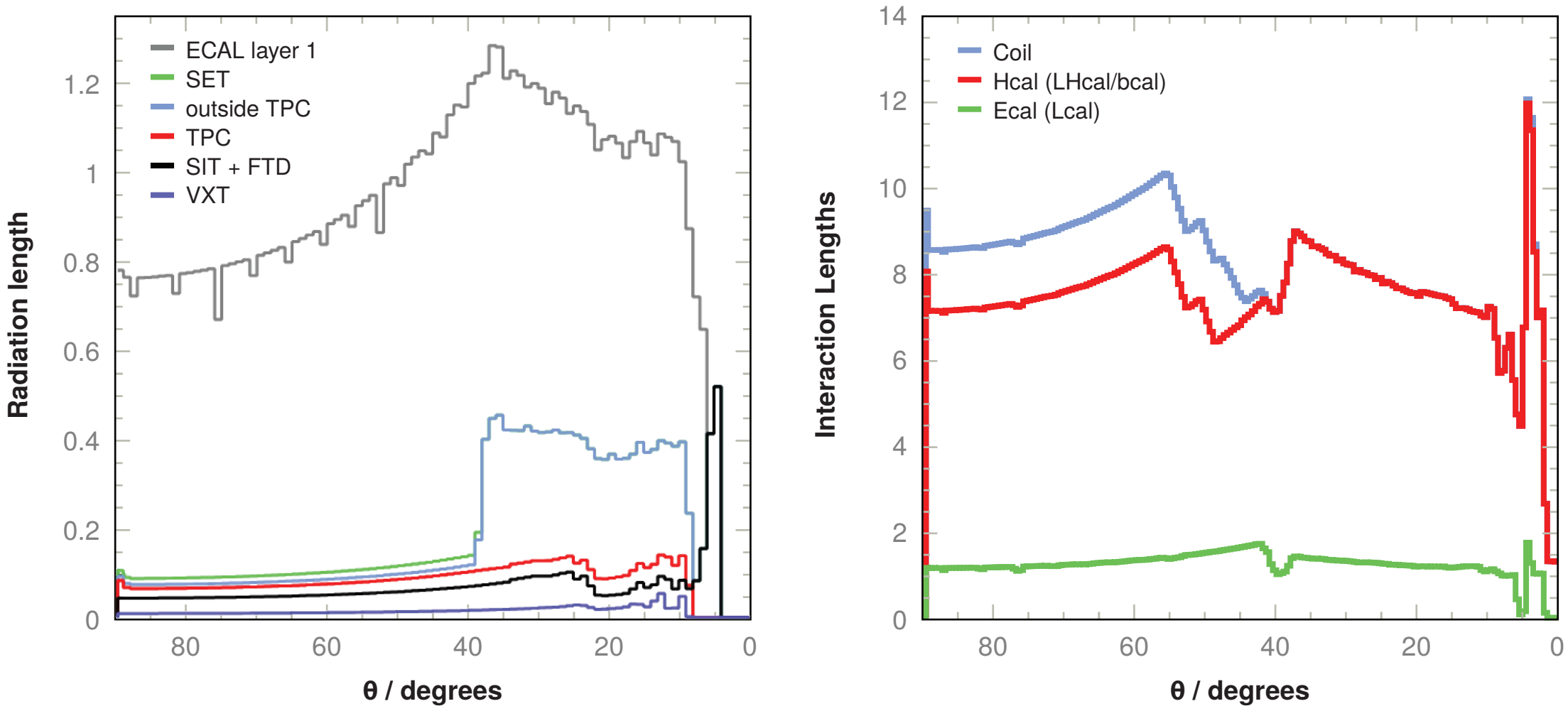

**Figure 4.7.** Left: Average total radiation length of the material in the tracking detectors as a function of polar angle. The red curve shows the material in the tracking system without the outer material in field cage or endplate, the grey line the total material up to the first active layer inside the ECAL is reached. Right: Total interaction length in the detector, up to the end of the calorimeter system (red line), and including the coil of the detector (light-blue line).

The performance of the tracking system can be summarised by its combined momentum resolution, shown in Fig. 4.8 (left). A resolution of $\sigma_{1/p_T} = 2 \times 10^{-5}$ GeV$^{-1}$ has been achieved for high momenta.

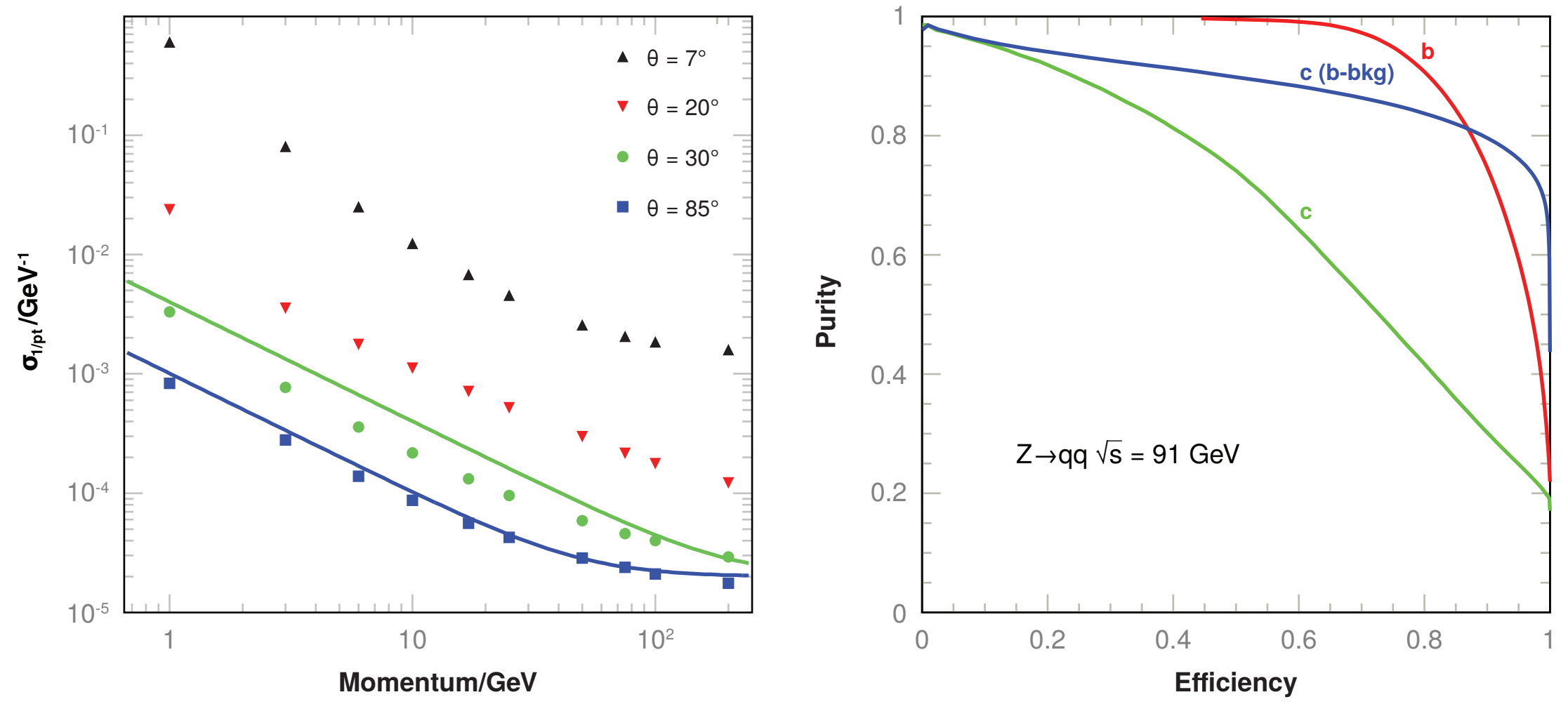

**Figure 4.8.** Left: Momentum resolution as a function of the transverse momentum of particles, for tracks with different polar angles. Also shown is the theoretical expectation. Right: Flavour tagging performance for $Z \to q\bar{q}$ samples at 91 GeV.

For many physics studies the tagging of long-lived particles is of key importance. The ability of ILD to reconstruct displaced vertices is shown in Fig. 4.8 (right).

Calorimeter system and tracking system enter into the particle flow performance together. The performance of the ILD detector for different energies and as a function of the polar angle is shown in Fig. 4.9. The jet-energy resolution stays below 4% for nearly the full solid angle, and is nearly independent of the centre-of-mass energy.

**Figure 4.9**
Fractional jet-energy resolution plotted against $|\cos\theta|$ where $\theta$ is the polar angle of the thrust axis of the event.

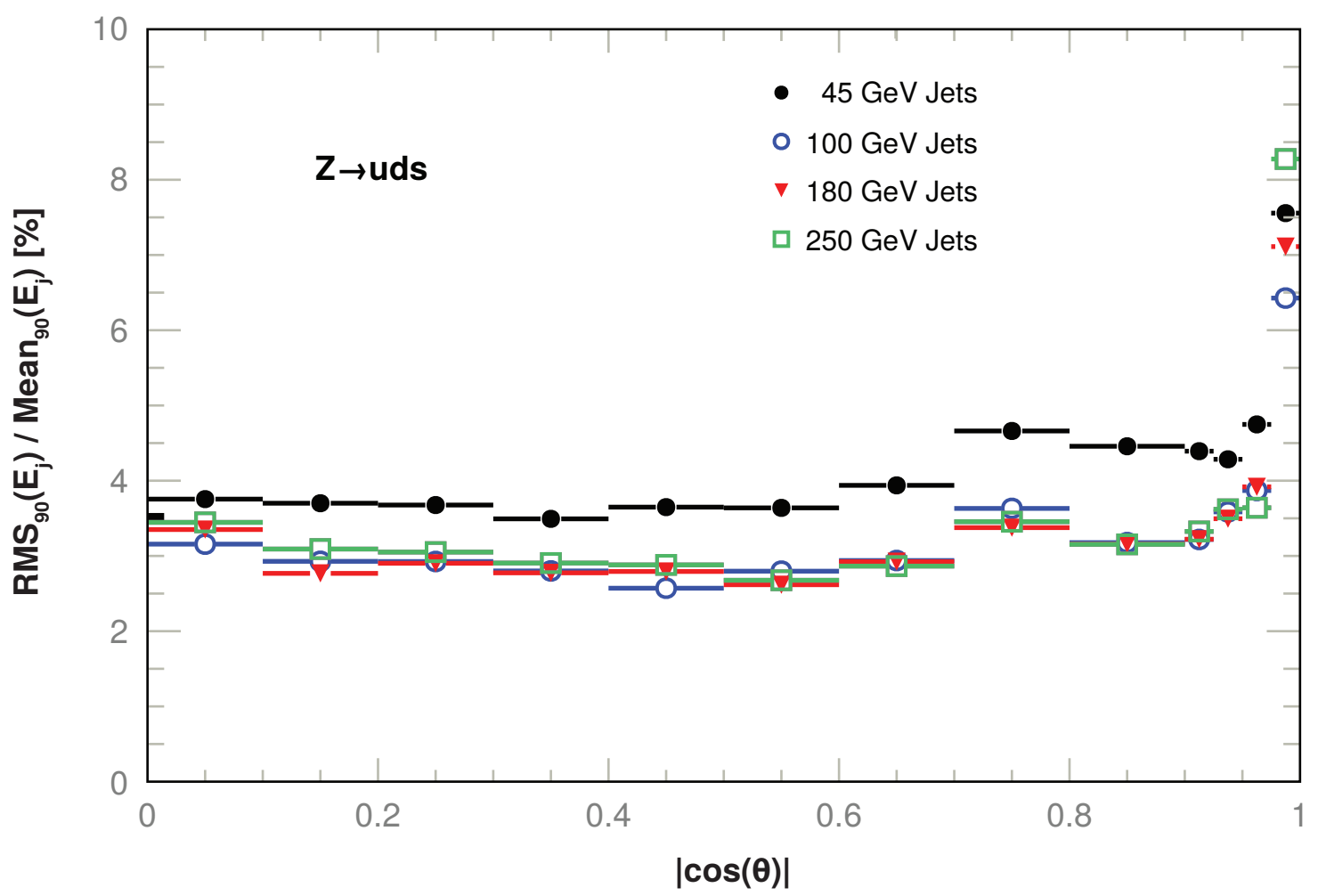

## 4.7.3 ILD costing

A significant effort has been undertaken to evaluate the cost of the ILD concept. In most cases the prototypes constructed over the past few years have given important input to validate the costs, and to provide guidance towards large-scale production of detector components. For the most relevant items the estimates are based on quotes from manufactures, and are often based on extensive discussions.

Depending on the option chosen, the cost of the ILD concept varies between 350 and 440 million ILCU. This cost does not include labour, or any contingency.

The distribution of costs among the different ILD systems is shown in Fig. 4.10 as a fraction of the total cost of the detector. As basis for this plot an average cost of the ILD concept of 392 million ILCU was used.

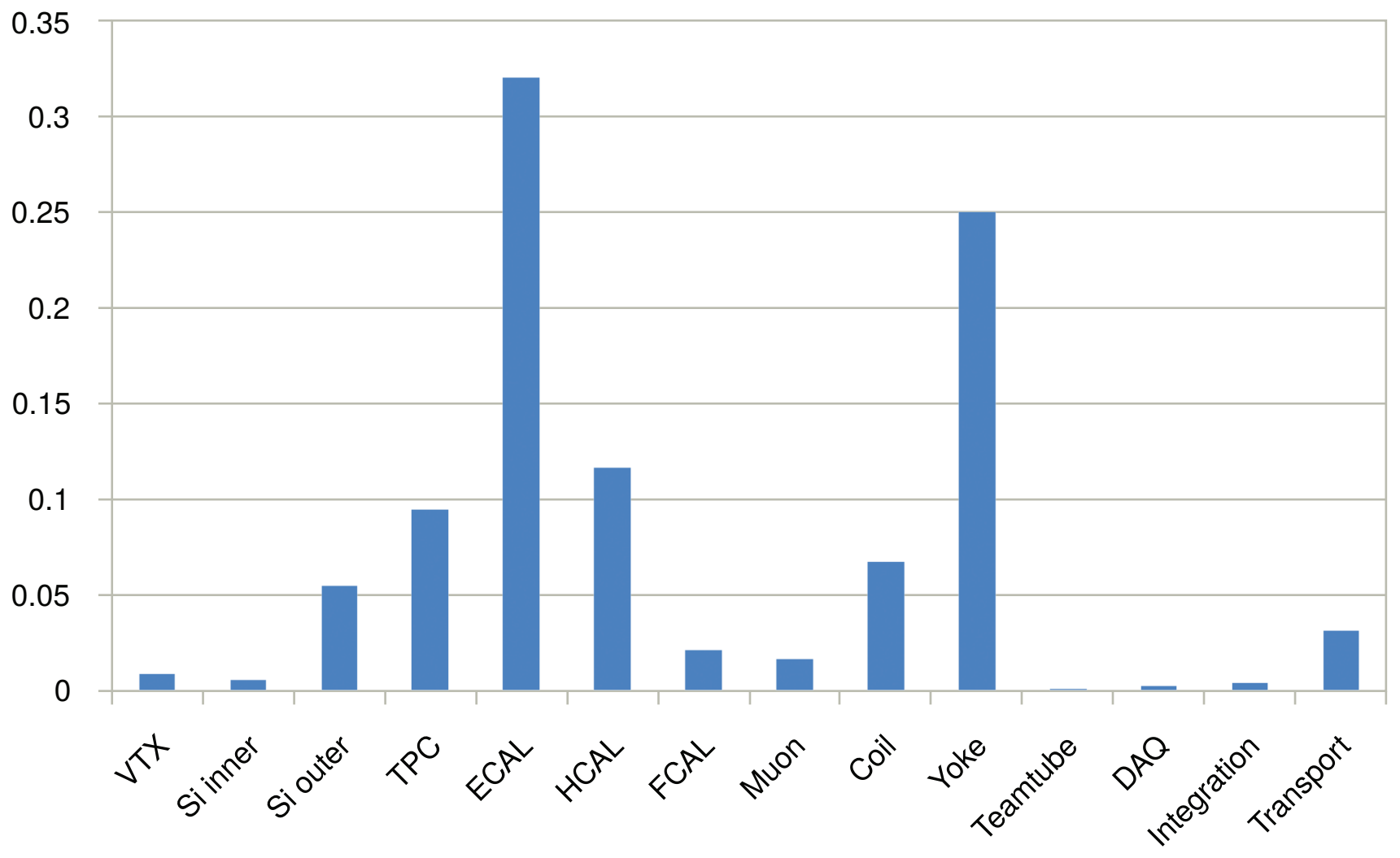

**Figure 4.10** Costs distribution of ILD (shown as a fraction of the total cost).

The costs are dominated by the coil/yoke system, and by the calorimeter system, in particular the electromagnetic calorimeter. While the scope for the reduction of the costs for the coil/yoke are limited and largely depend on the requirement of a very small magnetic stray field outside the detector, large efforts are ongoing to optimise the cost of the active detector elements. A large part of the cost of the electromagnetic calorimeter come from the sensors. Significant technological progress over the past years has already resulted in significantly reduced costs, and more progress might be expected in the future. The subdetector technology itself is also undergoing a vigorous optimisation, including the exploration of hybrid solutions from silicon- and scintillator-based systems. Re-optimisation of the ILD detector for the anticipated physics program, in view of the much improved understanding of technologies and costs, will be one of the challenges in the near future.

# Chapter 5
# Conclusion

The ILC accelerator project described in this Executive Summary and in the volumes of the *Technical Design Report* is technically mature and ready for construction. The elements of the project have been carefully costed and reviewed by teams of experts. The scientific case for constructing the ILC is extraordinarily strong; the discovery at the LHC of a 125 GeV boson displaying many of the properties expected of the Higgs boson provides a compelling scientific target for the ILC. In addition to studies of this new Higgs boson candidate, operating the ILC at top threshold at 350 GeV provides crucial precision in the study of top quark properties, surpassing LHC precision. Above 500 GeV the Higgs self-coupling, the top Yukawa coupling and longitudinal W scattering can be investigated with very high accuracy. Finally, the ILC provides new physics discovery potential through a variety of channels, including dark-matter searches and searches for new interactions at higher mass scales through precision studies of pair-production of quarks, leptons, and W and Z bosons. Detector designs have been developed, simulated, prototyped and validated to demonstrate the required high level detector performance is achievable.

The decades spent to prove that the ILC can be built and optimally exploited have now reached a culmination in the production of this *Technical Design Report*. The next steps are to choose a suitable site, allowing necessary site-dependent design details to be determined, while establishing a world-wide consensus to construct the project and setting up an appropriate international organisation. One of the most exciting scientific projects ever conceived is poised to begin.

# Contributing Institutes

Academia Sinica - 128 Sec. 2, Institute of Physics, Academia Rd., Nankang, Taipei 11529, Taiwan, R.O.C.
AGH University of Science and Technology, Akademia Gorniczo-Hutnicza im. Stanislawa Staszica w Krakowie, Al. Mickiewicza 30 PL-30-059 Cracow, Poland
Aichi Medical University, Nagakute, Aichi, 480-1195, Japan
Akita International University, Yuwa, Akita City, 010-1292, Japan
Albert-Ludwigs Universität Freiburg, Physikalisches Institut, Hermann-Herder Str. 3, D-79104 Freiburg, Germany
Ankara Üniversitesi Fen Fakültesi, Fizik Bölümü, Dögol Caddesi, 06100 Tandoĝan Ankara, Turkey
Argonne National Laboratory (ANL), 9700 S. Cass Avenue, Argonne, IL 60439, USA
Baylor University, Department of Physics, 101 Bagby Avenue, Waco, TX 76706, USA
Beijing University, Department of Physics, Beijing, China 100871
Benares Hindu University, Benares, Varanasi 221005, India
Borough of Manhattan Community College, The City University of New York, Department of Science, 199 Chambers Street, New York, NY 10007, USA
Brandenburg University of Technology, Postfach 101344, D-03013 Cottbus, Germany
Brookhaven National Laboratory (BNL), P.O.Box 5000, Upton, NY 11973-5000, USA
Brown University, Department of Physics, Box 1843, Providence, RI 02912, USA
Budker Institute for Nuclear Physics (BINP), 630090 Novosibirsk, Russia
Calcutta University, Department of Physics, 92 A.P.C. Road, Kolkata 700009, India
California Institute of Technology, Physics, Mathematics and Astronomy (PMA), 1200 East California Blvd, Pasadena, CA 91125, USA
California State University, Los Angeles, Dept. of Physics and Astronomy, 5151 State University Dr., Los Angeles, CA 90032, USA
Carleton University, Department of Physics, 1125 Colonel By Drive, Ottawa, Ontario, Canada K1S 5B6
Carnegie Mellon University, Department of Physics, Wean Hall 7235, Pittsburgh, PA 15213, USA
CEA Saclay, IRFU, F-91191 Gif-sur-Yvette, France
CEA Saclay, Service de Physique Théorique, CEA/DSM/SPhT, F-91191 Gif-sur-Yvette Cedex, France
Center for High Energy Physics (CHEP) / Kyungpook National University, 1370 Sankyuk-dong, Buk-gu, Daegu 702-701, Republic of Korea
Center for High Energy Physics (TUHEP), Tsinghua University, Beijing, China 100084
Center For Quantum Spacetime (CQUeST), Sogang University, 35 Baekbeom-ro, Mapo-gu, Seoul 121-742, Republic of Korea
Centre de Physique des Particules de Marseille (CPPM), Aix-Marseille Université, CNRS/IN2P3, 163, Avenue de Luminy, Case 902, 13288 Marseille Cedex 09, France
Centre de Physique Theorique, CNRS - Luminy, Universiti d"Aix - Marseille II, Campus of Luminy, Case 907, 13288 Marseille Cedex 9, France
Centre Lasers Intenses et Applications (CELIA), Université Bordeaux 1 - CNRS - CEA, 351 Cours de la Libération, 33405 Talence Cedex, France
Centro Brasileiro de Pesquisas Físicas (CBPF), Rua Dr. Xavier Sigaud, n.150 22290-180, Urca - Rio de Janeiro, RJ, Brazil
Centro de Investigaciones Energéticas, Medioambientales y Tecnológicas, CIEMAT, Avenida Complutense 22, E-28040 Madrid, Spain
Centro Nacional de Microelectrónica (CNM), Instituto de Microelectrónica de Barcelona (IMB), Campus UAB, 08193 Cerdanyola del Vallès (Bellaterra), Barcelona, Spain
CERN, CH-1211 Genève 23, Switzerland
Charles University, Institute of Particle & Nuclear Physics, Faculty of Mathematics and Physics, V Holesovickach 2, CZ-18000 Prague 8, Czech Republic
Chiba University of Commerce, 1-3-1 Konodai, Ichikawa-shi, Chiba, 272-8512, Japan
Chonbuk National University, Division of Science Education, Jeonju 561-756, Republic of Korea
Chonbuk National University, Physics Department, Jeonju 561-756, Republic of Korea
Chubu University, 1200 Matsumoto-cho, Kasugai-shi, Aichi, 487-8501, Japan
Chung Yuan Christian University, Department of Physics, 200 Chung Pei Rd., Chung Li 32023 Taiwan, R.O.C
Cockcroft Institute, Daresbury, Warrington WA4 4AD, UK
College of William and Mary, Department of Physics, Williamsburg, VA, 23187, USA
Columbia University, Department of Physics, New York, NY 10027-6902, USA
Cornell University, Laboratory for Elementary-Particle Physics (LEPP), Ithaca, NY 14853, USA
Czech Technical University in Prague, Institute of Experimental and Applied Physics (IEAP), Horska 3a/22, 12800 Prague 2, Czech Republic
Czech Technical University, Faculty of Nuclear Science and Physical Engineering, Brehova 7, CZ-11519 Prague 1, Czech Republic
Departamento de Física Teórica, Facultad de Ciencias, Módulo 15 (antiguo C-XI) y Módulo 8, Universidad Autónoma de Madrid, Campus de Cantoblanco, 28049 Madrid, Spain
Deutsches Elektronen-Synchrotron DESY, A Research Centre of the Helmholtz Association, Notkestrasse 85, 22607 Hamburg, Germany (Hamburg site)
Deutsches Elektronen-Synchrotron DESY, A Research Centre of the Helmholtz Association, Platanenallee 6, 15738 Zeuthen, Germany (Zeuthen site)
Durham University, Department of Physics, Ogen Center for Fundamental Physics, South Rd., Durham DH1 3LE, UK
École Normale Supérieure de Lyon, 46 allée d'Italie, 69364 Lyon Cedex 07, France
Ege University, Department of Physics, Faculty of Science, 35100 Izmir, Turkey
Enrico Fermi Institute, University of Chicago, 5640 S. Ellis Avenue, RI-183, Chicago, IL 60637, USA
ETH Zürich, Institute for Particle Physics (IPP), Schafmattstrasse 20, CH-8093 Zürich, Switzerland
ETH Zürich, Institute for Theoretical Physics (ITP), Wolfgang-Pauli-Str. 27, Zürich, Switzerland
European Spallation Source ESS AB, Box 176, 221 00 Lund, Sweden
Ewha Womans University, 11-1 Daehyun-Dong, Seodaemun-Gu, Seoul, 120-750, Republic of Korea
Excellence Cluster Universe, Technische Universität München, Boltzmannstr. 2, 85748 Garching, Germany
Fermi National Accelerator Laboratory (FNAL), P.O.Box 500, Batavia, IL 60510-0500, USA
Florida State University, Department of Physics, 77 Chieftan Way, Tallahassee, FL 32306-4350, USA

Fujita Gakuen Health University, Department of Physics, Toyoake, Aichi 470-1192, Japan
Fukui University of Technology, 3-6-1 Gakuen, Fukui-shi, Fukui 910-8505, Japan
Fukui University, Department of Physics, 3-9-1 Bunkyo, Fukui-shi, Fukui 910-8507, Japan
Gangneung-Wonju National University, 210-702 Gangneung Daehangno, Gangneung City, Gangwon Province, Republic of Korea
Georg-August-Universität Göttingen, II. Physikalisches Institut, Friedrich-Hund-Platz 1, 37077 Göttingen, Germany
Global Design Effort
Hanyang University, Department of Physics, Seoul 133-791, Republic of Korea
Harish-Chandra Research Institute, Chhatnag Road, Jhusi, Allahabad 211019, India
Helmholtz-Zentrum Berlin für Materialien und Energie (HZB), Wilhelm-Conrad-Röntgen Campus, BESSY II, Albert-Einstein-Str. 15, 12489 Berlin, Germany
Helsinki Institute of Physics (HIP), P.O. Box 64, FIN-00014 University of Helsinki, Finland
Henan Normal University, College of Physics and Information Engineering, Xinxiang, China 453007
High Energy Accelerator Research Organization, KEK, 1-1 Oho, Tsukuba, Ibaraki 305-0801, Japan
Hiroshima University, Department of Physics, 1-3-1 Kagamiyama, Higashi-Hiroshima, Hiroshima 739-8526, Japan
Hiroshima University, Graduate School of Advanced Sciences of Matter, 1-3-1 Kagamiyama, Higashi-Hiroshima, Hiroshima 739-8530, Japan
Hokkai-Gakuen University, 4-1-40 Asahimachi, Toyohira-ku, Sapporo 062-8605, Japan
Hokkaido University, Department of Physics, Faculty of Science, Kita, Kita-ku, Sapporo-shi, Hokkaido 060-0810, Japan
Humboldt Universität zu Berlin, Fachbereich Physik, Institut für Elementarteilchenphysik, Newtonstr. 15, D-12489 Berlin, Germany
Hyogo University of Teacher Education, 942-1 Shimokume, Kato-city, Hyogo 673-1494, Japan
Ibaraki National College of Technology, 866 Nakane, Hitachinaka, Ibaraki 312-8508, Japan
Ibaraki University, College of Technology, Department of Physics, Nakanarusawa 4-12-1, Hitachi, Ibaraki 316-8511, Japan
Imperial College, Blackett Laboratory, Department of Physics, Prince Consort Road, London, SW7 2BW, UK
Indian Association for the Cultivation of Science, Department of Theoretical Physics and Centre for Theoretical Sciences, Kolkata 700032, India
Indian Institute of Science Education and Research (IISER) Kolkata, Department of Physical Sciences, Mohanpur Campus, PO Krishi Viswavidyalaya, Mohanpur 741252, Nadia, West Bengal, India
Indian Institute of Science, Centre for High Energy Physics, Bangalore 560012, Karnataka, India
Indian Institute of Technology Indore, IET Campus, M-Block, Institute of Engineering and Technology (IET), Devi Ahilya Vishwavidyalaya Campus, Khandwa Road, Indore - 452017, Madhya Pradesh, India
Indian Institute of Technology, Bombay, Powai, Mumbai 400076, India
Indian Institute of Technology, Guwahati, Guwahati, Assam 781039, India
Indiana University, Department of Physics, Swain Hall West 117, 727 E. 3rd St., Bloomington, IN 47405-7105, USA
Institucio Catalana de Recerca i Estudis, ICREA, Passeig Lluis Companys, 23, Barcelona 08010, Spain
Institut de Physique Nucléaire de Lyon (IPNL), Domaine scientifique de la Doua, Bâtiment Paul Dirac 4, rue Enrico Fermi, 69622 Villeurbanne, Cedex, France
Institut für Experimentelle Kernphysik, KIT,Universität Karlsruhe (TH), Wolfgang-Gaede-Str. 1, Postfach 6980, 76128 Karlsruhe, Germany
Institut für Theoretische Physik (ITP), Karlsruher Institut für Technologie (KIT), Fakultät für Physik, Postfach 6980, 76049 Karlsruhe, Germany
Institut für Theoretische Teilchenphysik, Campus Süd, Karlsruher Institut für Technologie (KIT), 76128 Karlsruhe, Germany
Institut National de Physique Nucleaire et de Physique des Particules, 3, Rue Michel- Ange, 75794 Paris Cedex 16, France
Institut Pluridisciplinaire Hubert Curien, 23 Rue du Loess - BP28, 67037 Strasbourg Cedex 2, France
Institute for Chemical Research, Kyoto University, Gokasho, Uji, Kyoto 611-0011, Japan
Institute for Cosmic Ray Research, University of Tokyo, 5-1-5 Kashiwa-no-Ha, Kashiwa, Chiba 277-8582, Japan
Institute for Mathematics, Astrophysics and Particle Physics (IMAPP), P.O. Box 9010, 6500 GL Nijmegen, Netherlands
Institute for Nuclear Research, Russian Academy of Sciences (INR RAS), 60-th October Anniversary Prospect 7a, 117312, Moscow, Russia
Institute for Scintillation Materials (ISMA), 60 Lenina Ave, 61001, Kharkiv, Ukraine
Institute for studies in fundamental sciences (IPM), Niavaran Square, P.O. Box 19395-5746, Tehran, Iran
Institute of High Energy Physics - IHEP, Chinese Academy of Sciences, P.O. Box 918, Beijing, China 100049
Institute of Mathematical Sciences, Taramani, C.I.T. Campus, Chennai 600113, India
Institute of Particle Physics, Canada
Institute of Physics, ASCR, Academy of Science of the Czech Republic, Division of Elementary Particle Physics, Na Slovance 2, CZ-18221 Prague 8, Czech Republic
Institute of Physics, Vietnam Academy of Science and Technology (VAST), 10 Dao-Tan, Ba-Dinh, Hanoi 10000, Vietnam
Institute of Theoretical and Experimetal Physics, B. Cheremushkinskawa, 25, RU-117259, Moscow, Russia
Instituto de Fisica Corpuscular (IFIC), Centro Mixto CSIC-UVEG, Edificio Investigacion Paterna, Apartado 22085, 46071 Valencia, Spain
Instituto de Física da Universidade Federal do Rio Grande do Sul (UFRGS), Av. Bento Gonçalves 9500, Caixa Postal 15051, CEP 91501-970, Porto Alegre, RS, Brazil
Instituto de Fisica de Cantabria, (IFCA, CSIC-UC), Facultad de Ciencias, Avda. Los Castros s/n, 39005 Santander, Spain
Instituto de Física Teórica UAM/CSIC, C/ Nicolás Cabrera 13-15, Universidad Autónoma de Madrid, Cantoblanco, 28049 Madrid, Spain
Instituto Galego de Fisica de Altas Enerxias (IGFAE,USC) Facultad de Fisica, Campus Sur E-15782 Santiago de Compostela, Spain
Instituto Tecnológico de Aragón (ITA), C/ María de Luna 7-8, 50018 Zaragoza, Spain
Instituto Universitario de Física Fundamental y Matemáticas de la Universidad de Salamanca (IUFFyM), Casas del Parque, 37008 Salamanca, Spain
Inter-University Accelerator Centre, Aruna Asaf Ali Marg, Post Box 10502, New Delhi 110067, India
International Center for Elementary Particle Physics, University of Tokyo, Hongo 7-3-1, Bunkyo District, Tokyo 113-0033, Japan
International Institute of Physics, Federal University of Rio Grande do Norte, Av. Odilon Gomes de Lima, 1722 - Capim Macio - 59078-400 - Natal-RN, Brazil
Iowa State University, Department of Physics, High Energy Physics Group, Ames, IA 50011, USA
Istituto Nazionale di Fisica Nucleare (INFN), Laboratorio LASA, Via Fratelli Cervi 201, 20090 Segrate, Italy
Istituto Nazionale di Fisica Nucleare (INFN), Sezione di Firenze, Via G. Sansone 1, I-50019 Sesto Fiorentino (Firenze), Italy
Istituto Nazionale di Fisica Nucleare (INFN), Sezione di Milano Bicocca, Piazza della Scienza 3, I-20126 Milano, Italy
Istituto Nazionale di Fisica Nucleare (INFN), Sezione di Milano, Via Celoria 16, I-20133 Milano, Italy
Istituto Nazionale di Fisica Nucleare (INFN), Sezione di Napoli, Complesso Università di Monte Sant'Angelo,via, I-80126 Napoli, Italy
Istituto Nazionale di Fisica Nucleare (INFN), Sezione di Pavia, Via Bassi 6, I-27100 Pavia, Italy
Istituto Nazionale di Fisica Nucleare (INFN), Sezione di Perugia, Via A. Pascoli, 06123 Perugia, Italy
Istituto Nazionale di Fisica Nucleare (INFN), Sezione di Pisa, Edificio C - Polo Fibonacci Largo B. Pontecorvo, 3, I-56127 Pisa, Italy
Istituto Nazionale di Fisica Nucleare (INFN), Sezione di Roma, c/o Dipartimento di Fisica - Università degli Studi di Roma "La Sapienza", P.le Aldo Moro 2, I-00185 Roma, Italy
Istituto Nazionale di Fisica Nucleare (INFN), Sezione di Torino, c/o Università di Torino, facoltà di Fisica, via P Giuria 1, 10125 Torino, Italy
Istituto Nazionale di Fisica Nucleare (INFN), Sezione di Trieste, Padriciano 99, I-34012 Trieste (Padriciano), Italy
ITER Organization, Route de Vinon-sur-Verdon, 13115 St. Paul-lez-Durance, France
Iwate University, 4-3-5 Ueda, Morioka, Iwate, 020-8551, Japan

Jagiellonian University, Institute of Physics, Ul. Reymonta 4, PL-30-059 Cracow, Poland
Jamia Millia Islamia, Department of Physics, Jamia Nagar, New Delhi 110025, India
Japan Aerospace Exploration Agency, Sagamihara Campus, 3-1-1 Yoshinodai, Sagamihara, Kanagawa 220-8510 , Japan
Japan Atomic Energy Agency, 4-49 Muramatsu, Tokai-mura, Naka-gun, Ibaraki 319-1184, Japan
Japan Synchrotron Radiation Research Institute (JASRI), 1-1-1, Kouto, Sayo-cho, Sayo-gun, Hyogo 679-5198, Japan
Johannes Gutenberg Universität Mainz, Institut für Physik, 55099 Mainz, Germany
John Adams Institute for Accelerator Science at Royal Holloway University of London, Egham Hill, Egham, Surrey TW20 0EX, UK
John Adams Institute for Accelerator Science at University of Oxford, Denys Wilkinson Building, Keble Road, Oxford OX1 3RH, UK
Johns Hopkins University - Henry A. Rowland Department of Physics & Astronomy 3701 San Martin Drive, Baltimore, Maryland (MD) 21218, USA
Joint Institute for Nuclear Research (JINR), Joliot-Curie 6, 141980, Dubna, Moscow Region, Russia
Joint Institute for Power and Nuclear Research "Sosny" at National Academy of Sciences of Belarus, 99 Academician A.K.Krasin Str., Minsk BY-220109, Belarus
Jozef Stefan Institute, Jamova cesta 39, 1000 Ljubljana, Slovenia
Julius-Maximilians-Universität Würzburg, Fakultät für Physik und Astronomie, Am Hubland, 97074 Würzburg, Germany
Juntendo University, School of Medicine, Dept. of Physics, Hiraga-gakuendai 1-1, Inzai-shi, Chiba 270-1695, Japan
Justus-Liebig-Universität Gießen, II. Physikalisches Institut, Heinrich-Buff-Ring 16, 35392 Gießen, Germany
Kanazawa University, Institute for Theoretical Physics (KITP), School of Mathematics and Physics, College of Science and Engineering, Kakuma-machi, Kanazawa city, Ishikawa 920-1192, Japan
Kansas State University, Department of Physics, 116 Cardwell Hall, Manhattan, KS 66506, USA
Kavli Institute for the Physics and Mathematics of the Universe (Kavli IPMU), University of Tokyo, 5-1-5 Kashiwanoha, Kashiwa, 277-8583, Japan
King Saud University (KSU), Dept. of Physics, P.O. Box 2454, Riyadh 11451, Saudi Arabia
King's College London - Department of physics, Strand, London WC2R 2LS, London, UK
Kinki University, Department of Physics, 3-4-1 Kowakae, Higashi-Osaka, Osaka 577-8502, Japan
Kobe University, Department of Physics, 1-1 Rokkodai-cho, Nada-ku, Kobe, Hyogo 657-8501, Japan
Kogakuin University, Department of Physics, Shinjuku Campus, 1-24-2 Nishi-Shinjuku, Shinjuku-ku, Tokyo 163-8677, Japan
Konkuk University, 93-1 Mojin-dong, Kwanglin-gu, Seoul 143-701, Republic of Korea
Korea Advanced Institute of Science & Technology, Department of Physics, 373-1 Kusong-dong, Yusong-gu, Taejon 305-701, Republic of Korea
Korea Institute for Advanced Study (KIAS), School of Physics, 207-43 Cheongryangri-dong, Dongdaemun-gu, Seoul 130-012, Republic of Korea
Kyoto University, Department of Physics, Kitashirakawa-Oiwakecho, Sakyo-ku, Kyoto 606-8502, Japan
Kyoto University, Yukawa Institute for Theoretical Physics, Kitashirakawa-Oiwakecho, Sakyo-Ku, Kyoto 606-8502, Japan
Kyushu University, Department of Physics, 6-10-1 Hakozaki, Higashi-ku, Fukuoka 812-8581, Japan
L.P.T.A., UMR 5207 CNRS-UM2, Université Montpellier II, Case Courrier 070, Bât. 13, place Eugène Bataillon, 34095 Montpellier Cedex 5, France
Laboratoire Charles Coulomb UMR 5221 CNRS-UM2, Université Montpellier 2, Place Eugène Bataillon - CC069, 34095 Montpellier Cedex 5, France
Laboratoire d'Annecy-le-Vieux de Physique des Particules (LAPP) , Université de Savoie, CNRS/IN2P3, 9 Chemin de Bellevue, BP 110, F-74941 Annecy-Le-Vieux Cedex, France
Laboratoire d'Annecy-le-Vieux de Physique Theorique (LAPTH), Chemin de Bellevue, BP 110, F-74941 Annecy-le-Vieux Cedex, France
Laboratoire d'AstroParticules et Cosmologie (APC), Université Paris Diderot-Paris 7 - CNRS/IN2P3, Bâtiment Condorcet, Case 7020, 75205 Paris Cedex 13, France
Laboratoire de l'Accélérateur Linéaire (LAL), Université Paris-Sud 11, Bâtiment 200, 91898 Orsay, France
Laboratoire de Physique Corpusculaire de Clermont-Ferrand (LPC), Université Blaise Pascal, I.N.2.P.3./C.N.R.S., 24 avenue des Landais, 63177 Aubière Cedex, France
Laboratoire de Physique Nucléaire et des Hautes Energies (LPNHE), UPMC, UPD, IN2P3/CNRS, 4 Place Jussieu, 75005, Paris Cedex 05, France
Laboratoire de Physique Subatomique et de Cosmologie (LPSC), Université Joseph Fourier (Grenoble 1), CNRS/IN2P3, Institut Polytechnique de Grenoble, 53 rue des Martyrs, F-38026 Grenoble Cedex, France
Laboratoire de Physique Theorique, Université de Paris-Sud XI, Batiment 210, F-91405 Orsay Cedex, France
Laboratoire Leprince-Ringuet (LLR), École polytechnique – CNRS/IN2P3, Route de Saclay, F-91128 Palaiseau Cedex, France
Laboratoire Univers et Particules de Montpellier (LUPM) - UMR5299, Université de Montpellier II, Place Eugène Bataillon - Case courrier 72, 34095 Montpellier Cedex 05, France
Laboratori Nazionali di Frascati, via E. Fermi, 40, C.P. 13, I-00044 Frascati, Italy
Laboratório de Instrumentação e Física Experimental de Partículas (LIP LISBOA), Av. Elias Garcia 14 - 1°, 1000-149 Lisbon, Portugal
Lancaster University, Physics Department, Lancaster LA1 4YB, UK
Lawrence Berkeley National Laboratory (LBNL), 1 Cyclotron Rd, Berkeley, CA 94720, USA
Lawrence Livermore National Laboratory (LLNL), Livermore, CA 94551, USA
Lebedev Physical Institute, Leninsky Prospect 53, RU-117924 Moscow, Russia
Lomonosov Moscow State University, Skobeltsyn Institute of Nuclear Physics (MSU SINP), 1(2), Leninskie gory, GSP-1, Moscow 119991, Russia
Louisiana Tech University, Department of Physics, Ruston, LA 71272, USA
Ludwig-Maximilians-Universität München, Fakultät für Physik, Am Coulombwall 1, D - 85748 Garching, Germany
Lunds Universitet, Fysiska Institutionen, Avdelningen för Experimentell Högenergifysik, Box 118, 221 00 Lund, Sweden
L'Université Hassan II, Aïn Chock, "Réseau Universitaire de Physique des Hautes Energies" (RUPHE), Département de Physique, Faculté des Sciences Aïn Chock, B.P 5366 Maarif, Casablanca 20100, Morocco
Massachusetts Institute of Technology (MIT), Laboratory for Nuclear Science, 77 Massachusetts Avenue, Cambridge, MA 02139, USA
Max-Planck-Institut für Physik (Werner-Heisenberg-Institut), Föhringer Ring 6, 80805 München, Germany
McGill University, Department of Physics, Ernest Rutherford Physics Bldg., 3600 University Street, Montreal, Quebec, H3A 2T8 Canada
McMaster University, Department of Physics & Astronomy, 1280 Main Street West, Hamilton, ON, L8S 4M1, Canada
Meiji Gakuin University, Department of Physics, 2-37 Shirokanedai 1-chome, Minato-ku, Tokyo 244-8539, Japan
Michigan State University, Department of Chemical Engineering & Materials Science, 2527 Engineering Building East Lansing, MI 48824-1226, USA
Michigan State University, Department of Physics and Astronomy, East Lansing, MI 48824, USA
Middle East Technical University, Department of Physics, TR-06531 Ankara, Turkey
Miyagi Gakuin Women's University, Faculty of Liberal Arts, 9-1-1 Sakuragaoka, Aoba District, Sendai, Miyagi 981-8557, Japan
MSU-Iligan Institute of Technology, Department of Physics, Andres Bonifacio Avenue, 9200 Iligan City, Phillipines
Nagasaki Institute of Applied Science, 536 Abamachi, Nagasaki-Shi, Nagasaki 851-0193, Japan
Nagoya University, Department of Materials Science and Engineering, Furo-cho, Chikusa-ku, Nagoya, 464-8603, Japan
Nagoya University, Department of Physics, School of Science, Furo-cho, Chikusa-ku, Nagoya, Aichi 464-8602, Japan
Nagoya University, Kobayashi-Maskawa Institute for the Origin of Particles and the Universe (KMI), Furo-cho, Chikusa-ku, Nagoya Aichi 464-8602, Japan
Nanjing University, Department of Physics, Nanjing, China 210093
Nara Women's University, High Energy Physics Group, Kitauoya-Nishimachi, Nara 630-8506, Japan

National Central University, High Energy Group, Department of Physics, Chung-li, Taiwan 32001, R.O.C
National Centre of Nuclear Research (NCBJ), ul. Andrzeja Soltana 7, 05-400 Otwock-Swierk, Poland
National Cheng Kung University, Physics Department, 1 Ta-Hsueh Road, Tainan, Taiwan 70101, R.O.C
National Chiao-Tung University, Institute of Physics, 1001 Ta Hsueh Rd, Hsinchu, Taiwan 300, R.O.C.
National Institute of Chemical Physics and Biophysics (NICPB), Ravala pst 10, 10143 Tallinn, Estonia
National Institute of Physics and Nuclear Engineering "Horia Hulubei" (IFIN-HH), Str. Reactorului no.30, P.O. Box MG-6, R-76900 Bucharest - Magurele, Romania
National Research Centre "Kurchatov Institute", 1 Akademika Kurchatova pl., Moscow, 123182, Russia
National Science Center - Kharkov Institute of Physics and Technology (NSC KIPT), Akademicheskaya St. 1, Kharkov, 61108, Ukraine
National Scientific & Educational Centre of Particle & High Energy Physics (NCPHEP), Belarusian State University, M.Bogdanovich street 153, 220040 Minsk, Belarus
National Taiwan University, Physics Department, Taipei, Taiwan 106, R.O.C
Niels Bohr Institute (NBI), University of Copenhagen, Blegdamsvej 17, DK-2100 Copenhagen, Denmark
Niigata University, Department of Physics, Ikarashi, Niigata 950-218, Japan
Nikhef, National Institute for Subatomic Physics, P.O. Box 41882, 1009 DB Amsterdam, Netherlands
Nippon Dental University School of Life Dentistry at Niigata, 1-8 Hamaura-cho, Chuo-ku, Niigata 951-1500, Japan
North Carolina A&T State University, 1601 E. Market Street, Greensboro, NC 27411, USA
Northeastern University, Physics Department, 360 Huntington Ave, 111 Dana Research Center, Boston, MA 02115, USA
Northern Illinois University, Department of Physics, DeKalb, Illinois 60115-2825, USA
Northwestern University, Department of Physics and Astronomy, 2145 Sheridan Road., Evanston, IL 60208, USA
Novosibirsk State University (NGU), Department of Physics, Pirogov st. 2, 630090 Novosibirsk, Russia
Ochanomizu University, Department of Physics, Faculty of Science, 1-1 Otsuka 2, Bunkyo-ku, Tokyo 112-8610, Japan
Orissa University of Agriculture & Technology, Bhubaneswar 751003, Orissa, India
Osaka City University, Department of Physics, Faculty of Science, 3-3-138 Sugimoto, Sumiyoshi-ku, Osaka 558-8585, Japan
Osaka University, Department of Physics, 1-1 Machikaneyama, Toyonaka, Osaka 560-0043, Japan
Österreichische Akademie der Wissenschaften, Institut für Hochenergiephysik, Nikolsdorfergasse 18, A-1050 Vienna, Austria
Pacific Northwest National Laboratory, (PNNL), PO Box 999, Richland, WA 99352, USA
Panjab University, Chandigarh 160014, India
Pavel Sukhoi Gomel State Technical University, ICTP Affiliated Centre & Laboratory for Physical Studies, October Avenue, 48, 246746, Gomel, Belarus
Perimeter Institute for Theoretical Physics, 31 Caroline Street North, Waterloo, Ontario N2L 2Y5, Canada
Physical Research Laboratory, Navrangpura, Ahmedabad 380 009, Gujarat, India
Pohang Accelerator Laboratory (PAL), San-31 Hyoja-dong, Nam-gu, Pohang, Gyeongbuk 790-784, Republic of Korea
Pontificia Universidad Católica de Chile, Avda. Libertador Bernardo OHiggins 340, Santiago, Chile
Princeton University, Department of Physics, P.O. Box 708, Princeton, NJ 08542-0708, USA
Purdue University, Department of Physics, West Lafayette, IN 47907, USA
Queen Mary, University of London, Mile End Road, London, E1 4NS, United Kingdom
Rheinisch-Westfälische Technische Hochschule (RWTH), Physikalisches Institut, Physikzentrum, Otto-Blumenthal-Straße, 52056 Aachen
RIKEN, 2-1 Hirosawa, Wako, Saitama 351-0198, Japan
Royal Holloway, University of London (RHUL), Department of Physics, Egham, Surrey TW20 0EX, UK
Russian Academy of Science, Keldysh Institute of Applied Mathematics, Muiskaya pl. 4, 125047 Moscow, Russia
Rutgers, The State University of New Jersey, Department of Physics & Astronomy, 136 Frelinghuysen Rd, Piscataway, NJ 08854, USA
Saga University, Department of Physics, 1 Honjo-machi, Saga-shi, Saga 840-8502, Japan
Saha Institute of Nuclear Physics, 1/AF Bidhan Nagar, Kolkata 700064, India
Salalah College of Technology (SCOT), Engineering Department, Post Box No. 608, Postal Code 211, Salalah, Sultanate of Oman
Saudi Center for Theoretical Physics, King Fahd University of Petroleum and Minerals (KFUPM), Dhahran 31261, Saudi Arabia
Seikei University, Faculty of Science and Technology, 3-3-1 Kichijoji-Kitamachi, Musashino-shi, Tokyo 180-8633, Japan
Seoul National University, San 56-1, Shinrim-dong, Kwanak-gu, Seoul 151-742, Republic of Korea
Setsunan University, Institute for Fundamental Sciences, 17-8 Ikeda Nakamachi, Neyagawa, Osaka, 572-8508, Japan
Shanghai Jiao Tong University, Department of Physics, 800 Dongchuan Road, Shanghai, China 200240
Shinshu University, 3-1-1, Asahi, Matsumoto, Nagano 390-8621, Japan
Shiv Nadar University, Village Chithera, Tehsil Dadri, District Gautam Budh Nagar, 203207 Uttar Pradesh, India
Shizuoka University, Department of Physics, 836 Ohya, Suruga-ku, Shizuoka 422-8529, Japan
SLAC National Accelerator Laboratory, 2575 Sand Hill Road, Menlo Park, CA 94025, USA
Society for Applied Microwave Electronics Engineering and Research (SAMEER), I.I.T. Campus, Powai, Post Box 8448, Mumbai 400076, India
Sokendai, The Graduate University for Advanced Studies, Shonan Village, Hayama, Kanagawa 240-0193, Japan
Spallation Neutron Source (SNS), Oak Ridge National Laboratory (ORNL), P.O. Box 2008 MS-6477, Oak Ridge, TN 37831-6477, USA
State University of New York at Binghamton, Department of Physics, PO Box 6016, Binghamton, NY 13902, USA
State University of New York at Buffalo, Department of Physics & Astronomy, 239 Franczak Hall, Buffalo, NY 14260, USA
State University of New York at Stony Brook, Department of Physics and Astronomy, Stony Brook, NY 11794-3800, USA
STFC Daresbury Laboratory, Daresbury, Warrington, Cheshire WA4 4AD, UK
STFC Rutherford Appleton Laboratory, Chilton, Didcot, Oxon OX11 0QX, UK
Sungkyunkwan University (SKKU), Natural Science Campus 300, Physics Research Division, Chunchun-dong, Jangan-gu, Suwon, Kyunggi-do 440-746, Republic of Korea
Swiss Light Source (SLS), Paul Scherrer Institut (PSI), PSI West, CH-5232 Villigen PSI, Switzerland
Tata Institute of Fundamental Research, School of Natural Sciences, Homi Bhabha Rd., Mumbai 400005, India
Technical University of Lodz, Department of Microelectronics and Computer Science, al. Politechniki 11, 90-924 Lodz, Poland
Technische Universität Dresden, Institut für Kern- und Teilchenphysik, D-01069 Dresden, Germany
Tel-Aviv University, School of Physics and Astronomy, Ramat Aviv, Tel Aviv 69978, Israel
Texas A&M University, Physics Department, College Station, 77843-4242 TX, USA
The Henryk Niewodniczanski Institute of Nuclear Physics, Polish Academy of Sciences (IFJ PAN), ul. Radzikowskiego 152, PL-31342 Cracow, Poland
Thomas Jefferson National Accelerator Facility (TJNAF), 12000 Jefferson Avenue, Newport News, VA 23606, USA
Tohoku Gakuin University, Department of Business Administration, 1-3-1 Tsuchitoi, Aoba-ku Sendai, Miyagi 980-8511, Japan
Tohoku Gakuin University, Faculty of Technology, 1-13-1 Chuo, Tagajo, Miyagi 985-8537, Japan
Tohoku University, Department of Physics, Aoba District, Sendai, Miyagi 980-8578, Japan
Tohoku University, Research Center for Electron Photon Science, Taihaku District, Sendai, Miyagi 982-0826, Japan
Tohoku University, Research Center for Neutrino Science, Aoba District, Sendai, Miyagi 980-8578, Japan
Tokyo Institute of Technology, Department of Physics, 2-12-1 O-Okayama, Meguro, Tokyo 152-8551, Japan
Tokyo Metropolitan University, Faculty of Science and Engineering, Department of Physics, 1-1 Minami-Osawa, Hachioji-shi, Tokyo 192-0397, Japan
Tokyo University of Agriculture Technology, Department of Applied Physics, Naka-machi, Koganei, Tokyo 183-8488, Japan
Toyama Prefectural University, Department of Mathematical Physics, 5180 Kurokawa Imizu-shi, Toyama, 939-0398, Japan
TRIUMF, 4004 Wesbrook Mall, Vancouver, BC V6T 2A3, Canada

Universidad Autónoma de San Luis Potosí, Alvaro Obregon 64, Col. Centro, San Luis Potosí, S.L.P. 78000, México
Universidad de Granada, Departamento de Física Teórica y del Cosmos, Campus de Fuentenueva, E-18071 Granada, Spain
Universidad de los Andes, Faculty of Science, Department of Physics, Carrera 1 18A-10, Bloque Ip. Bogotá, Colombia
Universidad de Oviedo, Departamento de Física, Campus de Llamaquique. C/ Calvo Sotelo, s/n 33005 Oviedo, Spain
Universidad de Salamanca, Departamento de Física Fundamental, Plaza de la Merced, s/n., 37008 Salamanca, Spain
Universidad de Sevilla, Escuela Técnica Superior de Ingeniería, Departamento Ingeniería Electrónica, Camino de los Descubrimientos s/n, 41092 Sevilla, Spain
Universidad de Zaragoza - Departamento de Física Teórica, Pedro Cerbuna 12, E-50009 Zaragoza, Spain
Universidad Nacional Autónoma de México, Instituto de Física, Circuito de la Investigación Cientifica s/n, Ciudad Universitaria, CP 04510 México D.F., Mexico
Universidad Nacional de La Plata, Departamento de Física, Facultad de Ciencias Exactas, C.C. N 67, 1900 La Plata, Argentina
Universidade do Estado do Rio de Janeiro (UERJ), Rio de Janeiro, RJ - Brasil 20550-900, Brazil
Universidade Federal de Pelotas, Instituto de Física e Matemática, Campus Universitário, Caixa Postal 354, 96010-900 Pelotas, RS, Brazil
Universidade Federal do Rio de Janeiro (UFRJ), Instituto de Física, Av. Athos da Silveira Ramos 149, Centro de Tecnologia - Bloco A, Cidade Universitária, Ilha do Fundão, Rio de Janeiro, RJ, Brazil
Universitá degli Studi di Napoli "Federico II", Dipartimento di Fisica, Via Cintia, 80126 Napoli, Italy
Universitat Autònoma de Barcelona, Departament de Física, Edifici C, 08193 Bellaterra, Barcelona, Spain
Universitat Autònoma de Barcelona, Institut de Fisica d'Altes Energies (IFAE), Campus UAB, Edifici Cn, E-08193 Bellaterra, Barcelona, Spain
Universität Bonn, Physikalisches Institut, Nußallee 12, 53115 Bonn, Germany
Universität Heidelberg, Institut für Theoretische Physik, Philosophenweg 16, 69120 Heidelberg, Germany
Universität Heidelberg, Kirchhoff-Institut für Physik, Im Neuenheimer Feld 227, 69120 Heidelberg, Germany
Universitat Politècnica de Catalunya, Institut de Tècniques Energètiques, Campus Diagonal Sud, Edifici PC (Pavelló C). Av. Diagonal, 647 08028 Barcelona, Spain
Universitat Ramon Llull, La Salle, C/ Quatre Camins 2, 08022 Barcelona, Spain
Universität Rostock, 18051 Rostock, Germany
Universität Siegen, Naturwissenschaftlich-Technische Fakultät, Department Physik, Emmy Noether Campus, Walter-Flex-Str.3, 57068 Siegen, Germany
Universität Wien - Theoretische Physik Boltzmanngasse 5, A-1090 Vienna, Austria
Université catholique de Louvain, Centre for Cosmology, Particle Physics and Phenomenology (CP3), Institute of Mathematics and Physics, 2 Chemin du Cyclotron, 1348 Louvain-la-Neuve, Belgium
Université de Montréal, Département de Physique, Groupe de Physique des Particules, C.P. 6128, Succ. Centre-ville, Montréal, Qc H3C 3J7, Canada
Université de Strasbourg, UFR de Sciences Physiques, 3-5 Rue de l'Université, F-67084 Strasbourg Cedex, France
Université Libre de Bruxelles, Boulevard du Triomphe, 1050 Bruxelles, Belgium
Universittà di Catania, Dipartimento di Fisica e Astronomia, Via Santa Sofia 64, 95123 Catania, Italy
University College London (UCL), High Energy Physics Group, Physics and Astronomy Department, Gower Street, London WC1E 6BT, UK
University College, National University of Ireland (Dublin), Department of Experimental Physics, Science Buildings, Belfield, Dublin 4, Ireland
University de Barcelona, Facultat de Física, Av. Diagonal, 647, Barcelona 08028, Spain
University of Alberta - Faculty of Science, Department of Physics, 4-181 CCIS, Edmonton AB T6G 2E1, Canada
University of Arizona, Department of Physics, 1118 E. Fourth Street, PO Box 210081, Tucson, AZ 85721, USA
University of Bergen, Institute of Physics, Allegaten 55, N-5007 Bergen, Norway
University of Birmingham, School of Physics and Astronomy, Particle Physics Group, Edgbaston, Birmingham B15 2TT, UK
University of Bristol, H. H. Wills Physics Lab, Tyndall Ave., Bristol BS8 1TL, UK
University of British Columbia, Department of Physics and Astronomy, 6224 Agricultural Rd., Vancouver, BC V6T 1Z1, Canada
University of California (UCLA), Los Angleles, CA 90095, US
University of California Berkeley, Department of Physics, 366 Le Conte Hall, #7300, Berkeley, CA 94720, USA
University of California Davis, Department of Physics, One Shields Avenue, Davis, CA 95616-8677, USA
University of California Irvine, Department of Physics and Astronomy, High Energy Group, 4129 Frederick Reines Hall, Irvine, CA 92697-4575 USA
University of California Santa Cruz, Institute for Particle Physics, 1156 High Street, Santa Cruz, CA 95064, USA
University of Cambridge, Cavendish Laboratory, J J Thomson Avenue, Cambridge CB3 0HE, UK
University of Colorado at Boulder, Department of Physics, 390 UCB, University of Colorado, Boulder, CO 80309-0390, USA
University of Cyprus, Department of Physics, P.O.Box 20537, 1678 Nicosia, Cyprus
University of Delhi, Department of Physics and Astrophysics, Delhi 110007, India
University of Delhi, S.G.T.B. Khalsa College, Delhi 110007, India
University of Dundee, Department of Physics, Nethergate, Dundee, DD1 4HN, Scotland, UK
University of Edinburgh, School of Physics, James Clerk Maxwell Building, The King's Buildings, Mayfield Road, Edinburgh EH9 3JZ, UK
University of Florida, Department of Physics, Gainesville, FL 32611, USA
University of Ghent, Department of Subatomic and Radiation Physics, Proeftuinstraat 86, 9000 Gent, Belgium
University of Glasgow, SUPA, School of Physics & Astronomy, University Avenue, Glasgow G12 8QQ, Scotland, UK
University of Hamburg, Physics Department, Luruper Chaussee 149, 22761 Hamburg, Germany
University of Hawaii, Department of Physics and Astronomy, HEP, 2505 Correa Rd., WAT 232, Honolulu, HI 96822-2219, USA
University of Helsinki, Department of Physical Sciences, P.O. Box 64 (Vaino Auerin katu 11), FIN-00014, Helsinki, Finland
University of Illinois at Chicago, Department Of Physics, 845 W Taylor St., Chicago IL 60607, USA
University of Iowa, Department of Physics and Astronomy, 203 Van Allen Hall, Iowa City, IA 52242-1479, USA
University of Kansas, Department of Physics and Astronomy, Malott Hall, 1251 Wescoe Hall Drive, Room 1082, Lawrence, KS 66045-7582, USA
University of Liverpool, Department of Physics, Oliver Lodge Lab, Oxford St., Liverpool L69 7ZE, UK
University of Liverpool, Division of Theoretical Physics, Department of Mathematical Sciences, Chadwick Building, Liverpool L69 3BX, UK
University of Ljubljana, Faculty of Mathematics and Physics, Jadranska ulica 19, 1000 Ljubljana, Slovenia
University of Lodz, Faculty of Physics and Applied Informatics, Pomorska 149/153, PL-90-236 Lodz, Poland
University of Malaya, Faculty of Science, Department of Physics, 50603 Kuala Lumpur, Malaysia
University of Manchester, School of Physics and Astronomy, Schuster Lab, Manchester M13 9PL, UK
University of Maribor, Faculty of Chemistry and Chemical Engineering (FKKT), Smetanova ulica 17, 2000 Maribor, Slovenia
University of Maryland, Department of Physics and Astronomy, Physics Building (Bldg. 082), College Park, MD 20742, USA
University of Massachusetts - Amherst, Department of Physics, 1126 Lederle Graduate Research Tower (LGRT), Amherst, MA 01003-9337, USA
University of Melbourne, School of Physics, Victoria 3010, Australia
University of Michigan, Department of Physics, 500 E. University Ave., Ann Arbor, MI 48109-1120, USA
University of Minnesota, 148 Tate Laboratory Of Physics, 116 Church St. S.E., Minneapolis, MN 55455, USA
University of Mississippi, Department of Physics and Astronomy, 108 Lewis Hall, PO Box 1848, Oxford, Mississippi 38677-1848, USA

University of Missouri – St. Louis, Department of Physics and Astronomy, 503 Benton Hall One University Blvd., St. Louis Mo 63121, USA
University of New Mexico, New Mexico Center for Particle Physics, Department of Physics and Astronomy, 800 Yale Boulevard N.E., Albuquerque, NM 87131, USA
University of North Carolina at Chapel Hill, Department of Physics and Astronomy, Phillips Hall, CB #3255, 120 E. Cameron Ave., Chapel Hill, NC 27599-3255, USA
University of Notre Dame, Department of Physics, 225 Nieuwland Science Hall, Notre Dame, IN 46556, USA
University of Oklahoma, Department of Physics and Astronomy, Norman, OK 73071, USA
University of Oregon, Department of Physics, 1371 E. 13th Ave., Eugene, OR 97403, USA
University of Oxford, Particle Physics Department, Denys Wilkinson Bldg., Keble Road, Oxford OX1 3RH England, UK
University of Pavia, Department of Physics, via Bassi 6, I-27100 Pavia, Italy
University of Pennsylvania, Department of Physics and Astronomy, 209 South 33rd Street, Philadelphia, PA 19104-6396, USA
University of Pittsburgh, Department of Physics and Astronomy, 100 Allen Hall, 3941 O'Hara St, Pittsburgh PA 15260, USA
University of Regina, Department of Physics, Regina, Saskatchewan, S4S 0A2 Canada
University of Rochester, Department of Physics and Astronomy, Bausch & Lomb Hall, P.O. Box 270171, 600 Wilson Boulevard, Rochester, NY 14627-0171 USA
University of Science and Technology of China, Department of Modern Physics (DMP), Jin Zhai Road 96, Hefei, China 230026
University of Silesia, Institute of Physics, Ul. Uniwersytecka 4, PL-40007 Katowice, Poland
University of South Carolina, Department of Physics and Astronomy, 712 Main Street, Columbia, SC 29208, USA
University of Southampton, School of Physics and Astronomy, Highfield, Southampton S017 1BJ, England, UK
University of Southern California, Department of Physics & Astronomy, 3620 McClintock Ave., SGM 408, Los Angeles, CA 90089-0484, USA
University of Sydney, Falkiner High Energy Physics Group, School of Physics, A28, Sydney, NSW 2006, Australia
University of Tartu, Institute of Physics, Riia 142, 51014 Tartu, Estonia
University of Texas at Austin, Department of Physics, 1 University Station C1600, Austin, Texas 78712, USA
University of Texas at Dallas, Department of Physics, 800 West Campbell Road, Richardson, Texas 75080, USA
University of Texas, Center for Accelerator Science and Technology, Arlington, TX 76019, USA
University of Tokushima, Institute of Theoretical Physics, Tokushima-shi 770-8502, Japan
University of Tokyo, Department of Physics, 7-3-1 Hongo, Bunkyo District, Tokyo 113-0033, Japan
University of Toronto, Department of Physics, 60 St. George St., Toronto M5S 1A7, Ontario, Canada
University of Toyama, Department of Physics, 3190 Gofuku, Toyama 930-8555, Japan
University of Tsukuba, Faculty of Pure and Applied Sciences, 1-1-1 Ten'nodai, Tsukuba, Ibaraki 305-8571, Japan
University of Victoria, Department of Physics and Astronomy, P.O.Box 3055 Stn Csc, Victoria, BC V8W 3P6, Canada
University of Virginia, Department of Physics, 382 McCormick Rd., PO Box 400714, Charlottesville, VA
University of Warsaw, Institute of Experimental Physics, Ul. Hoza 69, PL-00 681 Warsaw, Poland
University of Warsaw, Institute of Theoretical Physics, Ul. Hoza 69, PL-00 681 Warsaw, Poland
University of Washington, Department of Physics, PO Box 351560, Seattle, WA 98195-1560, USA
University of Wisconsin, Physics Department, Madison, WI 53706-1390, USA
University of Wuppertal, Gaußstraße 20, D-42119 Wuppertal, Germany
Università degli Studi di Milano, Dipartimento di Fisica, Via Celoria 16, 20133 Milano, Italy
Università degli Studi di Roma "La Sapienza", Dipartimento di Fisica, Piazzale Aldo Moro 5, 00185 Roma, Italy
Università degli Studi di Trieste, Dipartimento di Fisica, via A. Valerio 2, I-34127 Trieste, Italy
Università dell'Insubria in Como, Dipartimento di Scienze CC.FF.MM., via Vallegio 11, I-22100 Como, Italy
Università di Milano-Bicocca, Dipartimento di Fisica"G. Occhialin", Piazza della Scienza 3, 20126 Milano, Italy
Università di Pisa, Departimento di Fisica "Enrico Fermi", Largo Bruno Pontecorvo 3, I-56127 Pisa, Italy
Universiy of Huddersfield, International Institute for Accelerator Applications, Queensgate Campus, Huddersfield HD1 3DH, UK
UPMC Univ. Paris 06, Faculté de Physique (UFR 925), 4 Place Jussieu, 75252 Paris Cedex 05, France
Vietnam National University, Laboratory of High Energy Physics and Cosmology, Faculty of Physics, College of Science, 334 Nguyen Trai, Hanoi, Vietnam
Vietnam National University, University of Natural Sciences, 227 Nguyen Van Cu street, District 5, Ho Chi Minh City, Vietnam
VINCA Institute of Nuclear Sciences, Laboratory of Physics, PO Box 522, YU-11001 Belgrade, Serbia
Virginia Commonwealth University, Department of Physics, P.O. Box 842000, 701 W. Grace St.,Richmond, VA. 23284-2000, USA
Virginia Polytechnic Institute and State University, Physics Department, Blacksburg, VA 2406, USA
Vrije Universiteit Brussel, Pleinlaan 2, 1050 Brussels, Belgium
Vrije Universiteit, Department of Physics, Faculty of Sciences, De Boelelaan 1081, 1081 HV Amsterdam, Netherlands
Warsaw University of Technology, The Faculty of Electronics and Information Technology, ul. Nowowiejska 15-19, 00-665 Warsaw, Poland
Waseda University, Advanced Research Institute for Science and Engineering, Shinjuku, Tokyo 169-8555, Japan
Wayne State University, Department of Physics, Detroit, MI 48202, USA
Weizmann Institute of Science, Department of Particle Physics, P.O. Box 26, Rehovot 76100, Israel
Yale University, Department of Physics, New Haven, CT 06520, USA
Yamagata University, 1-4-12 Kojirakawa-cho, Yamagata-shi, Yamagata, 990-8560, Japan
Yerevan Physics Institute, 2 Alikhanyan Brothers St., Yerevan 375036, Armenia
Yonsei University, Department of Physics, 134 Sinchon-dong, Sudaemoon-gu, Seoul 120-749, Republic of Korea

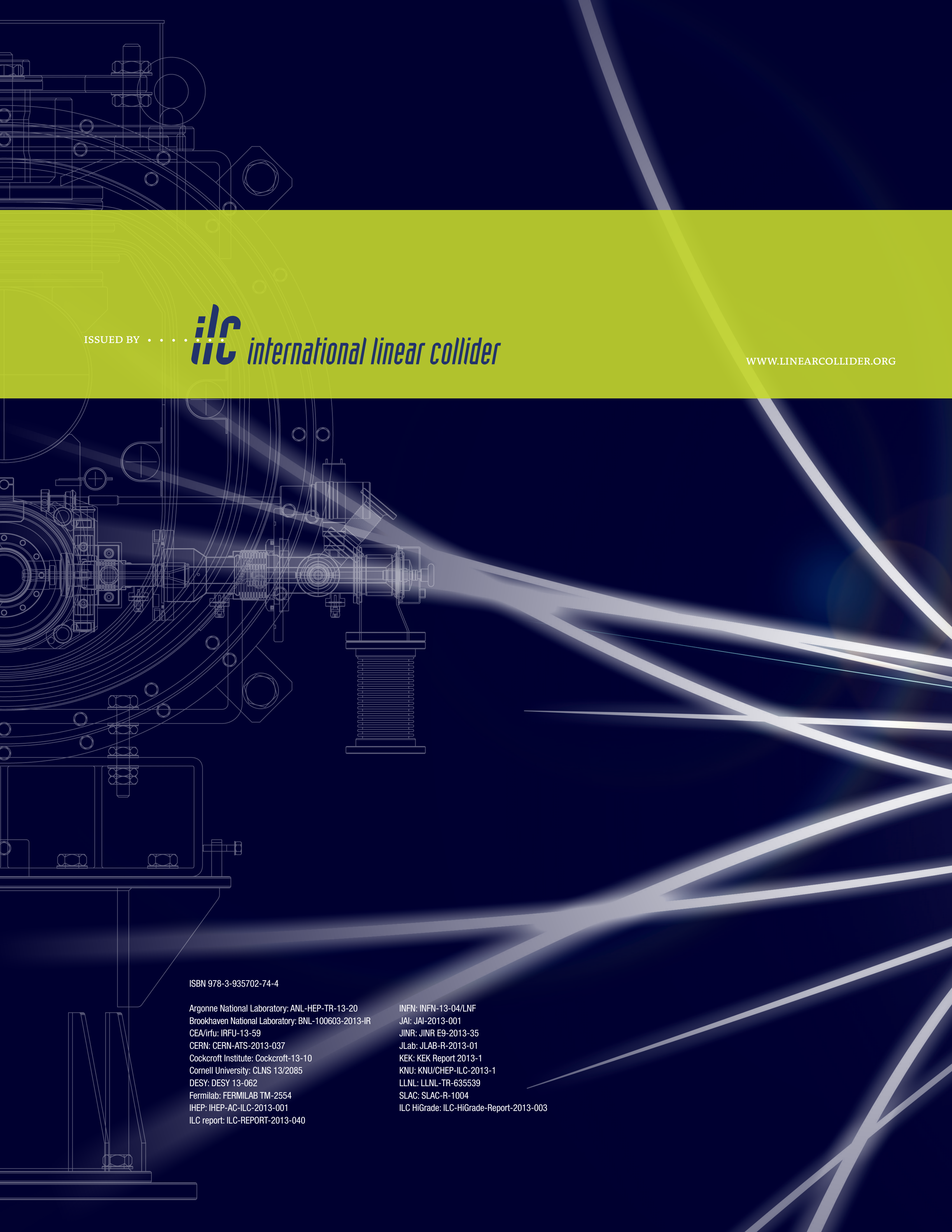

ISSUED BY • • • • • • • ilc international linear collider

WWW.LINEARCOLLIDER.ORG